\documentclass[draft]{agujournal2019}
\usepackage{url} 
\usepackage{lineno}
\usepackage{flafter}
\usepackage{rotating}
\usepackage[inline]{trackchanges} 
\usepackage{soul}
\usepackage{appendix}
\usepackage{array}
\usepackage{multirow}
\usepackage{amsfonts}
\usepackage{amsmath}
\usepackage{amssymb}
\usepackage{xr}
\DeclareMathOperator{\arcsinh}{arcsinh}
\usepackage{todonotes}

\newcommand{\Shmax}{S{\!}H_\mathrm{max}}
\newcommand{\Shmin}{S{\!}H_\mathrm{min}}
\newcommand{\Mw}{M_\mathrm{w}}

\draftfalse

\journalname{JGR: Solid Earth}

\begin{document}

\title{Rupture Dynamics of Cascading Earthquakes in a Multiscale Fracture Network}

\authors{Kadek Hendrawan Palgunadi\affil{1}, Alice-Agnes Gabriel\affil{2,4}, Dmitry Igor Garagash\affil{3}, Thomas Ulrich\affil{4}, Paul Martin Mai\affil{1}}

\affiliation{1}{Physical Science and Engineering, King Abdullah University of Science and Technology, Thuwal, Saudi Arabia}
\affiliation{2}{Institute of Geophysics and Planetary Physics, Scripps Institution of Oceanography, University of California, San Diego, CA, USA}
\affiliation{3}{Dalhousie University, Department Civil Resource Engineering, Halifax, Canada}
\affiliation{4}{Department of Earth and Environmental Sciences, Geophysics, Ludwig-Maximilians-Universit\"{a}t M\"{u}nchen, Munich, Germany}
\correspondingauthor{Authors}{kadek.palgunadi@kaust.edu.sa}

\begin{keypoints} 
\item We perform 3D dynamic rupture simulations in a network of $ >800$ fractures, a listric main fault, assuming scale-dependent fracture energy.
\item Cascading rupture is possible under high fracture connectivity if a subset of fractures is favorably oriented with respect to the ambient stress.
\item We demonstrate the feasibility of dynamic rupture cascade irrespective of hypocenter locations within the fracture network.
\end{keypoints}

\begin{abstract} 
Fault-damage zones comprise multiscale fracture networks that may slip dynamically and interact with the main fault during earthquake rupture. Using 3D dynamic rupture simulations and scale-dependent fracture energy, we examine dynamic interactions of more than 800 intersecting multiscale fractures surrounding a listric fault, emulating a major fault and its damage zone. We investigate ten distinct orientations of maximum horizontal stress, probing the conditions necessary for sustained slip within the fracture network or activating the main fault. Additionally, we assess the feasibility of nucleating dynamic rupture earthquake cascades from a distant fracture and investigate the sensitivity of fracture network cascading rupture to the effective normal stress level. We model either pure cascades or main fault rupture with limited off-fault slip. We find that cascading ruptures within the fracture network are dynamically feasible under certain conditions, including: (i) the state-evolutional distance scales with fracture and fault size, (ii) favorable relative pre-stress of fractures within the ambient stress field, and (iii) close proximity of fractures. We find that cascading rupture within the fracture network discourages rupture on the main fault. Our simulations suggest that favorable relative pre-stress fractures within a fault damage zone may lead to cascading earthquake rupture reaching off-fault moment magnitudes up to $\Mw \approx 5.6$, shadowing the main fault slip. Our findings offer fundamental insights into physical processes governing cascading earthquake dynamic rupture within multiscale fracture networks. Our results have implications for the seismic hazard of naturally activated fracture or fault networks and earthquakes induced in geo-energy exploitation activities.
\end{abstract}

\section*{Plain Language Summary} 
Large geological faults are surrounded by many small fractures of different sizes and orientations, forming a fracture network around the main fault. The characteristics of an earthquake and its size may depend on the orientation of the main fault (and surrounding fractures) within the ambient stress field. Can a small (initial) rupture within a fracture network start a domino-like (cascading) earthquake across the entire fault network? How does the compounded earthquake process depend on fracture orientation, local stress, and rupture starting point?
To address these questions, we study earthquake rupture physics in a complex fault-zone model comprising over 800 multiscale fracture planes surrounding a main fault.
Using 3D dynamic rupture simulations and supercomputing, we explore how different ambient stress orientations, earthquake hypocenters, and levels of fault loading affect earthquake dynamics in fracture networks. The simulations demonstrate that cascading rupture in a fault zone is possible under certain conditions. Our results provide important insights into the physics of cascading earthquakes in multiscale fracture systems and hence are important for advancing seismic hazard assessment for both natural and induced earthquakes.

\section{Introduction}
\begin{figure}
    \centering
    \includegraphics[width=0.6\textwidth]{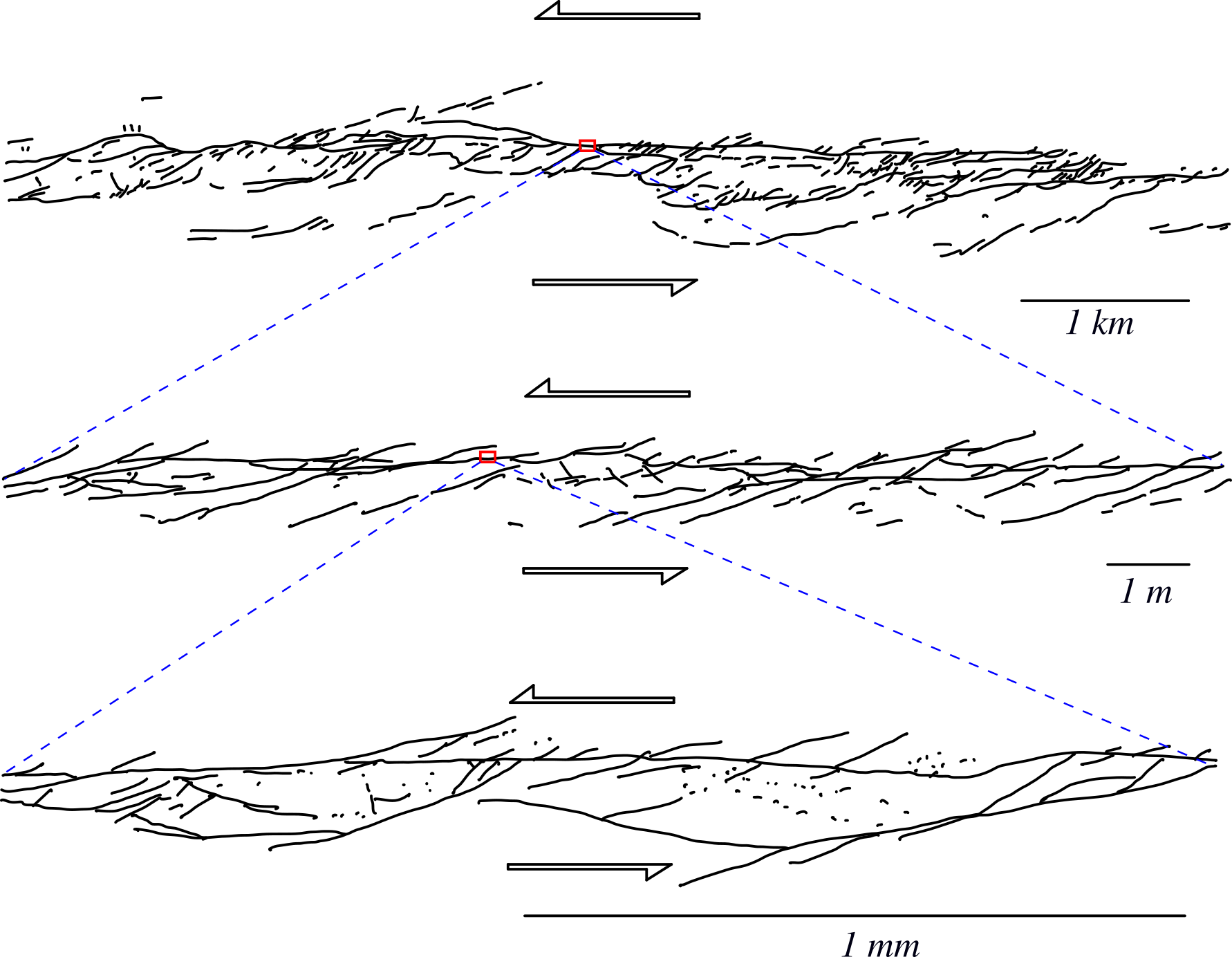}
    \caption{Schematic view of multiscale fault structure and its associated damage zone that accommodates continued deformation in strike-slip faulting (modified after \citeA{tchalenko1970similarities}).}
    \label{fig:fault_complexity}
\end{figure}

Faults zones are geometrically complex deformation zones that may consist of intersecting segments and fractures at multiple scales (Figure \ref{fig:fault_complexity}). They are constituted of nested low- and high-strain surfaces, forming a fault core surrounded by a damage zone that includes subsidiary faults, highly fractured material, and distributed macro-fractures \cite{faulkner2010review}. \citeA{chester1986implications} proposed a fault-zone model based on field observations of the Punchbowl fault, Southern California, with a damage zone generated by a combination of co-seismic events \cite{chester1993internal, mitchell2009nature, mitchell2012towards} and aseismic/quasi-static processes \cite{faulkner2011scaling, griffith2012coseismic}. The damage zone comprises brittle, heterogeneous, anisotropic, discontinuous materials, and multiscale fractures oriented in different strike and dip directions \cite{schulz2000mesoscopic, faulkner2010review}. 
These fracture networks may have formed under different tectonic stress conditions in the past. The ensemble of multiscale fractures of different ages and orientations generates strong spatial variations in mechanical properties \cite{ostermeijer2020damage}. The properties of the fault-damage zone play an essential role in fault mechanics and earthquake rupture dynamics, which in turn generates co-seismic off-fault damage and geometrical complexity \cite{andrews2005rupture, dunham2011earthquake, gabriel2013source, okubo2019dynamics,cappa2014off}. However, dynamic source characteristics of earthquake rupture within a fracture network are largely unexplored. 

Several fault segments may rupture in a single large earthquake, like during the  1992 $\Mw$ 7.3 Landers,  the 2012 $\Mw$ 8.6 off-shore Sumatra, the 2016 $\Mw$ 7.8 Kaik{\=o}ura, or the 2019 $\Mw$ 6.4, and $\Mw$ 7.1 Ridgecrest earthquakes. For large magnitude multi-segment earthquakes ($\Mw > 7$), dynamic rupture branching or ``jumping'' across fault segments has been inferred from observations or in simulations (e.g., \citeA{hauksson19931992, harris1993dynamics, oglesby2008rupture, meng2012earthquake, wollherr2019landers, ross2019hierarchical, taufiqurrahman2023dynamics}). The characteristics of such cascading multi-segment rupture include relatively closely spaced segments \cite{harris1993dynamics}, some fault segments being optimally oriented and/or  critically stressed within the ambient stress field (e.g., \citeA{ulrich2019dynamic}). 
For example, rupture nucleation on a favorably oriented fault segment may create a cascading earthquake, while nucleation on a less favorably oriented segment may lead to early rupture arrest (e.g., \citeA{oglesby2012fault, kyriakopoulos2019dynamic, lozos2020dynamic}). In this process, co-seismic stress transfer plays a pivotal role, altering the imminent stress conditions of adjacent fault segments by decreasing or increasing local stresses. Correspondingly, further rupture propagation may be promoted, or impeded by pre-stress fault conditions, depending on fault orientation within the ambient stress field.

In this study, we explore smaller-scale cascading earthquake ruptures, for example, in fracture networks of a fault damage zone or forming a geo-reservoir. Generally, a fracture network comprises more than one fracture population; each population of fractures is then characterized by its size distribution and its dominant strike direction, which depends on the stress orientations and tectonic conditions at the time of fracture formation \cite{zoback2019unconventional}. The prescribed fracture size, density, strike, and dip orientation can be described statistically \cite{dershowitz2019fracman} and constrained from observations \cite{mitchell2009nature, savage2011collateral}.

In particular, we investigate the potential for cascading earthquakes on multiscale fracture networks with a very large number of fractures. Instead of considering only several fault segments (as in the above-mentioned studies for $\Mw > 7$), we include in our simulations over 800 intersected fractures of different sizes surrounding a listric main fault. 
We examine how dynamic earthquake rupture cascades may occur in such a fracture network if a subset of fractures is favorably oriented. We also explore mechanisms by which a rupture cascade starting on a fracture may promote or suppress rupture on the main fault. Pre-stress conditions may or may not lead to a sustained dynamic rupture cascade within the fracture network, or to the activation of the main fault. Therefore, we investigate dynamic rupture processes within a complex fracture network under varying pre-stress conditions, rupture nucleation points, and levels of fluid overpressure. To facilitate comparisons with observations, we then analyze earthquake source parameters (i.e., magnitude, centroid moment tensor, stress drop, rupture speed) and general aspects of seismic wave radiation.

\section{Model Setup}\label{modeling_section}
We model dynamic earthquake rupture scenarios with a principal listric fault and many off-fault fractures. Our model represents a typical fault structure in a sedimentary basin near a subsurface reservoir (Figure \ref{fig:1}a; e.g., \citeA{gibbs1984clyde, hardman1991significance,  ward2016reservoir}), as we are also interested in examining general aspects of induced earthquakes in geo-reservoirs. In deep sedimentary basins, listric faults often extend to basement depth and form traps for natural resources near geological reservoirs (e.g., \citeA{withjack2002rift, onajite2013seismic, dixon2019geological}).

Dynamic rupture simulations are governed by initial stress conditions, including fault strength, fault and fracture geometry, subsurface material properties, the nucleation procedure, and the friction law. In this study, we vary the direction of maximum horizontal stress and constrain other parameters from observations. In the following, we explain how we construct the fracture network, considering both field observations and statistical estimates. We summarize the applied material and fault properties and how the initial stress and fault strength conditions are specified.

\subsection{Fracture Network Construction}\label{modeling_setup}
Numerous studies have characterized the fault damage zone focusing on its width and fracture distribution, as these are critical for fluid transport, naturally in the context of fault-valving effects and fault healing, but also in industrial applications like oil $\&$ gas exploration, geothermal production, or waste-water injection \cite{rice1992fault,faulkner2010review,faulkner2011scaling}. Numerical simulations and field observations suggest that the formation of off-fault fractures is affected by factors like fault geometry and co-seismic displacement, tectonic environment and ambient stress state \cite{faulkner2011scaling, okubo2019dynamics, wu2019fracture, gabriel2021unified, sainoki2021numerical}. A key property of off-fault fractures is that their density decreases with increasing distance from the fault core, typically following a power-law scaling that depends on long-term stress evolution, rock type, and fault maturity \cite{sainoki2021numerical}. 

\begin{figure}
    \centering
    \includegraphics[width=0.8\textwidth]{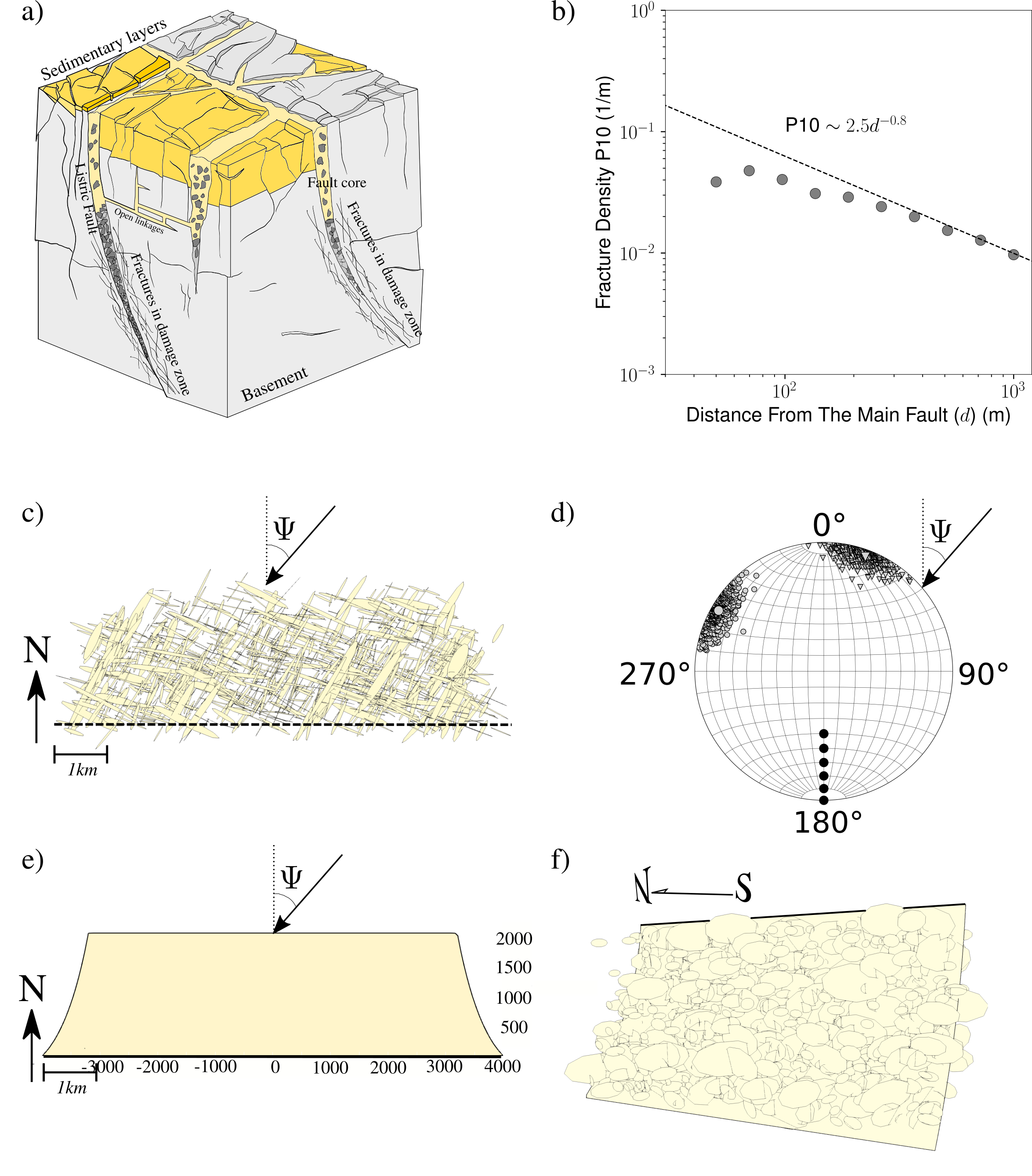}
    \caption{Geometric constraints on the 3D dynamic rupture model setup including 854 multiscale fractures and a listric main fault. a) Geological interpretation of fractures in a damage zone near a listric fault (modified from \citeA{mccaffrey2020basement}). 
    b) Measured fracture density at varying distances from the main fault (grey dots) compared with constraint from field observation (dashed black line) \cite{savage2011collateral}. c) Spatial distribution of our fracture network in map view. The black dashed line represents the shallow top of the main fault, while the black arrow indicates the direction $\Psi$ of maximum horizontal stress measured clockwise relative to North. d) Polar view of fractures and the main fault in a lower hemisphere projection (based on local fault normal orientation). Grey dots and triangles represent two different fracture families, striking on average N120E (dip $84^\circ$, grey triangles) and N20E (dip $84^\circ$, big grey dots). Black dots illustrate the depth-dependent dip of the listric fault. e) Map view of the listric fault, with the thick black line indicating the top of the fault. f) Perspective view of the listric fault geometry surrounded by 854 multiscale fractures. The view direction is east-south-east (ESE). The thick black line indicates the top of the listric fault. }
    \label{fig:1}
\end{figure}

Using a statistical approach, we generate two approximately conjugate fracture families to achieve a high degree of fracture connectivity (henceforth referred to as fracture family 1 and family 2). The conceptual geometry is illustrated in Figure \ref{fig:1}a. The fracture network is generated using the commercial software FRACMAN \cite{dershowitz2019fracman}, whereby we consider four key quantitative properties: fracture density, fracture size distribution, fracture orientation distribution, and fracture shape. We discuss these properties and our assumptions in the following section, but state already here that the resulting fracture network comprises in total 854 fracture surfaces, with 423 and 431 fracture surfaces in family 1 and 2, respectively. $78\%$ of all fractures are connected to more than one other fracture, $3\%$ fractures are connected to only one other fracture, and $19\%$ are unconnected. 

\subsubsection{Fracture Density}
Field observations suggest a power law decay of fracture density with the distance from the fault core \cite{mitchell2009nature,savage2011collateral}. They use the number of fractures per unit length ($P_{10}$) to quantify fracture density. However, $P_{10}$ is inefficient for describing a volumetric fracture density distribution because it only specifies fracture distribution along one line. To overcome this limitation, we adopt fracture density based on $P_{32}$, defined as the ratio between the area of all fractures and the rock volume. However, observing $P_{32}$ in nature is difficult. We thus constrain $P_{32}$ using an inferred $P_{10}$ value based on a multi-dimensional intensity metric \cite{wang2005stereological}. The particular relation between $P_{32}$ and $P_{10}$ is implemented in FRACMAN, which we apply to generate the fracture network. To validate the $P_{32}$-constrained fracture network, we measure $P_{10}$ of the modeled fracture network using an average number from linear transects. The such defined average fracture density $P_{10}$ in our model is well approximated by a power law decay,  with higher fracture density near the main fault, and with a field-observation-consistent decay exponent $m$ (Figure \ref{fig:1}b) as

\begin{equation}
    \label{eq:P10}
    P_{10} = C d^{-m} \,.
\end{equation}
The exponential decay follows $m = 0.8$ and the fault-specific constant is $C = 2.5$, whereby $d$ is the fracture distance  \cite{savage2011collateral}. The nearest neighbor method is used to populate the volumetric fault zone with fractures. As the fracture-network generation is statistically performed, small fluctuations in the final fracture density are expected, hence the lower slope at distances $ d < 200$ m (Figure \ref{fig:1}b), indicating that the FRACMAN-generated number of smaller fractures is slightly lower than given by Eq.\ref{eq:P10}.
 
\subsubsection{Fracture Size Distribution and Main Fault Geometry}
Fracture lengths within a fault damage zone range from micrometer- to kilometer-scales for mature faults \cite{tchalenko1970similarities}. The maximum fracture length is limited by the width of the fault damage zone  \cite{mitchell2009nature}. Various factors influence fracture size, including fault length, fault roughness, in-situ stress conditions, maximum fault displacement, and distance from the fault core \cite{perrin2016off, sainoki2021numerical}. In studies of natural fracture networks, a power-law relation is commonly used to characterize the fracture-size distribution within assumed maximum and minimum fracture length \cite{panza2018discrete}. 
Our chosen fracture size distribution follows a power-law with fracture lengths between 500~m and 100~m (Figures \ref{fig:1}c and \ref{fig:fracture_size}). A similar distribution is described by \citeA{lavoine2019density} who develop a theoretical approximation based on natural fracture networks in geological reservoirs considering a power law exponential decay of 2 which we use as well (Figure \ref{fig:fracture_size}). The maximum fracture size is determined by our assumed damage zone thickness of 1~km, the shortest fracture length of 100~m is due to computational considerations. 

Our main fault extends 8~km along-strike and 4~km along-dip. For reference, we define the strike direction as N270E with respect to the Cartesian system in Figure \ref{fig:1}c. Fractures are confined to a volumetric damage zone of 10~km $\times$ 1~km $\times$ 6~km in the $x$ (along-strike of the main fault, East-West), $y$ (normal to the main fault strike, North-South), and $z$ (depth, positive upward) direction, respectively. 
The listric main fault is embedded within the fracture network (Figure \ref{fig:1}e), spanning the depth range from 1- 5~km, with a smoothly varying dip between 80$^\circ$ and 30$^\circ$. We choose the main fault's dimensions similar to rupture areas of the largest ($\Mw$ $5.5$ - $6.0$) observed induced earthquakes (i.e., 2011 $\Mw$ 5.7 Prague, 2016 $\Mw$ 5.8 Pawnee, and 2017 $\Mw$ 5.5 Pohang earthquakes) based on a fault-area scaling relation (\citeA{thingbaijam2017new}).

\subsubsection{Fracture Orientation}
We prescribe fracture orientations based on field observations and numerical models. For example, the observed orientation of large-scale fractures and conjugate faults in the damage zone of the Caleta Coloso strike-slip fault occur around  $25^\circ-30^\circ$ with respect to the main fault \cite{mitchell2009nature}. 
2D dynamic rupture simulations reveal that small-scale and meso-scale discrete off-fault fractures spontaneously group into two sets of conjugate fracture orientations separated by $\sim 100^\circ$ \cite{dalguer2003simulation, ando2007effects, okubo2019dynamics, gabriel2021unified}. 

Our model comprises two fracture populations. The first one, fracture family 1, has an average strike of N120E, while fracture family 2 has an average strike of N20E (Figure \ref{fig:1}d). Both fracture families form an average angle of $100^\circ$ to each other and an angle of $30^\circ$ and $70^\circ$ with respect to the main fault's strike.
For both families, we consider a $10^\circ$ standard deviation in the strike and dip angle. Distributions in both strike and dip of the fractures follow a Fisher distribution \cite{fisher1995statistical}.

\subsubsection{Fracture Shape}
Detailed knowledge of 3D fracture geometry in nature is scarce due to incomplete outcrop data and limited resolution in 3D reflection-seismic data. The surface of small fractures is commonly assumed to be a penny-shaped disk of infinitesimal thickness \cite{priest1981estimation, laslett1982censoring, piggott1997fractal, berkowitz1998stereological}. The main control on fracture aspect ratio may be the mechanical anisotropy due to rock layering \cite{nicol1996shapes}. The aspect ratio of fractures is typically assumed to vary from 0.5 - 3.5, irrespective of slip direction. For simplicity, we assume elliptical fractures with aspect ratio 2. We assume pure shear fractures (mode II and mode III slip) and do not allow for fracture opening (mode I).

\subsection{Material Properties}
We assume uniform material properties to model fractures embedded in a 3D elastic half-space. The main fault and fractures are buried and do not intersect the free surface. This setup allows us to focus on interactions of fractures and the main fault without rupture and wave-propagation complexities introduced by material interfaces. We choose a Poisson solid with $v_p = 6000$~m/s, $v_s = 3464$~m/s, $\rho = 2670$~km/m$^3$, and $\lambda = \mu = 32$~GPa.

\subsection{Initial Stresses, Friction Law, and Fault Strength}\label{subsection: initial stress loading}
We assume a transitional strike-slip to normal faulting stress regime. Local stress conditions for each fracture and the main fault are modulated by their geometry and orientation within the regional
A transitional stress regime may promote different faulting styles during a single compounded rupture (e.g., 1992 Landers and 2016 Norcia, 2016 Kaikōura earthquakes\cite{hauksson1994state, wollherr2018off,tinti2021constraining, ulrich2019dynamic}, seen also for induced earthquakes \cite{schoenball2014change, palgunadi2020dynamic}).

\subsubsection{Initial Stresses}\label{subsection:initial_stress}

We use a Cartesian initial stress tensor to load the fault and all fractures. We constrain the stress regime by 
the relative stress magnitude $A_\phi$ (\citeA{simpson1997quantifying} given as

\begin{equation}
    A_\phi = (n + 0.5) + (-1)^n(\phi - 0.5)\,,
\end{equation}
where $n$ indicates the faulting style ($n=0$ for normal faulting, $n=1$ for strike-slip faulting, and $n=2$ for reverse faulting), and $\phi$ denotes the stress shape ratio given by

\begin{equation}
    \phi = \frac{\sigma_2 - \sigma_3}{\sigma_1 - \sigma_3}\,.
\end{equation}
Here, $\sigma_1$, $\sigma_2$, and $\sigma_3$ are the maximum, intermediate, and minimum principal stress magnitudes, respectively. We consider a stress shape ratio of $\phi = 0.9$ and a predominantly normal stress regime ($n=0$, \cite{simpson1997quantifying}), and hence enforce $A_\phi = 0.9$. This transitional stress regime implies that two faulting styles, normal and strike-slip, may be activated due to $ S_v \geq \Shmax \gg \Shmin $, where $ S_v=\sigma_1, \Shmax=\sigma_2 $, and $\Shmin=\sigma_3$ are vertical overburden stress, maximum horizontal stress, and minimum horizontal stress, respectively.

Our model includes a depth-dependent effective normal stress, with pore pressure $P_f(z)=\gamma \rho g z$, where $g$ is the gravitational force, $z$ is depth, and $\gamma$ is the fluid pressure ratio \cite{ulrich2019dynamic}, which value is given by $\rho_{water}/\rho = 0.37$ when pore fluid pressure is hydrostatic. The pore pressure counteracts the fault normal stress in the effective normal stress $\sigma_n'=\sigma_n - P_f$. We vary $\gamma$ in the range $\gamma = [0.37; 0.80]$ to explore varying fluid-overpressure scenarios.

\subsubsection{Friction law and fault strength} 
The fault strength ($\tau_p$) is defined by the relation between friction law and effective normal stress:

\begin{equation}
    \tau_p = f(V,\theta)\sigma'_n\,.
\end{equation}
We adopt the laboratory-based rate-and-state friction law with rapid velocity weakening \cite{lapusta2000, rice2006heating, noda2009, dunham2011earthquake}:

\begin{equation}
  f(V, \theta) = a \arcsinh \left( \frac{V}{2V_0} \exp \left({\frac{\theta}{a}} \right) \right)\,,
\end{equation}
where $V$ is slip rate, $V_0$ is the reference slip rate, and $a$ is the direct effect.
The state variable $\theta$ evolves according to:

\begin{equation}
    \frac{d \theta}{d t} = -\frac{V}{L}(\theta - \theta_{SS})\,,
\end{equation}
where $L$ is the state-evolution slip distance for rate and state friction law. $\theta_{SS}$ at steady-state is:

\begin{equation}
    \theta_{SS}(V) = a \ln \left( \frac{2V_0}{V} \sinh \left( \frac{f_{SS}(V)}{a} \right) \right)\,,
\end{equation}
where $f_{SS}(V)$ is the friction coefficient at steady-state:

\begin{equation}
    f_{SS}(V) = f_w + \frac{f_{LV}(V) - f_w}{\left( 1 + (V/V_w)^8 \right)^{1/8}}\,,
\end{equation}
where the low-velocity steady-state friction $f_{LV}(V)$ is defined as:

\begin{equation}
    f_{LV}(V) = f_0 - (b-a)\ln(V/V_0)\,,
\end{equation}
with the evolutional effect given by $b$.

Friction parameters are chosen to generate realistic stress drops and frictional resistance. Aside from the state-evolution slip distance ($L$, see section \ref{RSParameterL}), all frictional parameters are constant (Table \ref{tab:A1}). We assume that the main fault and all fractures are frictionally unstable ($b-a > 0$). We use a characteristic weakening velocity $V_w = 0.1$ m/s based on the experimentally observed onset of rapid decay of the effective friction coefficient, ranging between 0.01 - 1 m/s (e.g., \citeA{rice2006heating,beeler2008constitutive,di2011fault}), consistent with the value used in community dynamic rupture benchmarks \cite{harris2018suite}. 
We choose a steady-state friction coefficient of $f_0=0.6$ at reference slip velocity value $V_0=10^{-6}$~m/s. The steady-state weakened friction coefficient $f_w$ is set to 0.1, similar to \citeA{rice2006heating}. 

\begin{table}
\caption{Friction parameters of  the rate-and-state friction law with rapid velocity weakening  assumed in this study.}
\centering
 \begin{tabular}{l c r} 
 \hline
 \textbf{Parameter} & \textbf{Symbol} & \textbf{Value}  \\
 \hline
 Direct effect parameter & $a$ & 0.01 \\  
 Evolution effect parameter & $b$ & 0.014 \\
 Reference slip velocity & $V_0$ & $10^{-6} m/s$ \\
 Steady-state friction coefficient at $V_0$ & $f_0$ & 0.6 \\
 Weakening slip velocity & $V_w$ & $0.1 m/s$ \\
 Fully weakened friction coefficient & $f_w$ & 0.1 \\
 Initial slip rate & $V_{ini}$ & $10^{-16}m/s$ \\
 State Evolution Slip Distance & $L$ & $0.002 - 0.01m$ \\
 \hline
 \end{tabular}
 \label{tab:A1}
\end{table}

Earthquake rupture dynamics are largely controlled by the relative pre-stress ratio $\mathcal{R}$ that describes the ratio of the maximum possible stress drop and frictional strength drop \cite{aochi20031999}

\begin{equation}\label{formula: R}
    \mathcal{R} = \frac{\Delta \tau_d}{\tau_p - \tau_d} = \frac{\tau_0 - f_w \sigma_n'}{(f_p - f_w)\sigma_n'}\,.
\end{equation}

In Eq. \ref{formula: R}, $\tau_0$ is the initial shear traction, $\tau_d=f_w \sigma'_n$ is dynamic stress, and $\Delta\tau_d$ is dynamic stress drop. $\tau_p=f_p\sigma'_n$ is the peak dynamic stress, and $\tau_p-\tau_d$ is the maximum dynamic strength reduction. The peak value of the friction coefficient ($f_p$) depends on the rupture dynamics \cite{garagash2021fracture} but is approximated in evaluating Eq. \ref{formula: R} by the reference value $f_0$. The value of $f_p$ in simulations varies along the fault and fractures and may exceed $f_0$ but rarely falls below it. $\mathcal{R}_0$ is the maximum possible value of $\mathcal{R}$ for a fracture at the most-optimal orientation. Hence, the fault-local $\mathcal{R}$ is always smaller than or equal to $\mathcal{R}_0$. We prescribe the maximum pre-stress ratio, $\mathcal{R}_0$, to be constant across the model to constrain the initial stress state (Section \ref{subsection:initial_stress}). $\mathcal{R}_0$ depends on the proximity to failure of an optimally oriented fault ($\mathcal{R}_0 = 1$ corresponds to the maximum degree of stress criticality, i.e., an optimally oriented fault is at ``failure" for the given initial stress state). We consider $\mathcal{R}_0=0.8$ in all simulations.
In nature, faults and fractures are typically not optimally oriented. In a dynamic rupture scenario, only a small part of a fault or fracture must reach failure to initiate a sustained rupture.

\subsection{Fault-Size-Dependent State Evolution Slip Distance ($L$) and Fracture Energy ($G$)}\label{RSParameterL}
Dynamic rupture modeling across different fault or stress-heterogeneity scales may use variable characteristic slip distances \cite{bizzarri2003slip, ando2007effects, galvez2021multicycle, ulrich2022stress}. The critical nucleation size required to initiate spontaneous or runaway rupture (i.e., ruptures across the entire main fault \cite{galis2019initiation}), scale with characteristic slip distance $L$ under rate-and-state friction \cite{rubin2005earthquake}, or with critical slip distance $D_c$ under linear slip-weakening friction \cite{andrews1976rupture, galis2015initiation}. To model dynamic rupture on fractures of different sizes, we adopt a fracture-scale dependent $L$. Our scaling of $L$ \cite{garagash2022fracture} is linked to the scale-dependence of the fracture energy emerging from seismological observations (e.g., \citeA{abercrombie2005can}) and earthquake physics.

The elastic energy, stored in the rock volume and available to be released and dissipated as frictional heat, fracture energy $G$, expended to propagate the rupture, and radiated in seismic waves, scales with the fault size $R$ (e.g., \citeA{madariaga1976dynamics}). Thus, fracture energy has to scale with fault size for the fault to be able to host dynamic rupture \cite{garagash2022fracture}. Such linear scaling of $G$ with fault size $R$ has been previously proposed in relation to fault growth \cite{scholz1993fault} and co-seismic rupture with off-fault plasticity \cite{andrews2005rupture}. Recently, \citeA{garagash2022fracture} inferred from a multi-weakening fracture energy decomposition applied to seismologic estimates of fracture energy \cite{abercrombie2005can, tinti2005kinematic, mai2006fracture, causse2014variability, viesca2015ubiquitous} that:

\begin{equation}
    G_{c}(R) \approx 400 [\text{Pa}] \times R\,.
\end{equation}
Here, we assume $G \approx G_{c}(R)$ and apply an equivalent scaling of the state evolution slip distance $L$ following an analytical approximation of fracture energy for a rate-and-state governed fault \cite{garagash2021fracture}:

\begin{equation}
    G_c = (f_p - f_w) \sigma'_n L\,.
\end{equation}
We use the reference value $f_0$ to approximate the peak friction $f_p$ at the rupture front, and $\sigma'_n \sim 40$~MPa representative of lithostatic and hydrostatic gradients at the median fault depth ($\sim 3$~km) to define:

\begin{equation}
    L \approx 2 \times 10^{-5} R\,.
\end{equation}

\section{Results}\label{section:result}
\begin{figure}
    \centering
    \includegraphics[width=1\textwidth]{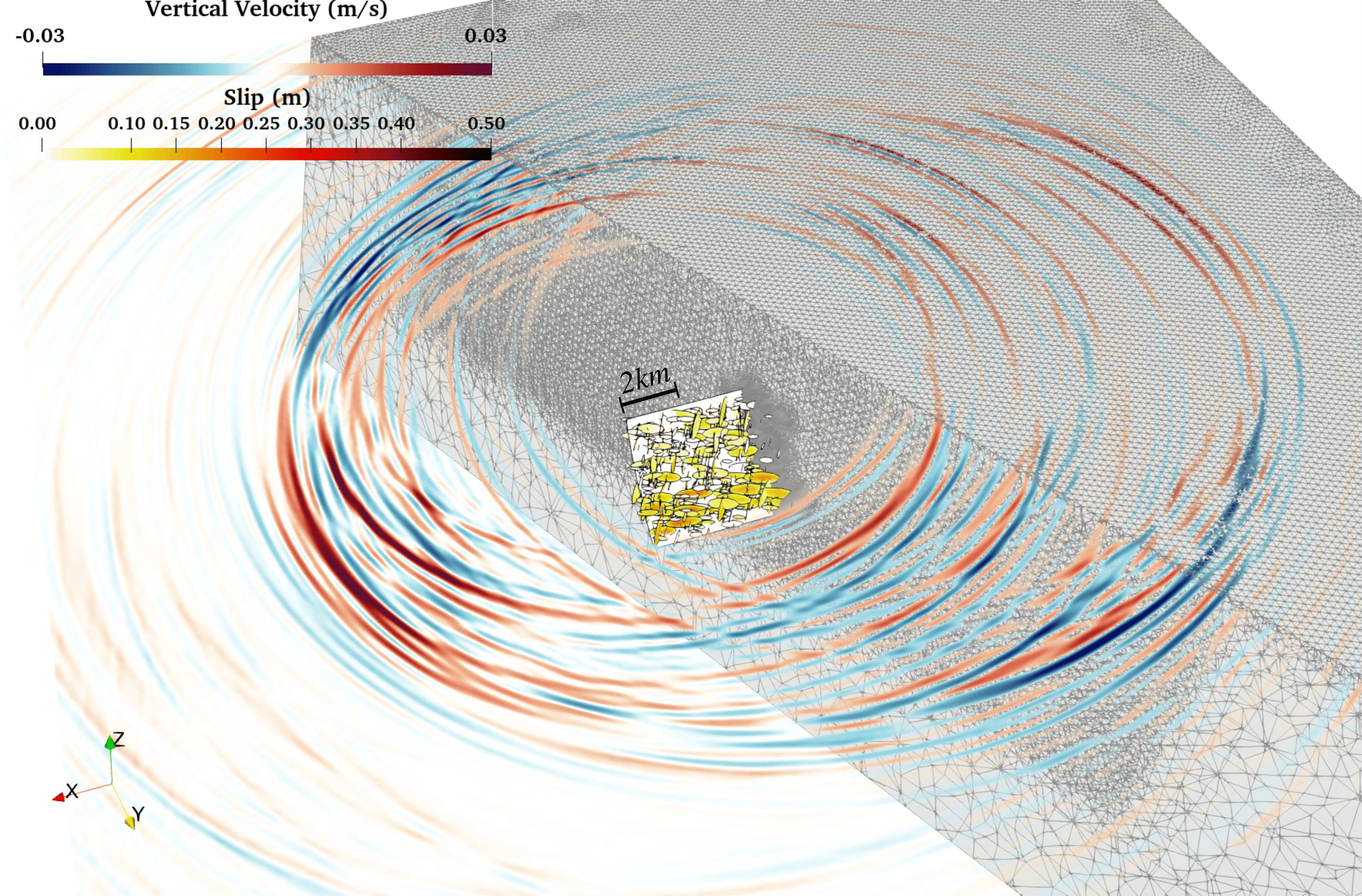}
    \caption{Rendering of the high-resolution 3D unstructured tetrahedral computational mesh of this study. The model combines a listric main fault and a network of 854 fractures of varying size. We show a snapshot of slip on the fault and fractures (hot colors, in [m]) and the radiated seismic waves (vertical particle velocity, in [m/s]) at $t=6$~s simulation time. Arrows in the bottom left mark the directions of $X$ (along-strike of the main fault, East-West), $Y$ (North-South), and $Z$ (depth, positive upward).}
    \label{fig:2}
\end{figure}

In the following, we present eighteen 3D dynamic rupture simulations, each on the same multiscale fracture network with an embedded listric main fault. All simulations are carried out using the open-source software SeisSol (Section \ref{section:data_resources} and \ref{section:numerical_method}). Ten simulations are scenarios under varying pre-stress conditions. In five simulations, we shift the rupture nucleation location within the fault damage zone and vary fluid overpressure levels. Finally, in three scenarios, we use higher-resolution simulations for generating reliable high-frequency seismic waveforms (\ref{mesh_generation}). Figure \ref{fig:2} shows the computational high-resolution mesh overlain by a snapshot of absolute slip and vertical particle velocity at time $t=6$~s for a pure rupture cascade (explained below). To improve visibility and aid interpretation, 3D views of our results are presented in an ``exploded" view, generated by displacing each fracture centroid from coordinates 
($x,y,z$) to ($3.5x$, $3.5y$, $z$). 

We first analyze fault and fracture strength for varying maximum horizontal stress ($\Shmax=\sigma_2$) orientations without performing dynamic rupture simulations (Section \ref{subsec:results:static}). In Section \ref{subsec:results:dynamic}, we describe ten dynamic rupture simulations for variable $\Shmax$ orientation ($\Psi$), and analyze cascading dynamic rupture within the fracture network and on the main fault. In Section \ref{subsec:results:kinematic}, we describe the earthquake kinematics of all fracture network-main fault rupture scenarios. Next, we compare two scenarios of cascading ruptures initiated at different locations and with different mechanisms in Section \ref{subsec:results:triggering}: (i) a dynamic rupture cascade initiating on a fracture due to overstress at $\sim$1~km away from the main fault, and (ii) a dynamic rupture cascade initiating at the same location as (i) assuming an elevated (in excess of hydrostatic) pore fluid pressure regime within the fracture network. Finally, in Section \ref{subsec:results:waves}, we describe seismic waveform characteristics.

\subsection{Fault and Fracture Strength for Variable Maximum Horizontal Stress Orientation}
\label{subsec:results:static}
We examine different orientations of maximum horizontal stress, ranging from $\Psi = 40^\circ -120^\circ$ with respect to North (normal to the main fault strike; Figure \ref{fig:1}e). Changes in $\Psi$ gradually alter the criticality of the main fault and all smaller fractures, ranging from favorably to unfavorably oriented fractures with respect to ambient stress conditions. Recall that favorably oriented fractures or fault segments are characterized by higher values of $\mathcal{R}$, whereby $\mathcal{R}$ depends on how the normal and shear tractions are resolved on fractures with varying orientations within the ambient stress field.

The static slip tendency analyses to examine how variations in $\Psi$ result in favorably or unfavorably oriented dynamic rupture planes (fractures, main fault, or both) do not yet involve dynamic rupture simulations, but still provide a valuable preliminary assessment of possibly mechanically viable conditions for earthquake rupture \cite{palgunadi2020dynamic}.
They consist of quantifying the relative pre-stress ratio $\mathcal{R}$ on the main fault and on the fractures for a given $\mathcal{R}_0$. However, the static slip tendency analyses cannot fully anticipate the dynamic processes or potential interactions between multiscale fractures, particularly if fractures are unfavorably oriented. As we show later, in several cases, dynamic stress interactions between multisegmented fractures facilitate sustained rupture even on unfavorable oriented fractures.

\begin{figure}
    \centering
    \includegraphics[width=0.9\textwidth]{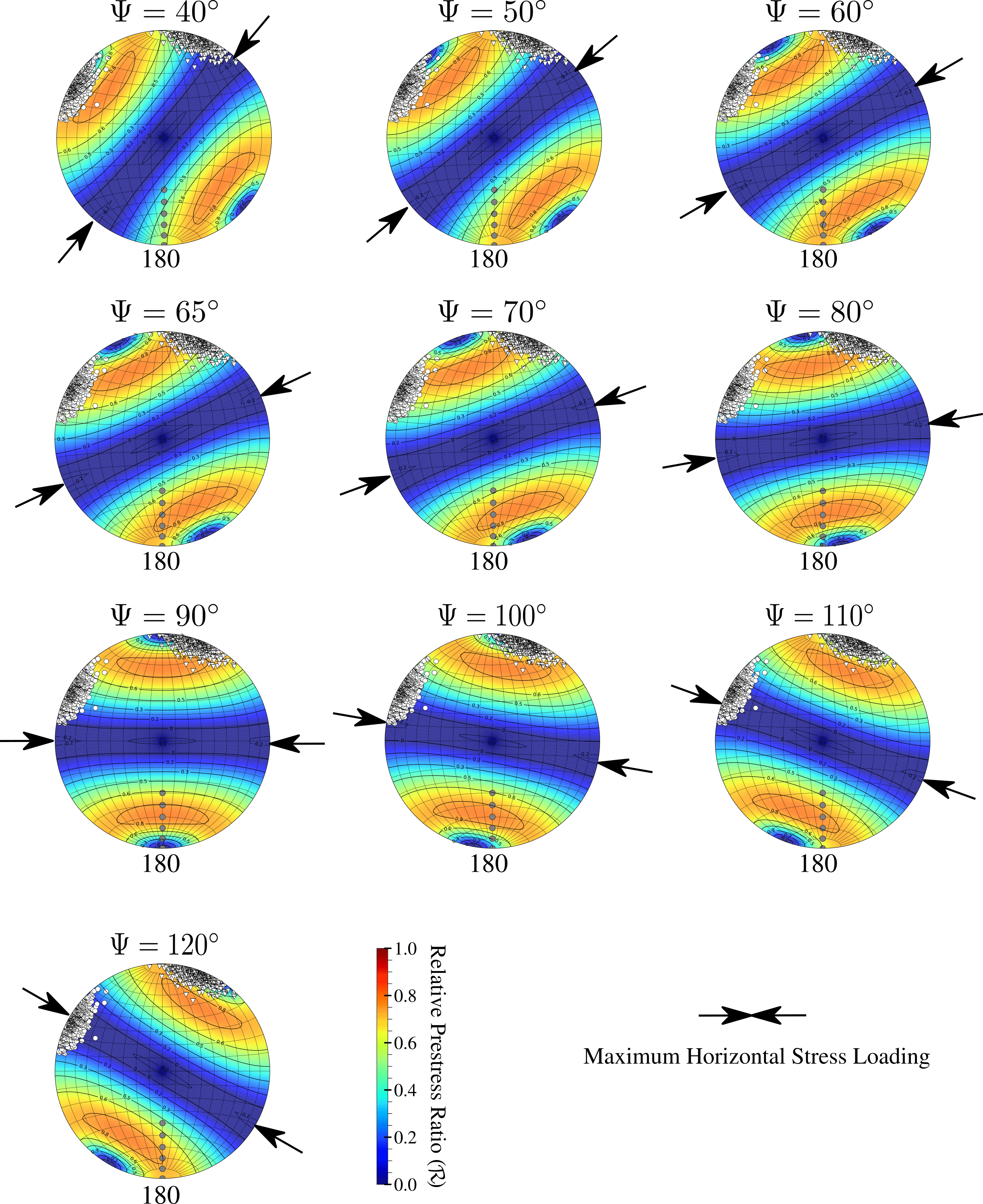}
    \caption{Static analysis considering varying $\Shmax$ orientation ($\Psi$). Stereoplots show the fault or fracture-local relative pre-stress ratio ($\mathcal{R}$) in lower hemisphere projection, illustrating optimal and non-optimal orientations with respect to the ambient pre-stress for different $\Psi$ values. The counteracting arrows represent the orientation of the most compressive horizontal pre-stress loading. White triangles mark fracture family 1 (average strike: $120^\circ$)), while white dots represent fracture family 2 (average strike: $20^\circ$) in polar projection. Grey dots illustrate the main listric fault. Hot and cold colors represent areas with favorably and unfavorably oriented fractures, respectively.}
    \label{fig:3}
\end{figure}

For the assumed $A_\phi = 0.9$ and using Eq. \ref{formula: R} with $f_0 = 0.6$ and $f_w = 0.1$, we characterize and display the failure propensity of fractures and the main fault for different $\Psi$ (Figure \ref{fig:3}). Based on this analysis, we identify three prominent cases for subsequent dynamic rupture modeling: \textit{Case 1} has unfavorably oriented fractures and main fault plane for $\Psi = 40^\circ$; \textit{Case 2} has favorably oriented fractures and unfavorably oriented main fault for $\Psi = 65^\circ$; \textit{Case 3} has unfavorably oriented fractures but favorably oriented main fault plane for $\Psi = 120^\circ$.
Considering the variability in strike and dip orientation directions of the small-scale fractures, some fractures will remain unfavorably oriented even under generally favorable orientation (e.g., $\Psi = 65^\circ$).

\subsection{Cascading dynamic rupture in multiscale fracture networks}
\label{subsec:results:dynamic}
To investigate the conditions under which cascading dynamic rupture in the fracture network may occur, we conduct ten 3D dynamic rupture simulations including the three cases (\textit{Case} 1-3) defined above. The seven additional simulations are defined as ``rupture on unfavorably to favorably oriented fractures but unfavorably oriented main fault" for $\Psi = 50^\circ$ (\textit{Case 1a}) and $\Psi = 60^\circ$ (\textit{Case 1b}) and  ``rupture on unfavorably to well oriented main fault plane" for $\Psi = 70^\circ, 80^\circ, 90^\circ, 100^\circ,110^\circ$ (\textit{Case 2a}). We prescribe a common hypocenter at $x=-800$~m, $y=600$~m, and $z=-2800$~m. For all cases in this section, we initiate rupture inside a 400~m-radius sphere centered on the main fault that includes a subset of fractures. This radius is larger than the critical nucleation sizes on fractures (see \ref{Appendix: Rupture Initiation}).

\subsubsection{Case 1, $\Psi = 40^\circ$, unfavorably oriented fractures and main fault}\label{section:case1}
We refer to this case as ``failed rupture nucleation''.
For $\Psi = 40^\circ$, the static analysis (Figure \ref{fig:3}) shows unfavorable orientation for the two fracture families and the main fault.
We observe that no self-sustained rupture is generated (Figure \ref{fig:S1}), as the slip rate remains confined within the nucleation volume (Movies S1a, S1b). Increasing the nucleation radius (we tested radii of up to 600~m) and increasing nucleation overstress (up to $\mathcal{R}_0$=5) do not lead to self-sustained rupture initiation.
Interestingly, during the nucleation phase, rupture branching occurs at the intersection of two fractures (red arrow in Figure \ref{fig:S1}). However, rupture is arrested subsequently, and slip remains small with an average of less than 0.02~m.

\subsubsection{Case 1a, $\Psi = 50^\circ$, less unfavorably oriented fractures and main fault than Case 1}\label{section:case1a}

\begin{figure}
    \centering
    \includegraphics[width=0.95\textwidth]{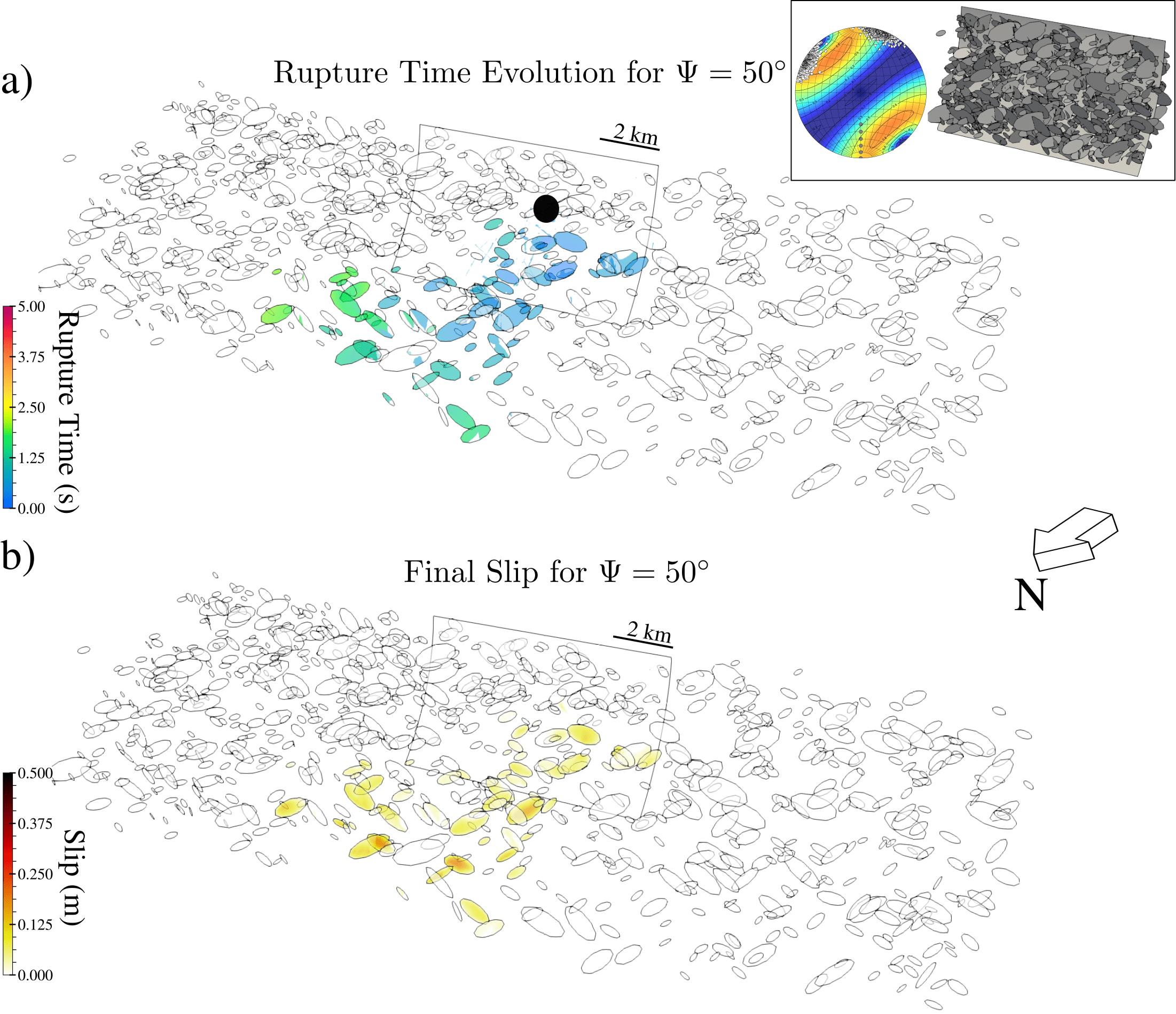}
    \caption{\textit{Case 1a}: Unfavorably oriented fractures and main fault at $\Psi=50^\circ$. The inset in the top-right corner shows the stereonet plot of the relative pre-stress ratio $\mathcal{R}$ in lower hemisphere projection, along with the original fracture network model. Panel (a) shows the exploded view of the rupture time evolution (hypocenter marked by a black circle). Panel (b) shows the final slip on all fractures.}
    \label{fig:4}
\end{figure}

For $\Psi = 50^\circ$, we observe dominantly cascading rupture within the fracture network. Small, localized slip on the main fault at fracture-fault intersections (Figure \ref{fig:S_main_fault_slip}) is dynamically induced by slip across the fracture network but quickly self-arrests on the main fault. 
Dynamic rupture is successfully initiated and propagates as a rupture cascade outside the nucleation area, activating 74 fractures, as shown in the rupture times (Figure \ref{fig:4}) and the evolution of slip velocity (Movies S2a, S2b). The total rupture duration is approximately $t=1.5$~s. Slip occurs predominantly on fractures located within the main fault’s hanging wall ($\sim 85\%$ of the total slipped fractures) and is distributed predominantly toward the North with respect to the main fault (Figure \ref{fig:4}). Although rupture branching between neighboring connected fractures is observed, rupture ``jumping''  (i.e., dynamic triggering) between distant fractures does not occur.
The relative pre-stress ratio of those fractures that dynamically slip are $\mathcal{R} \geq 0.3$ (Figure \ref{fig:4}). However, nine activated fractures initially have $\mathcal{R} < 0.3$. Dynamic rupture does not develop towards the E and W directions and stops spontaneously. At rupture termination, slip on the last fractures (with $\mathcal{R}=$ 0.2 - 0.3) is limited to approximately $40\%$ of the total fracture area and slip rate decays smoothly. The average slip across all slipped fractures is 0.04~m.

\subsubsection{Case 1b, $\Psi = 60^\circ$, dynamic rupture on unfavorably to favorably oriented fractures}\label{section:case1.2}
For $\Psi = 60^\circ$, fractures are more favorably oriented than in \textit{Case 1a}. 
Like for \textit{Case 1a}, we observe sustained cascading rupture within the fracture network and only small, non-sustained slip on the main fault near fault-fracture intersections (Figure \ref{fig:S_main_fault_slip}). Slip occurs more widespread across the fracture network and dynamic rupture propagates simultaneously in multiple directions (see evolution of rupture time and slip velocity in Figure \ref{fig:S1}; Movies S3a and S3b). We identify multiple backward-propagating slip episodes within the fracture network at the footwall side toward the nucleation area. The total rupture duration is  $t \approx 3$~s. Late-stage rupture on fractures occurs mainly in the fracture network surrounding the eastern part of the main fault (green to yellowish colors in Figure \ref{fig:S1}). The dynamic rupture cascade terminates spontaneously on different fractures via three mechanisms: (1) abrupt rupture cessation at individual fracture boundaries, despite connected neighboring fractures (2) smooth rupture stopping after reaching unfavorably oriented connected fractures ($\mathcal{R} = 0.1 - 0.25$), and (3) rupture termination on isolated fractures due to relatively large spacing between fractures that precludes dynamic rupture jumping (here $> 80$~m). 

Maximum slip is 0.4~m, located on the deepest fractures at a depth of 5~km (Figure \ref{fig:S2}), and average slip across all slipped fractures is 0.04~m. This scenario exhibits sustained cascading rupture or pure cascade behavior, which may be expected from the static analysis (Figure \ref{fig:3}) due to the favorable orientation of most fractures. We term the cascading rupture characteristics as ``sub-optimal'', since only $\sim 65\%$, that is, 536 out of 854 fractures dynamically slipped (Figure \ref{fig:S3}). Slip is distributed on fractures within both the hanging wall and footwall fault zone of the main fault plane. Interestingly, fractures on the hanging wall slip first, followed by those on the footwall. However, the dynamic rupture cascade does not activate fractures located in the western region of the main fault footwall. 

We highlight two fracture-to-fracture cascading mechanisms observed in the simulation with $\Psi=60^\circ$: (i) dynamic rupture transfer via rupture branching across connected fractures, or (ii) rupture jumping across unconnected fractures. We observe that rupture branching dominates the cascading process, activating approximately $\sim 94\%$ of the slipped fractures. The remaining $\sim 6\%$ of slipped fractures are in close proximity to the evolving cascade (with an average distance of $< 80$ m between fractures) and are activated by rupture jumping.

\subsubsection{Case 2, $\Psi = 65^\circ$, dynamic rupture on favorably oriented fractures and an unfavorably oriented main fault}\label{case:65}
\begin{figure}
    \centering
    \includegraphics[width=0.9\textwidth]{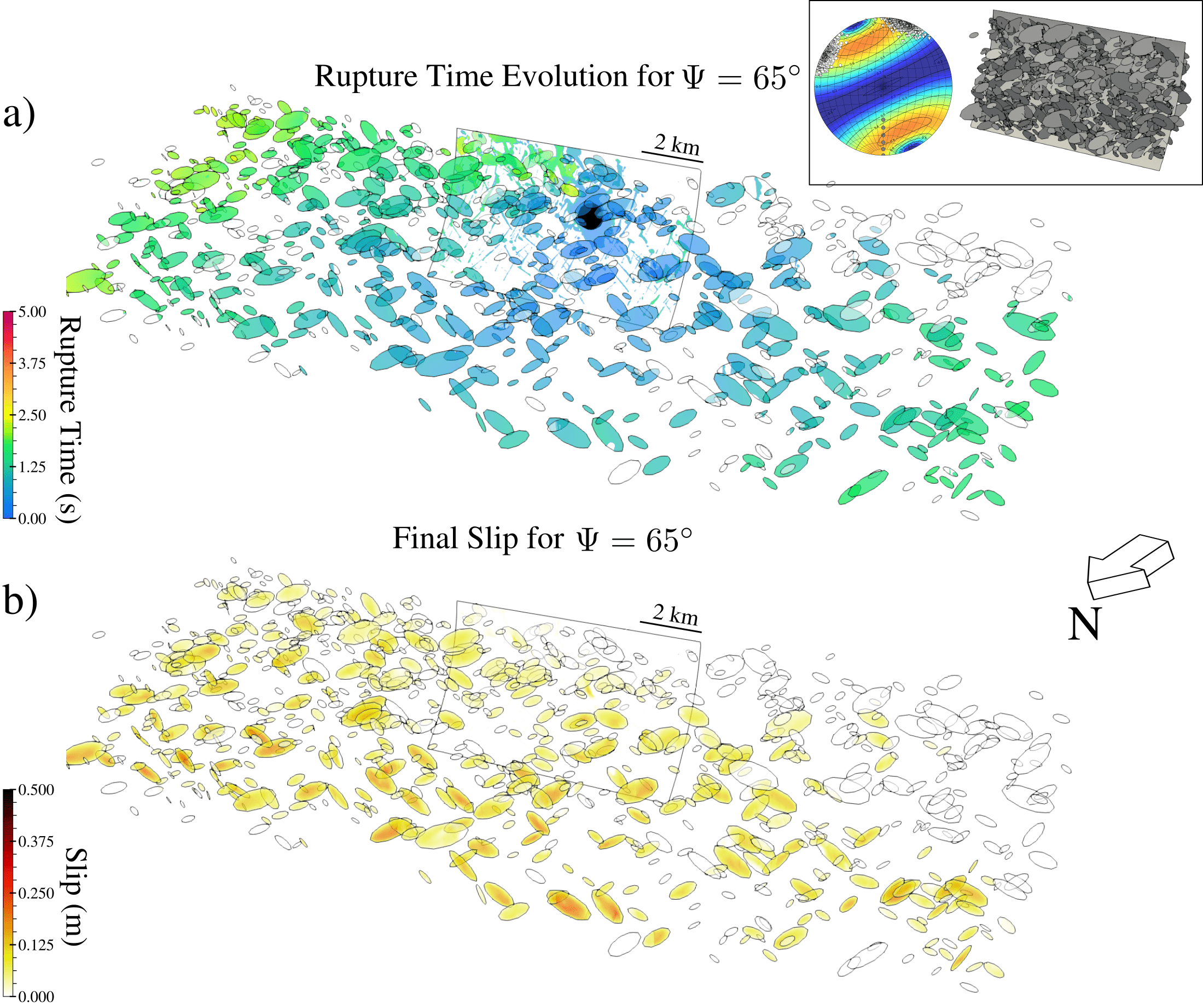}
    \caption{\textit{Case 2}: Pure cascade. Favorably oriented fractures and a poorly oriented main fault with $\Psi=65^\circ$. The top right corner panel shows the stereonet plot of the relative pre-stress ratio ($\mathcal{R}$) in lower hemisphere projection, side-by-side with the original model. Panel (a) shows the exploded view of the rupture time evolution, with the black circle illustrating the rupture nucleation area (hypocenter). Panel (b) shows the exploded view of the final slip.}
    \label{fig:5}
\end{figure}

We refer to this case as a ``pure cascade''.
Most fractures are favorably oriented for $\Psi = 65^\circ$ and sub-critically stressed ($\tau/\sigma_n'$ in the range $0.22-0.5$), but the main fault is unfavorably oriented. Likewise, six fractures have a low relative pre-stress ratio ($\mathcal{R} < 0.3$; Figure \ref{fig:3}), which are fewer than in the case of $\Psi=60^\circ$ but enough to eventually arrest a dynamic rupture cascade. In this simulation, dynamic rupture propagates as a sustained cascade, generating dynamic slip across the entire fracture network (Figure \ref{fig:5}). The main fault experiences only small, localized slip despite several fault-fracture intersections (see rupture time evolution in Figure \ref{fig:5} and Figure \ref{fig:S_main_fault_slip}).  

Dynamic rupture propagates in a zigzaging pattern, activating neighboring fractures of both fracture families. The cascade progressively evolves from one fracture to another, favoring those that are connected or located in close proximity (Movies S4a, S4b) via dynamic rupture branching and jumping. Macroscopically, rupture propagates bilaterally to the East and West, as shown in the rupture time evolution in Figure \ref{fig:5}a. Late rupture occurs in the fracture network surrounding the main fault's eastern part and then back-propagates towards the nucleation area (green to yellow in Figure \ref{fig:5}a). The dynamic rupture cascade stops after $t=3.5$~s due to the same three mechanisms as in \textit{Case 1b}. Four out of six unfavorably oriented fractures are responsible to stop the entire cascading rupture process.

In the \textit{Case 2} simulation, $561$ fractures slipped ($\sim 70\%$, out of 854) (Figure \ref{fig:5}). Slip is distributed predominantly in the fracture network surrounding the eastern part of the main fault (Figure \ref{fig:S3_1}), similar to $\Psi = 60^\circ$. The average slip across all slipped fractures is 0.04~m. We observe separation into left- and right lateral slip (Figure \ref{fig:S3_2}) hosted by the two fracture families, respectively. 
The main fault is unfavorably oriented, and the evolving pure cascade within the fracture network fails to trigger self-sustained dynamic rupture anywhere on the main fault. However, we observe dynamic interactions of the cascade and the main fault in the form of small self-arresting rupture on the main fault imprinting in ``rupture time'' on the main fault in Figure \ref{fig:5}a. The inability of fractures to activate the main fault is likely due to the fracture energy of the main fault being $\sim 10$ times larger than that of the largest fractures. 

\subsubsection{Case 2a, $\Psi = 70^\circ - 110^\circ$, dynamic rupture for an unfavorably to favorably oriented main fault}\label{case2.3}

In cases $\Psi=70^\circ$, $\Psi=80^\circ$, and $\Psi=90^\circ$, cascading rupture within the fracture network still occurs but is less widespread. For $\Psi=70^\circ$ and $\Psi=80^\circ$ , the cascade evolution is similar to \textit{Case 2}: bilateral propagation away from the hypocenter, triggering fractures located within the hanging wall first, followed by later rupture of fractures surrounding the eastern part of the main fault (Figure \ref{fig:S1}; Movies S5a, S5b, S6a, S6b). For both $\Psi = 70^\circ$ and $\Psi=80^\circ$, we observe slipped fractures confined to the western main fault's footwall (Figures \ref{fig:S2} and \ref{fig:S3}: $69\%$ and $66\%$ of all fractures slip, respectively). Total rupture duration is $t=2.8$~s ($\Psi=70^\circ$) and $t=3.1$~s ($\Psi=80^\circ$), respectively.

The main fault becomes more favorably oriented as $\Psi$ changes to a larger azimuth. However, sustained dynamic rupture on the main fault is not triggered for $\Psi=70^\circ$ nor for $\Psi=80^\circ$, and main-fault slip remains localized near the nucleation area and at several fracture-fault intersections (Figure \ref{fig:S_main_fault_slip}).
The transition from pure cascades within the fracture network to dynamic main fault activation occurs at $\Psi=90^\circ$ (Figures \ref{fig:S1} and \ref{fig:S2}). The rupture duration for this case is shorter than before, $t=1.8$~s, and fewer fractures slip, accounting for only $22\%$ of the total number of fractures. Figure \ref{fig:S2} shows that slip within the fracture network is concentrated on the western hanging wall, while there is almost no slip on fractures in the eastern part of the network. The three cases show the same average slip across the fractures which is 0.042~m.

Slip on the main fault is limited and constrained near the hypocenter (black circle in Figure \ref{fig:S2}, Movies S7a and S7b). Dynamic rupture on the main fault stops spontaneously for two reasons. Firstly, shallower parts of the main fault (with larger dip) are unfavorable for slip, as indicated by corresponding values $\mathcal{R} < 0.3$ for many main fault locations (``grey dots'' in Figure \ref{fig:3}) for $\Psi = 90^\circ$. Secondly, interconnected fractures with unfavorable orientations act as barriers, thereby inhibiting dynamic cascading rupture.

For the cases $\Psi=100^\circ$ and $\Psi=110^\circ$, more than $50\%$ of fractures are unfavorably oriented (Figure \ref{fig:3}), while all parts of the main fault are optimally oriented. After nucleation, the rupture progresses mainly on the main fault and activates a subset of fractures within the nucleation volume at simulation time $t=0.1$~s. Rupture then propagates bilaterally to the eastern and western parts of the main fault and stops abruptly at the main fault boundary (Figure \ref{fig:S1}). Dynamic rupture terminates at $t=2.6$~s. The intersecting fractures alter the space-time evolution of the main fault rupture (Movies S8a, S8b, S9a, and S9b). A complex rupture pattern with multiple rupture fronts contrasts with simple rupture dynamics on planar faults without small-scale pre-stress heterogeneity (e.g., \citeA{ramos2019transition}). Depending on the orientation of the activated fractures and their activation timing, activated fractures may facilitate or hinder the rupture evolution on the main fault, as evidenced by local variations in rupture fronts contours (Figure \ref{fig:S5}). 

The slip velocity evolution is mildly complex but not as heterogeneous as in dynamic rupture simulations on rough faults (e.g., \citeA{shi2013rupture, mai2018accounting, taufiqurrahman2022broadband}). Dynamic rupture on the main fault produced large enough stressing
to activate overall unfavorably oriented connected and unconnected fractures ($\mathcal{R} = 0.1 - 0.3$) located close to the main fault (Figure \ref{fig:S1}). The up to a magnitude smaller critical nucleation sizes and fracture energy of fractures also promote their activation.
We observe more slipped fractures on the dilatational side of the main fault's rupture direction, namely the western side at the hanging wall and the eastern side at the footwall relative to the hypocenter (Figure \ref{fig:S2}). Dynamic rupture on fractures directly connected to the main fault activates fractures further away by rupture branching and jumping. However, direct cascading between fractures not connected to the main fault is rare.

The case $\Psi = 110^\circ$ shows almost the same dynamic rupture behavior as $\Psi = 100^\circ$: rupture propagates bilaterally and activates off-fault fractures. While in case of $\Psi=100^\circ$, $63\%$ of all fractures slip, for $\Psi=110^\circ$, $57\%$ of fractures slip (Figure \ref{fig:S3}), corresponding to a slightly higher number of favorably oriented fractures ($\mathcal{R} > 0.6$, $\sim\,15$ fractures). 

For both $\Psi = 100^\circ$ and $\Psi = 110^\circ$, direct rupture branching is the predominant mechanism of dynamic cascading from the main fault to the fracture network. Rupture jumping from the main fault to the fracture network is mostly present near the fault boundary in the eastern and western parts of the main fault. Abrupt rupture termination creates locally more widespread fracture-network slip at the main fault boundary. Rupture jumping also occurs between parallel fractures adjacent to the main fault. 

\subsubsection{Case 3, $\Psi=120^\circ$, dynamic rupture on a favorably oriented main fault} 
We refer to this case as dynamic rupture with off-fault fracture slip. Based on the static analysis, most fractures are unfavorably oriented but the main fault is favorably oriented (Figure \ref{fig:3}). 

\begin{figure}
    \centering
    \includegraphics[width=0.9\textwidth]{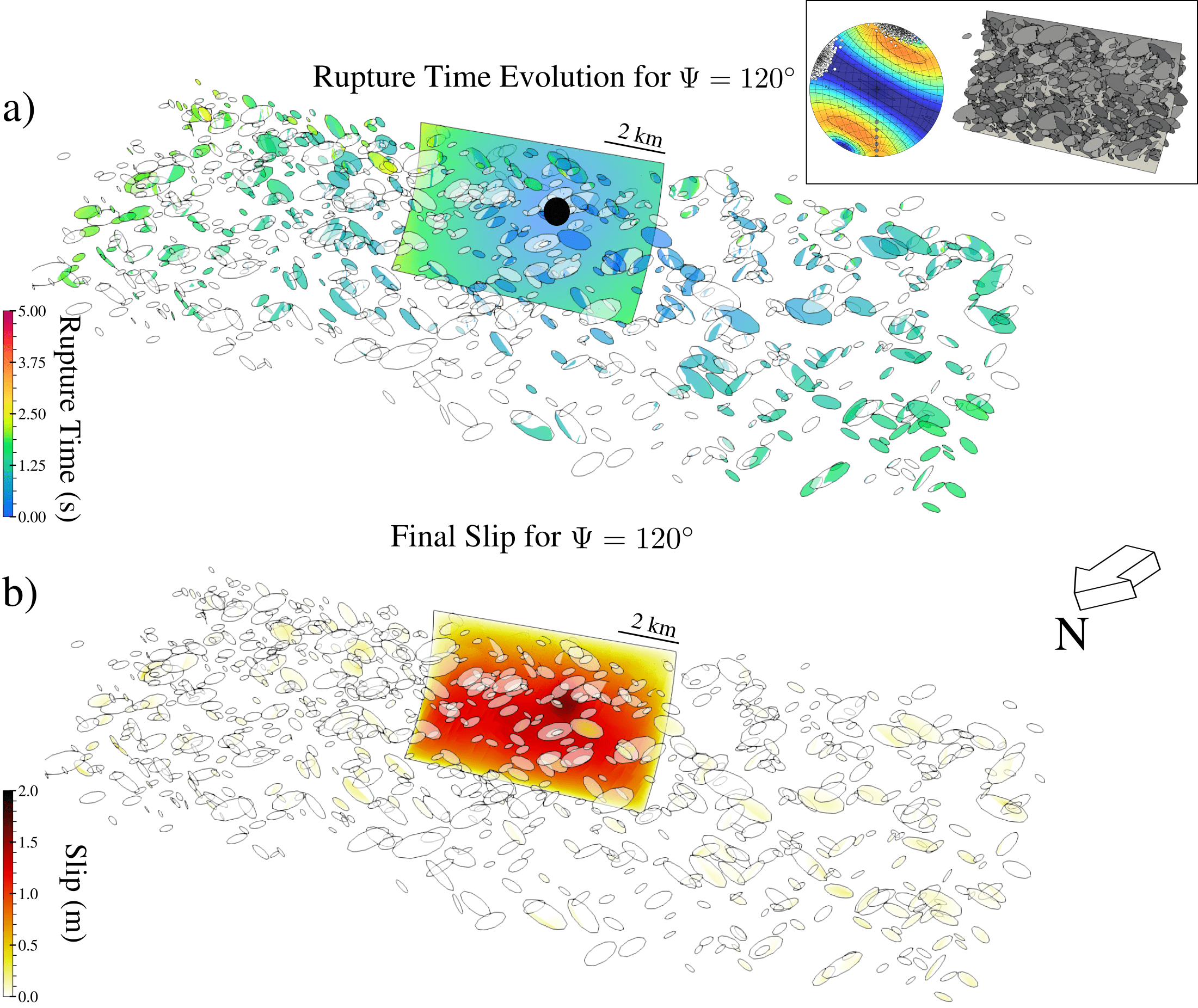}
    \caption{\textit{Case 3}: Rupture with off-fault fracture slip, unfavorably oriented fractures, and a well-oriented main fault plane at $\Psi=120^\circ$. The top right corner panel shows the stereonet plot of the relative pre-stress ratio ($\mathcal{R}$) in lower hemisphere projection, side-by-side with the original model. Panel (a) presents an exploded view of the rupture time evolution. The black circle illustrates the rupture nucleation area (hypocenter). Panel (b) exhibits an exploded view of final slip. Note that the maximum slip across the main fault (up to 2~m) is larger than in the previous cases (Figures \ref{fig:4} and \ref{fig:5}).}
    \label{fig:6}
\end{figure}

Dynamic rupture initiates on the main fault and a subset of fractures, and nucleates at $t=0.04$~s. It then propagates bilaterally on the main fault and causes a limited number of off-fault fractures to slip (Figures \ref{fig:6}a, \ref{fig:6}b, Movies S10a, S10b). Similar to the previous main fault rupture cases ($\Psi=100^\circ$ and $\Psi=110^\circ$), main fault rupture is heterogeneous due to the dynamic main fault-fracture interactions. The rupture on the main fault abruptly terminates at the fault boundary after $t=2.6$~s.

Similar to the case $\Psi=110^\circ$, slip on the fractures that are directly connected to the main fault may trigger dynamic rupture on the more distant and unconnected fractures to the main fault, especially for fractures with an acute angle to the deeper parts of the main fault in fracture family 1. Cascading rupture along fractures is mainly due to rupture branching. Again, the main fault rupture activates unfavorably oriented fractures ($\mathcal{R} \leq 0.3 $). 

Slip on fractures unconnected to the main fault is small and quickly self-arrests (Figure \ref{fig:6}b). The percentage of slipped fractures is $\sim 58\%$, i.e., 477 fractures (Figures \ref{fig:S3} and \ref{fig:S3_2}). In this case, slipping fractures are located preferably at the dilatational side of the main fault, that is, at the hanging wall in the eastern and the footwall in the western main fault (Figures \ref{fig:6} and \ref{fig:S3_2}).

\subsection{Rupture kinematics}
\label{subsec:results:kinematic}
In the following, we analyze the kinematic rupture properties of all ten dynamic rupture simulations with varying $\Psi$, including equivalent point source moment tensor estimates, seismic moment (magnitude), moment rate function, average rupture speed on individual fractures or the main fault, and stress drop ($\Delta \tau$). Inspecting the kinematic parameters for cascading earthquakes may help to identify observable signatures of cascading rupture episodes in real fault zones.

\subsubsection{Equivalent point source moment tensor}\label{subsection:source-mechanism}

We examine the apparent far-field source mechanism by calculating an equivalent moment tensor for a point-source representation of each dynamic rupture model (\ref{section:moment_tensor_solution}).
The equivalent moment tensor of our scenarios changes from strike-slip to thrust faulting as the rupture mode changes from the pure rupture cascade to the main fault rupture with increasing $\Psi$. The equivalent point-source moment tensor's strike for all cascading fracture network ruptures disagrees with the main fault's strike (Table \ref{table:ruptureKinematic}). The cascading rupture scenarios mostly appear as double couple (DC) strike-slip faulting. For the cases $\Psi = 100^\circ - 120^\circ$, the equivalent moment tensor has the same strike direction as the main fault (N270E).

The case $\Psi = 90^\circ$, which is a mixed case between a fracture network cascade and main fault rupture shows noticeable non-double couple (non-DC) components. Using moment tensor decomposition following \citeA{vavryvcuk2015moment}, we obtain $-23\%$ of non-DC, $77\%$ of DC, and $0\%$ for the isotropic components.  The fact that we observe thrust-faulting on the main fault mixed with strike-slip fracture network slip contributes to this increase in non-DC components. 
For $\Psi = 100^\circ - 120^\circ$, we find limited non-DC components ($-6\%$ up to $10\%$) in the equivalent moment tensors despite the complicated rupture processes on the main fault and slip on off-fault fractures.

\subsubsection{Moment magnitude}
The total moment magnitudes of cascading ruptures summing all slip within the fracture network and on the main fault fall in the range $\Mw = 4.9 - 5.58$. For $\Psi=50^\circ$ to $\Psi=80^\circ$, the moment magnitudes of cascading earthquakes located only in the fracture network gradually increase due to the successively larger number of slipped fractures (Table \ref{table:ruptureKinematic}). The failed nucleation for $\Psi = 40^\circ$ still generates an $\Mw = 2.78$ event. 
Moment magnitude decreases to $\Mw = 5.17$ for $\Psi=90^\circ$ due to limited fracture network rupture and limited slip on the main fault. The moment magnitudes of the scenarios with sizeable main fault slip are slightly larger than in the pure cascade scenarios ($\Mw = 6.0$ for $\Psi=100^\circ - 120^\circ$). We find that regardless of whether or not the main fault experiences runaway rupture, the fracture network in the damage zone releases a sizeable seismic moment ($\Mw = 4.9 - 5.6$).

\begin{table}
  \caption{Summary of rupture kinematic with different $\Shmax$ orientations ($\Psi=40^\circ - 120^\circ$). $v_{R,ave}$ is the surface-averaged rupture speed across all planes. $v_{R,tot}$ is calculated as the ratio of the distance of slipped fractures relative to the hypocenter and the duration of the rupture. $c_s$ denotes shear wave speed. $v_C$ is the cascading speed. $\Delta \tau$ is the average stress drop (see Section \ref{section_Stress_drop}). $M_{w,f}$ is the moment magnitude of slip in the fracture network. $M_{w,F}$ is the moment magnitude of slip on the main fault. $\Mw$ is the overall moment magnitude. MTS stands for equivalent moment tensor.}\label{table:ruptureKinematic}
  \centering
  \resizebox{\columnwidth}{!}{%
  \begin{tabular}{  c c c c c c c  c  c  c  c }
    \hline
    \multirow{2}{*}{\boldmath{$\Psi$}} & \multirow{2}{*}{\boldmath{$\frac{v_{R,ave}}{c_s}$}} & \multirow{2}{*}{\boldmath{$\frac{v_{R,tot}}{c_s}$}} & \multirow{2}{*}{\boldmath{$\frac{v_C}{c_s}$}} & \multirow{2}{*}{\boldmath{$\Delta\tau$} (MPa)} & \multirow{2}{*}{\boldmath{$M_{w,f}$}} & \multirow{2}{*}{\boldmath{$M_{w,F}$}} & \multirow{2}{*}{\boldmath{$\Mw$}} & \multirow{2}{*}{\textbf{MTS}} & \multicolumn{2}{c}{\textbf{Strike/Dip/Rake (\boldmath{$^\circ$})}}\\ 
    \cline{10-11} 
     & & & & & & & & & Plane 1 & Plane 2 \\
    \hline

    $40^\circ$ & - & 0.92 & - & 1.7 & 2.78 & 1.42 & 2.78 &
    \begin{minipage}{.1\textwidth}
    \centering
        \includegraphics[width=0.5\linewidth]{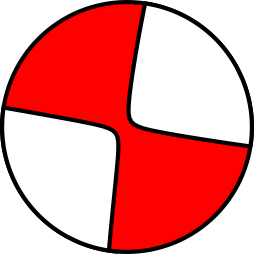}
    \end{minipage}
    & 105/86/0 & 15/90/179
    \\
    
    $50^\circ$ & 0.90 &  0.80 & 0.88 & 7.7 & 4.90 & 3.2 & 4.91 &
    \begin{minipage}{.1\textwidth}
    \centering
      \includegraphics[width=0.5\linewidth]{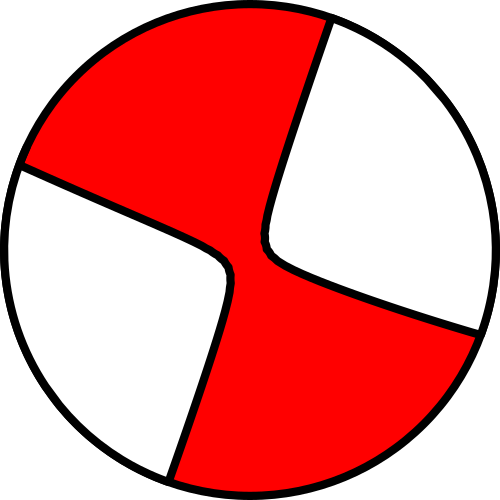}
    \end{minipage}
    & 110/86/0 & 20/90/176
    \\ 
    
    $60^\circ$ & 0.90 &  0.73 & 0.68 & 8.9 & 5.48 & 3.76 & 5.49 &
    \begin{minipage}{.1\textwidth}
    \centering
      \includegraphics[width=0.5\linewidth]{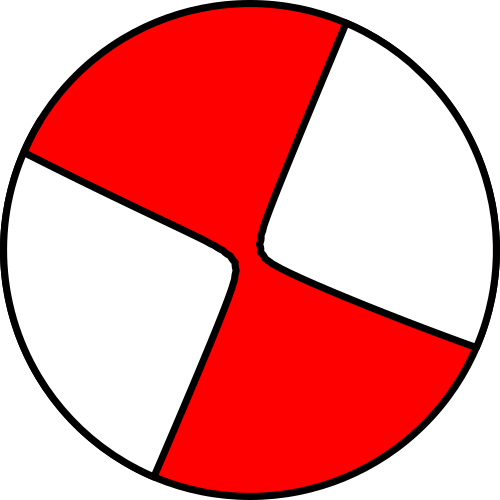}
    \end{minipage}
    & 113/86/0 & 23/90/176
    \\ 
    
    $65^\circ$ & 0.90 &  0.63 & 0.65 & 9.6 & 5.51 & 3.92 & 5.52 &
    \begin{minipage}{.1\textwidth}
    \centering
      \includegraphics[width=0.5\linewidth]{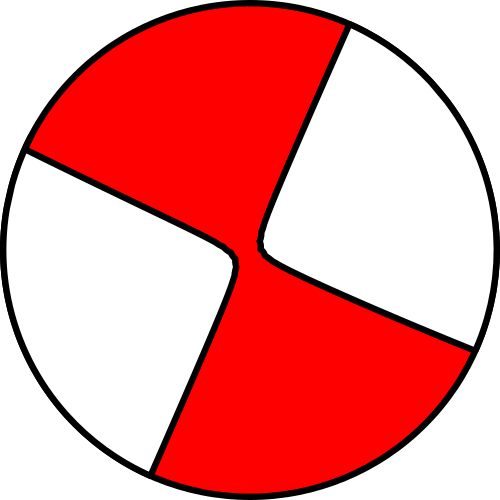}
    \end{minipage}
    & 114/88/0 & 24/90/178
    \\ 
    
    $70^\circ$ & 0.92 &  0.75 & 0.75 & 9.6 & 5.57 & 4.56 & 5.58 &
    \begin{minipage}{.1\textwidth}
    \centering
      \includegraphics[width=0.5\linewidth]{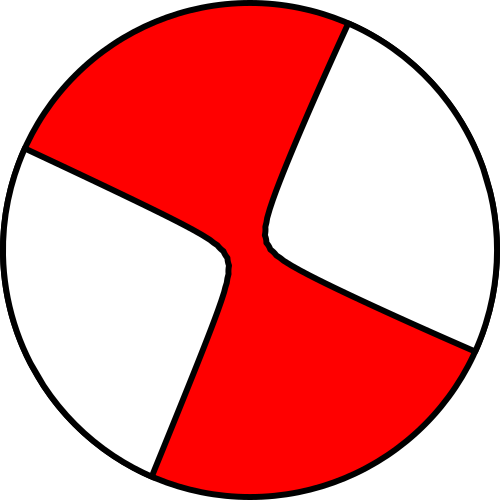}
    \end{minipage}
    & 114/89/0 & 24/89/179
    \\ 
    
    $80^\circ$ & 0.90 &  0.74 & 0.75 & 8.8 & 5.50 & 4.5 & 5.52 &
    \begin{minipage}{.1\textwidth}
    \centering
      \includegraphics[width=0.5\linewidth]{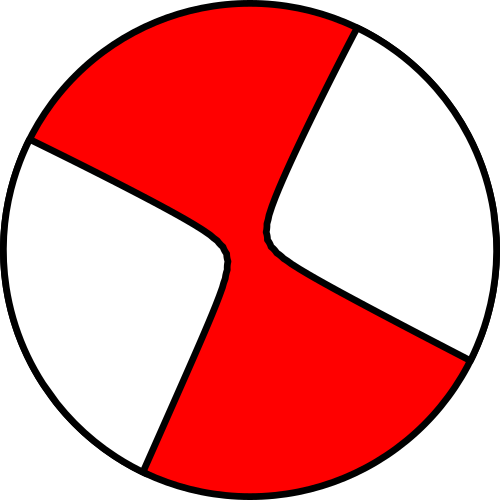}
    \end{minipage}
    & 116/89/2 & 26/88/179
    \\ 
    
    $90^\circ$ & 0.85 &  0.74 & 0.72 & 7.1 & 5.12 & 4.66 & 5.17 &
    \begin{minipage}{.1\textwidth}
    \centering
      \includegraphics[width=0.5\linewidth]{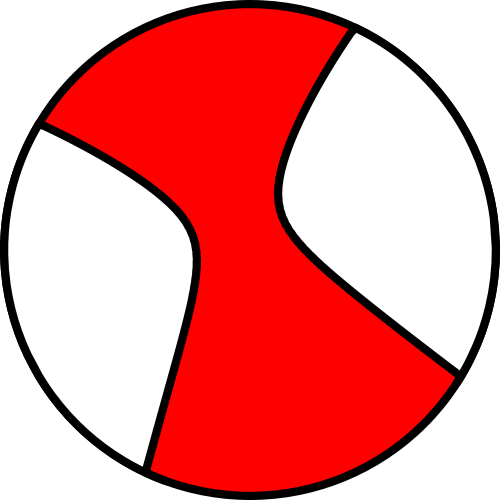}
    \end{minipage}
    & 119/82/12 & 27/78/171
    \\ 
    
    $100^\circ$ & 0.74 &  0.78 & 0.72 & 5.9 & 5.29 & 5.89 & 6.0 &
    \begin{minipage}{.1\textwidth}
    \centering
      \includegraphics[width=0.5\linewidth]{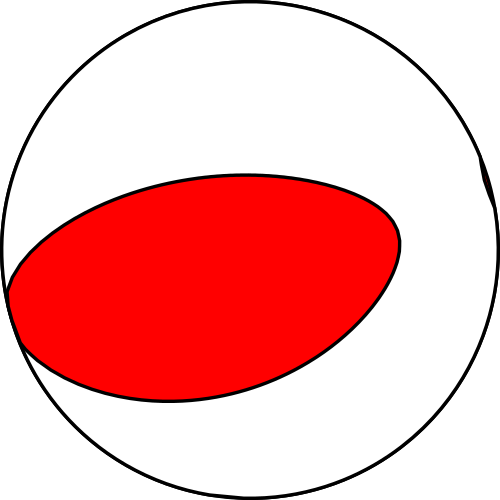}
    \end{minipage}
    & 269/57/107 & 59/37/65
    \\ 
    
    $110^\circ$ & 0.74 &  0.79 & 0.74 & 5.9 & 5.14 & 5.99 & 6.0 &
    \begin{minipage}{.1\textwidth}
    \centering
      \includegraphics[width=0.5\linewidth]{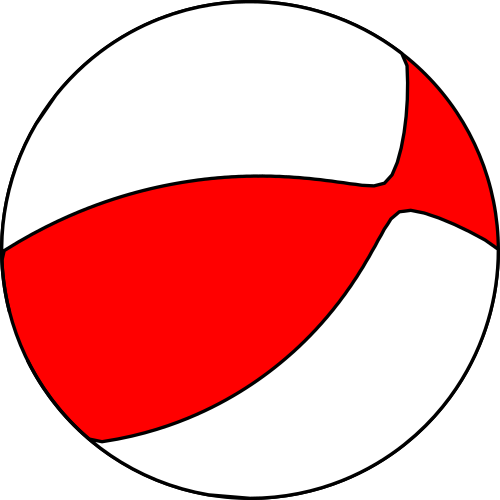}
    \end{minipage}
    & 269/57/123 & 39/45/50
    \\ 
    
    $120^\circ$ & 0.74 &  0.78 & 0.74 & 5.9 & 5.22 & 5.97 & 6.0 &
    \begin{minipage}{.1\textwidth}
    \centering
      \includegraphics[width=0.5\linewidth]{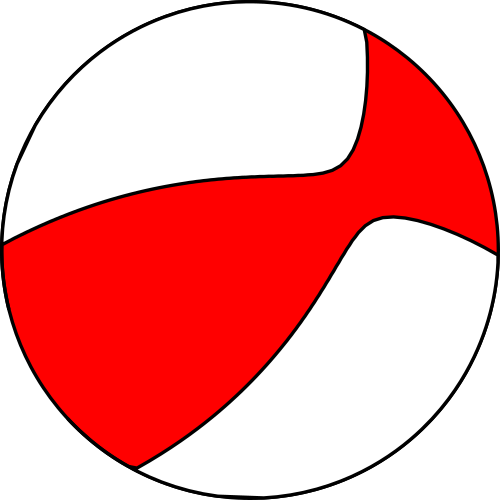}
    \end{minipage}
    & 271/58/136 & 28/54/41
    \\ 
    
    \hline
  \end{tabular}
  }
\end{table}

\subsubsection{Moment rate function}\label{section:MRF}
\begin{figure}
    \centering
    \includegraphics[width=0.95\textwidth]{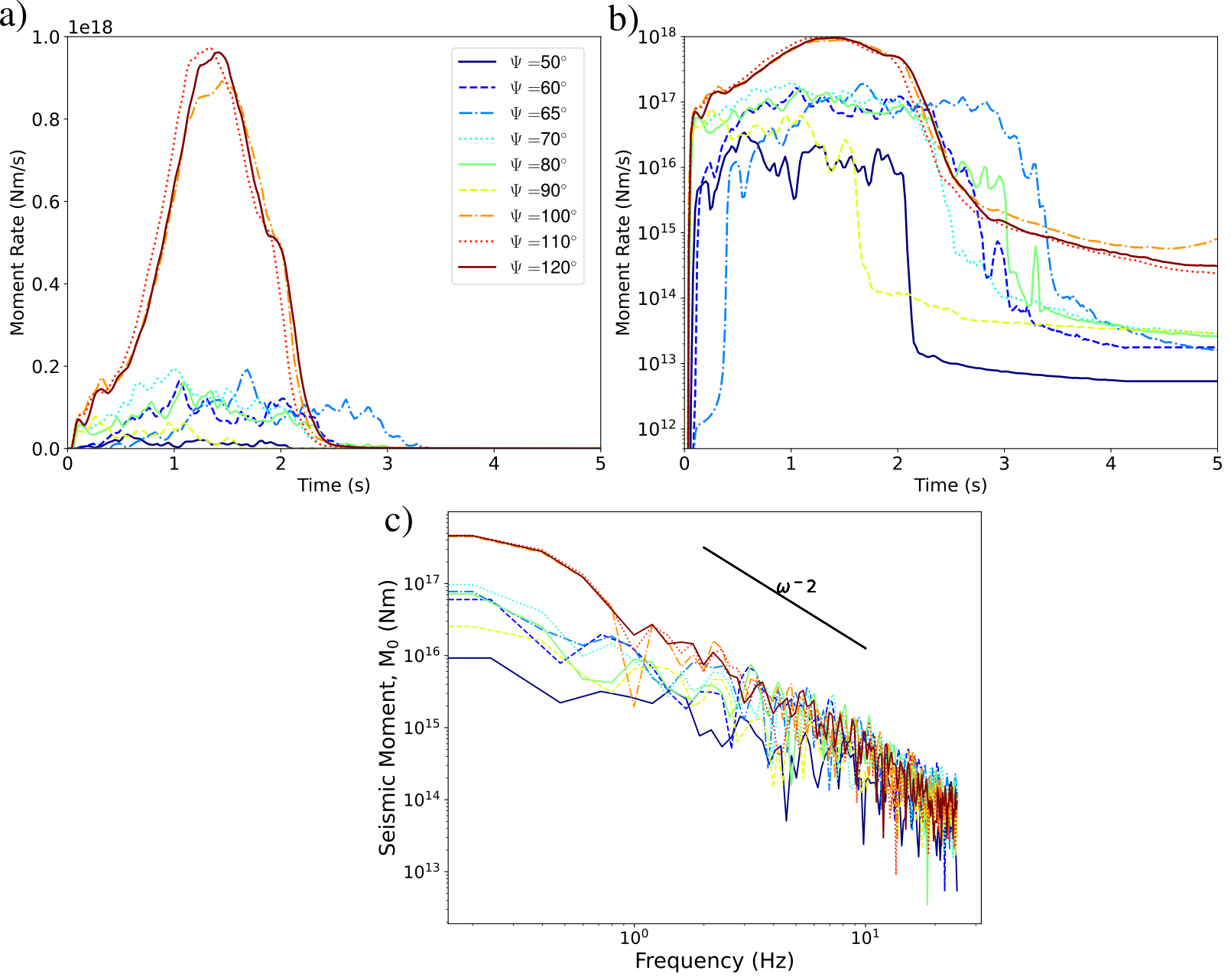}
    \caption{Moment rate functions (MRFs) for nine dynamic rupture simulations with different $\Psi$ values (blue to red colors). The failed nucleation \textit{Case 1}, $\Psi = 40^\circ$ is not included. a) Moment rate functions. b) The same as panel (a), but using a logarithmic scale. c) The corresponding seismic moment spectra illustrate differences in magnitude and small variations in spectral decay. The black line shows an $\omega^{-2}$ spectral decay.}
    \label{fig:momentRate}
\end{figure}

We analyze the moment rate functions (MRFs) and their amplitude spectra for different simulations (Figures \ref{fig:momentRate}a and \ref{fig:momentRate}b). Cases with large $\Psi=100^\circ - 120^\circ$ that include runaway rupture on the main fault show a simple triangle-shaped moment rate function 
\cite{meier2017hidden}, despite the activated off-fault fractures ($\Psi = 100^\circ - 120^\circ$ at $t<0.5s$ in Figure \ref{fig:momentRate}b). These MRFs include small amplitude variations shortly after rupture initiation, which can be observed more clearly using a logarithmic scale (Figure \ref{fig:momentRate}b). 

MRFs of cascading ruptures constrained to the fracture network are more complex and characterized by multiple peaks. For \textit{Case 2} ($\Psi=65^\circ$), larger fluctuations in MRF amplitudes occur after the first 1~s of the cascading rupture (cyan dashed line in Figures \ref{fig:momentRate}a and \ref{fig:momentRate}b). Such complex moment rate release is inferred at a larger scale for earthquakes including more than one fault, such as the 2016 $\Mw$ 7.8 Kaik\={o}ura earthquake \cite{bai2017two, zhang2017imaging, ulrich2019dynamic}.
The MRF complexity disappears in simulations in which the main fault slips sustainably. A runaway rupture on the main fault produces higher seismic moment release at lower-frequency spectra, overprinting the energy release generated by the slipped fractures. Figure \ref{fig:momentRate}c shows our modeled seismic moment spectra that follow a classical spectral decay with $\omega^{-2}$ at high frequencies (e.g., \citeA{aki1967scaling, vallee2011scardec}). 

\subsubsection{Average rupture speeds and cascading speed}
\begin{figure}
    \centering
    \includegraphics[width=0.9\textwidth]{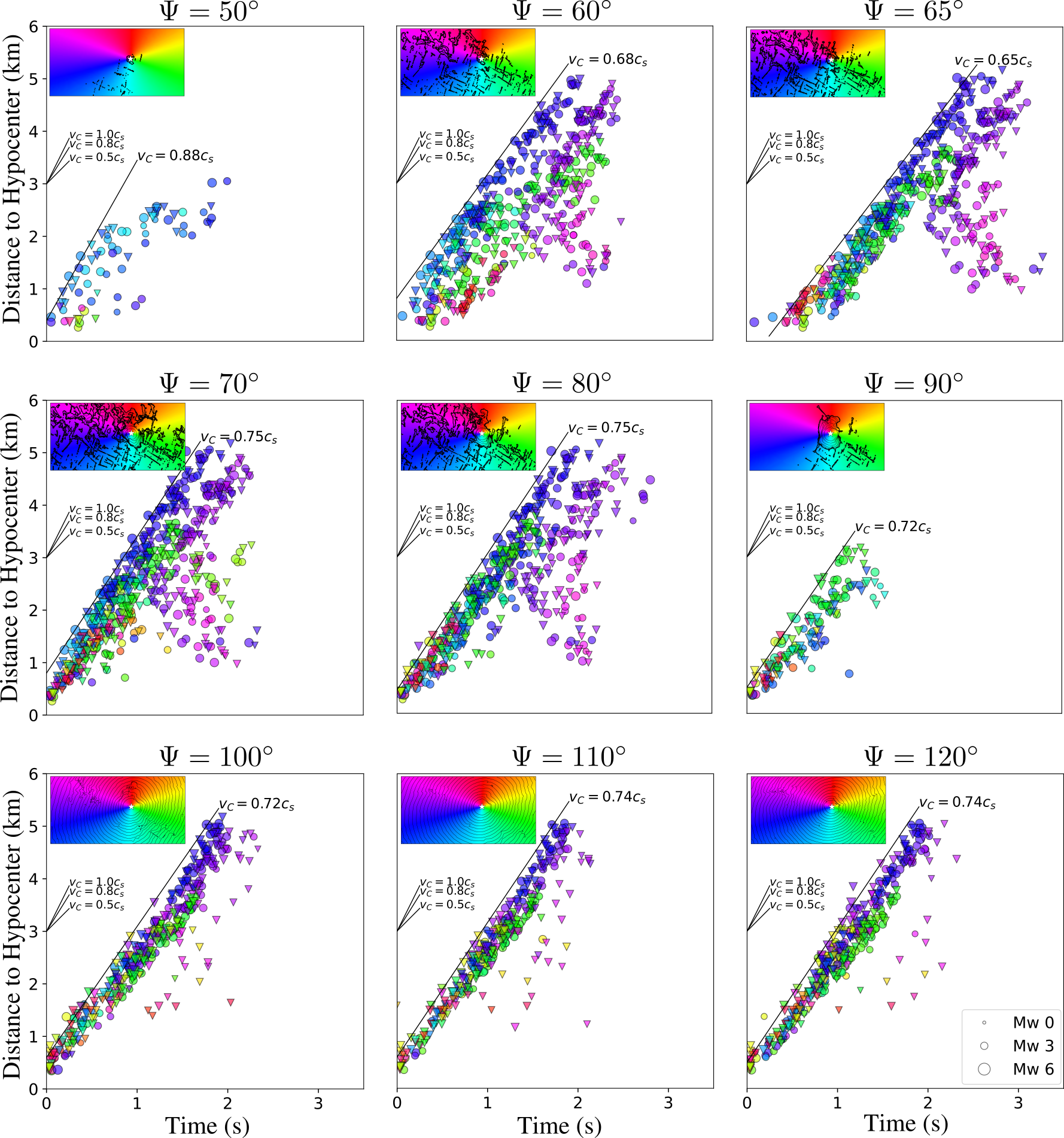}
    \caption{Cascading speed ($v_C$) for dynamic rupture simulations with different $\Psi$. Triangles and dots represent both fracture families with strike N20E and N120E, respectively. The marker size indicates the individual moment magnitude, while the color represents the azimuth of the fracture position relative to the hypocenter's position. The insets in the top left of each panel display the rupture time contours along the main fault in a 2D projection (fault normal facing out of paper) with a time increment of $0.08s$. The background color of the inset is the azimuth of the fractures, and a white dot indicates the hypocenter location. The black line is the cascading speed ($v_C$) measured as the activated fracture distance to the hypocenter versus its rupture onset time. Three black lines are fractions of the shear wave speed ($c_s$) for reference.}
    \label{fig:cascadingSpeed}
\end{figure}

Analyzing rupture speed in complex 3D models is challenging \cite{bizzarri2013calculation}. We define average rupture speeds in three different ways: (1) the overall average rupture speed for the entire process or seismologically-inferrable average rupture speed ($v_{R,tot}$), (2) the surface-averaged rupture speeds across all fractures and the main fault ($v_{R,ave}$), and (3) the ``cascading speed" ($v_C$). These three different rupture speeds are useful to distinguish rupture speeds for cascading and non-cascading ruptures.
$v_{R,tot}$ (1) is calculated as the ratio of the distance of slipped fractures and the main fault relative to the hypocenter and the duration of the rupture.
The cascading speed ($v_C$) quantifies the apparent rupture speed along the fracture network and is determined from the slope of a scatter plot showing the distance to the hypocenter versus rupture onset time at each slipped fracture (Figure \ref{fig:cascadingSpeed}). 
We manually determine the cascading speed for each $\Psi$-case. We ignore Case 1, $\Psi=40^\circ$, where slip occurs only within the nucleation volume.
Table \ref{table:ruptureKinematic} shows that $v_{R,tot}$ ranges from $0.63c_s$ to $0.80c_s$, where $c_s$ is the shear wave speed. For $\Psi < 90^\circ$, cascading rupture across the fracture network results in lower $v_{R,tot}$, with $v_{R,ave} > v_{R,tot}$.  
We observe supershear rupture ($c_s \leq v_{R,f} < v_p$) in parts of the fracture network, for instance for $\Psi = 65^\circ$ (see Figure \ref{fig:S_ruptureSpeed}a), 
resulting in higher $v_{R,ave}$. 

Supershear rupture occurs on smaller fractures after rupture branches or jumps onto them from larger fractures or from the main fault during runaway rupture (Figure \ref{fig:S_ruptureSpeed}b,c,d).  Several larger fractures also host supershear rupture speeds. 
Although inferences of supershear rupture speeds are rare for natural earthquakes, recent studies suggest that supershear rupture may frequently occur at a local scale, i.e., over a small part of a fault (e.g., \citeA{dunham2007conditions, passelegue2013sub, huang2016potential, bruhat2016rupture, bao2019early, bao2022global}). 

Figure \ref{fig:cascadingSpeed} plots distance to hypocenter vs rupture time for all fracture and simulations, which allows evaluating cascading speeds for all simulations with $\Psi$ in the range $50^\circ -120^\circ$. Markers coded by azimuth relative to the hypocenter and fracture family allow a detail analyse of the cascading rupture evolution along the fracture network. For instance, for $\Psi=65^\circ$, the dynamic rupture cascade propagates across both fracture families (shown as triangles and dots) and across all azimuths from the hypocenter (colors in the inset figure). 
The cases $\Psi = 60^\circ, 65^\circ, 70^\circ,$ and $80^\circ$ include back-propagating rupture cascades toward the nucleation area as seen in the decreasing hypocentral distance of rupture onset in late rupture stages ( $t \approx 1.5$~s). For all cases, we observe that as rupture progresses, the two fracture families are activated consecutively for rupture branching and jumping (Figure \ref{fig:cascadingSpeed}). We observe no rupture jumping and branching within fractures belonging to the same fracture family.

The inset in each panel of Figure \ref{fig:cascadingSpeed} shows rupture-time contours on the main fault. Ruptures of fractures intersecting the main fault but not activating sustained main fault rupture appear as localized densely spaced contours. The transition with increasing $\Psi$ from a cascading rupture within the fracture network to the main-fault runaway rupture on the main fault occurs around $\Psi=90^\circ$. We see mild complexity in the rupture front due to dynamic interaction with intersecting fractures for $\Psi > 100^\circ$.

Compared to the surface-averaged rupture speeds on all fractures and the main fault $v_{R,ave}$, the value of the cascading speed $v_C$ is consistently lower for cascades propagating within the fracture network, reflecting the delayed rupture propagation due to rupture branching and jumping. When considering $\Psi=65^\circ$, which involves the greatest number of slipped fractures, $v_C$ is notably smaller than $v_{R,ave}$, at $v_C=0.65c_s$ (Table \ref{table:ruptureKinematic}). For $\Psi=70^\circ$ and $\Psi=80^\circ$, $v_C$ increases to $v_C=0.75c_s$. For fracture network ruptures, $v_C$ is almost equivalent to $v_{R,tot}$. In the mixed case of $\Psi=90^\circ$ with limited cascading rupture, $v_C$ is comparable to $v_{R,ave}$. If cascading rupture across the fracture network does not occur (as for $\Psi=100^\circ - 120^\circ$), $v_C$ is comparable to $v_{R,ave}$ and $v_{R,tot}$.

\subsubsection{Average stress drop}
\label{section_Stress_drop}
We define the spatial-averaged stress drop as the vector integral over all points that ruptured and experienced a dynamic stress drop \cite{noda2013comparison}. The stress drop is computed from the difference between the initial and final shear stress after the rupture terminates. Our simulations produce an average stress drop $\Delta \tau$ comparable to observations of crustal earthquakes \cite{huang2017stress}. $\Delta \tau$ for (sub-shear, non-cascading) main fault rupture is lower than 6~MPa (e.g., $\Psi = 120^\circ$, $\Delta \tau = 5.9$~MPa) while cascading ruptures generate higher $\Delta \tau$ (e.g., $\Psi = 65$, $\Delta \tau = 9.6$~MPa, Table \ref{table:ruptureKinematic}). Higher $\Delta \tau$ can be associated with overall favorably oriented fractures (Figure \ref{fig:3}), and with the occurrence of supershear $v_{R,f}$ on several fractures, comparable to observations in laboratory experiments \cite{passelegue2013sub}. Lower $\Delta \tau$ when rupture on the main fault is activated could also be a consequence of the activation of unfavorably oriented fractures connected to the main fault, decreasing the surface averaged stress drop.

In our model, a relatively high stress drop facilitates rupture transfer across the fracture network. Although considerable uncertainties in calculating stress drop from seismic observations exist \cite{abercrombie2021resolution}, relatively high stress drop may be observed in earthquakes across geometrically complex fault systems, for example, for the 1992 $\Mw$ 7.3 Landers earthquake, a high stress drop (averaging $\Delta \tau > 10$~MPa) is observed and modeled \cite{kanamori1992initial, wollherr2019landers} or for induced seismic events (e.g., \citeA{lengline2014fluid, huang2017stress, abercrombie2021resolution}.

\subsection{Fracture network cascading rupture dynamics assuming a different hypocenter location and varying pore fluid pressure ratios}
\label{subsec:results:triggering}
In this section, we analyze five dynamic rupture scenarios of cascading ruptures initiated within the fracture network, with a larger distance to the main fault. These additional simulations are motivated by fluid injection scenarios and showcase the viability of the fracture network cascading rupture dynamics under varying initial conditions. 

We consider rupture initiation at $\sim$1~km fault-normal distance to mimic an off-main fault disturbance at deeper hypocentral depth than in earlier cases. The hypocenter is changed to $x = -45$~m, $y = 1900$~m, and $z = -3430$~m. We initiate rupture on a single prescribed fracture which has multiple intersections with neighboring fractures. To initiate the rupture, we use a nucleation radius of $r_{nuc}=150$~m, smaller than in previous cases (Section \ref{subsec:results:dynamic}). We examine five scenarios based on the same 1~km distant off-main fault nucleation: (1) a reference scenario with the same hydrostatic pore fluid pressure condition (pore fluid pressure ratio $\gamma=0.37$), as before, and (2) and four scenarios testing higher pore fluid pressure ratio($\gamma>0.37$), leading to decreased effective normal stress. 

\subsubsection{Case 4: rupture initiation on a fracture away from the main fault}
\begin{figure}
    \centering
    \includegraphics[width=0.9\textwidth]{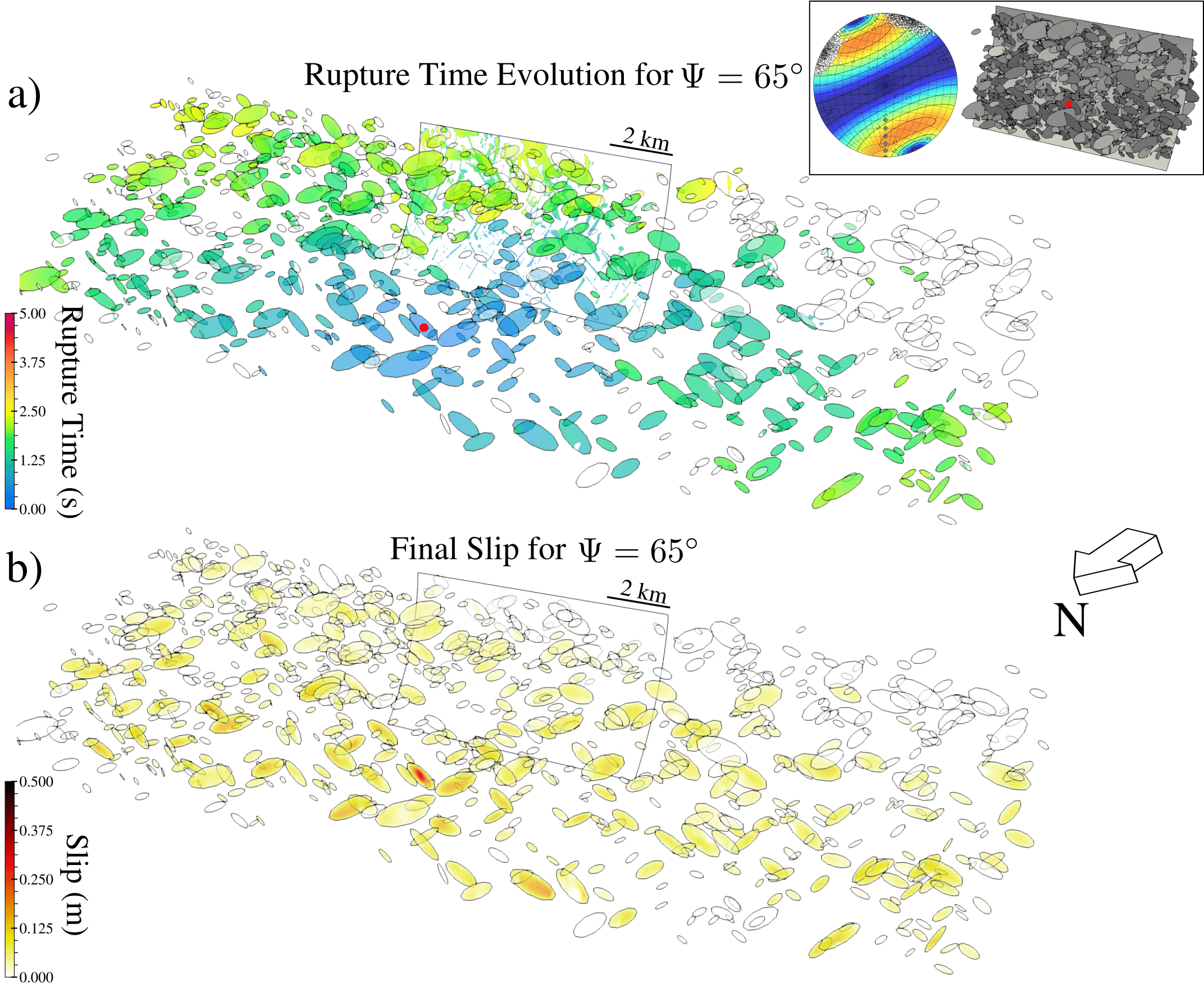}
    \caption{Rupture time (a) and final slip (b) for \textit{Case 4} in an exploded view. This scenario is similar to \textit{Case 2}, but the rupture nucleates on a single fracture in the damage zone,  (indicated by the red circle in the top-right inset) at 1 km distance from the main fault. The top right corner panel shows the stereonet plot of the relative pre-stress ratio ($\mathcal{R}$) in a lower hemisphere projection side-by-side with the original model.}
    \label{fig:farSource}
\end{figure}

The initial stress and fault/fracture strength conditions follow \textit{Case 2} (section \ref{case:65}, $\Psi=65^\circ$). The only difference is the location of the hypocenter which is now placed on a hanging wall fracture at $\sim 1$~km distance normal to the main fault (red circle and star in Figure \ref{fig:farSource}a). As in \textit{Case 2}, we observe a pure rupture cascade that does not lead to a runaway rupture on the main fault (Figure \ref{fig:farSource}b). The rupture duration is $t=3$~s, hence slightly shorter than \textit{Case 2}. However, the number of slipped fractures is higher (584 fractures slip), with more slipped fractures are located at the main fault's footwall.

\subsubsection{Case 5: rupture initiation within the fracture network with elevated pore fluid pressure}\label{section:S2}

\begin{figure}
    \centering
    \includegraphics[width=1\textwidth]{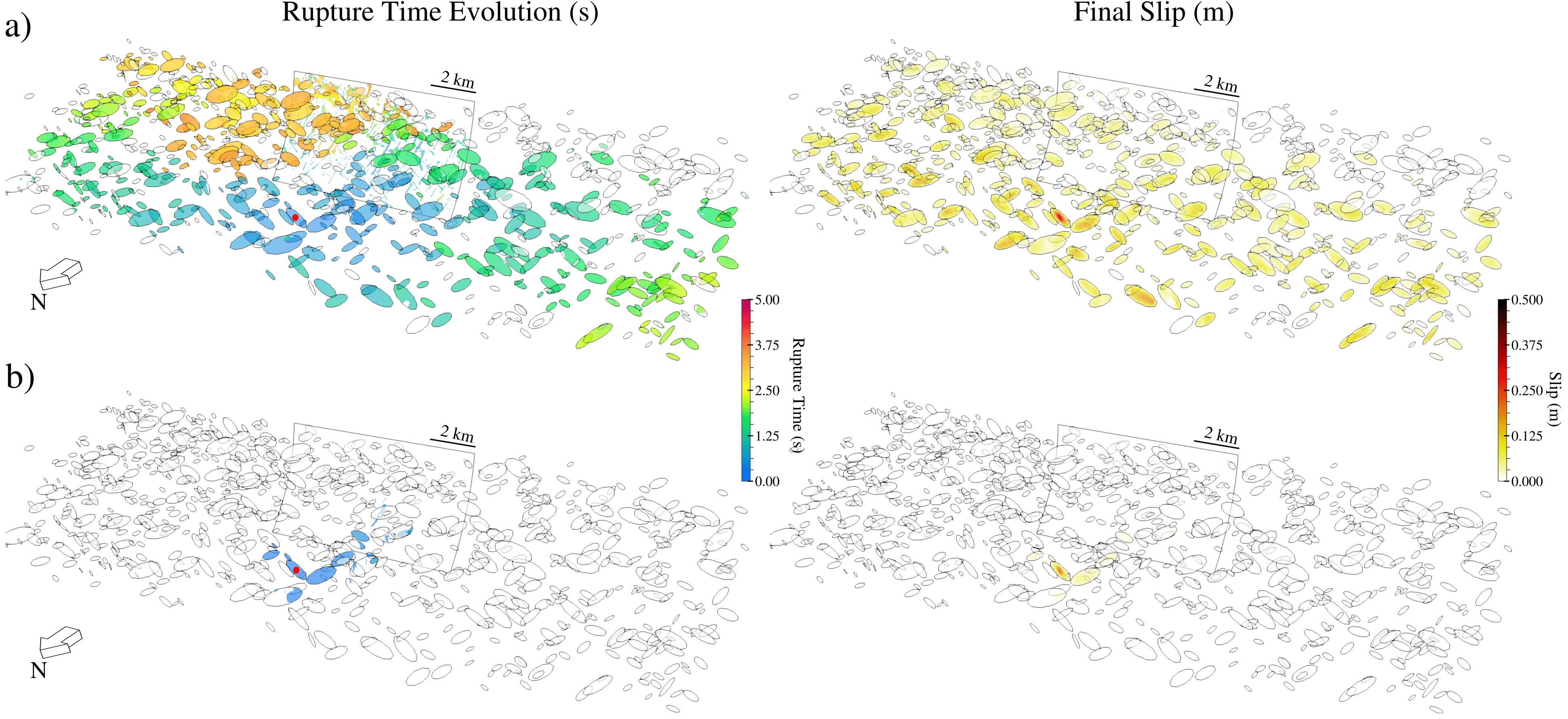}
    \caption{Exploded view of the rupture time evolution and final slip of \textit{Case 5} for two dynamic rupture cascades with pore fluid pressures larger than hydrostatic: a) $\gamma = 0.5$ and b) $\gamma=0.8$. Cases for $\gamma = 0.6$ and $\gamma = 0.7$ are shown in Figures \ref{fig:S6}a and \ref{fig:S6}b, respectively.}
    \label{fig:scenarioOverPressure}
\end{figure}

In this scenario, we apply initial stress similar to \textit{Case 2} (Section \ref{case:65}) but increase the fluid pressure ratio $\gamma$ to above hydrostatic levels, hence, effectively decreasing normal stress. We note that increasing fluid pressure does not increase fault criticality in this framework, as the maximum pre-stress ratio $\mathcal{R_0}$ is kept constant across simulations. We assume uniformly distributed $\gamma$ within the fracture network, across all fractures and the main fault. We consider four cases of overpressurized pore fluids: (1) $\gamma = 0.5$, (2) $\gamma = 0.6$, (3) $\gamma = 0.7$, and (4) $\gamma = 0.8$. 
When assuming $\gamma = 0.5$, the spatial slip distribution is comparable to using $\gamma = 0.37$ in \textit{Case 4} (Figure \ref{fig:scenarioOverPressure}a), however there a fewer slipped fractures (569 out of 854, i.e. $\sim 70\%$). Rupture activation is delayed within the footwall in the western part of the main fault at $t = 2.5$~s (Figure \ref{fig:scenarioOverPressure}a, Movie S12), leading to overall longer rupture duration ($t=4$~s) compared to \textit{Case 4}.  

As pore-fluid pressure increases, dynamic rupture cascade becomes less viable, because the radiated waves are of lower amplitude, and because the critical (re-)nucleation size inversely depends on effective normal stress, and thus, the number of slipped fractures decreases. Assuming $\gamma=0.6$, there are 241 slipped fractures, which is $\sim 29\%$ of all fractures (Figure \ref{fig:S6}a). $\gamma=0.7$ generates dynamic rupture only on $\sim 18\%$ of fractures (Figure \ref{fig:S6}b). If the pore-fluid pressure further increases to $\gamma = 0.8$, only about $3\%$ of fractures slip (Figure \ref{fig:scenarioOverPressure}b). Assuming increasing pore-fluid pressure, slip predominantly occurs within the fracture network on the hanging wall and close to the hypocenter.

None of the cascading rupture scenarios with overpressured pore fluids triggers a runaway rupture on the main fault.
Rupture kinematics of all scenarios are similar to our results for \textit{Case 2} (See \ref{section:rupture_kinematics_cases45}). Interestingly, \textit{Case 5} with $\gamma=0.5$ (Figure \ref{fig:MomentRateFluidInjection}) produces the longest rupture duration, because of the delayed triggering of a fracture, which also imprints a pronounced late MRF peak at $t=2.8 - 4.0$~s. The average stress drop also decreases as $\gamma$ increases.

\subsection{High-resolution seismic waveforms and spectral characteristics}
\label{subsec:results:waves}
Cascading and non-cascading ruptures are expected to result in different seismic waveform characteristics, which if observable, may provide insight into cascading rupture processes. We apply time-domain and Fourier spectral analyses to assess if there are notable differences in seismic-radiation properties.

\begin{figure}
    \centering
    \includegraphics[width=0.95\textwidth]{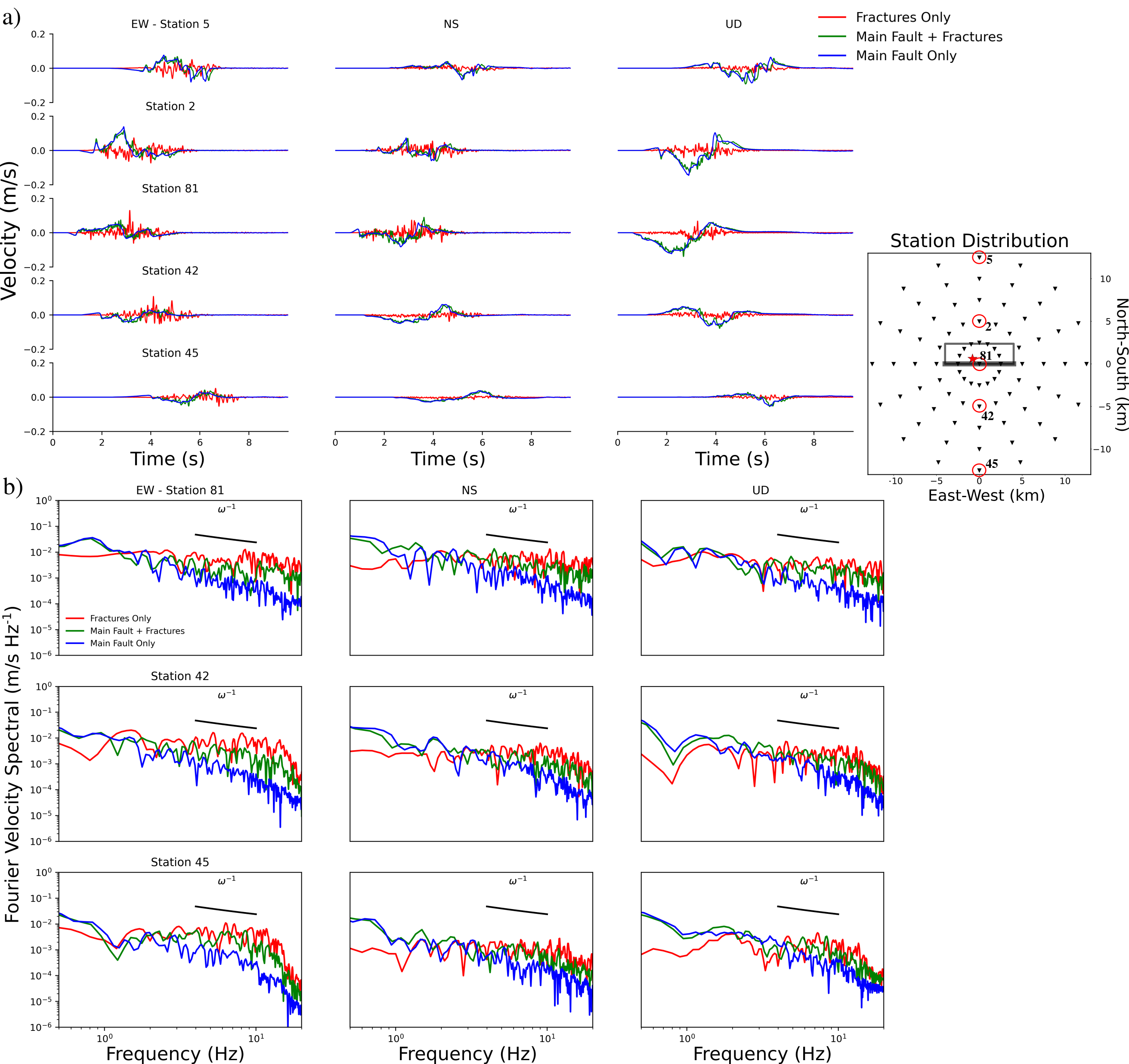}
    \caption{Waveform and Fourier amplitude spectra for selected stations (red circles in top-right inset of figure a). a) Velocity waveforms at selected stations for three cases: cascading rupture within the fracture network (fractures only, \textit{Case 6}) in red, rupture with off-fault fracture slip (main fault + fractures, \textit{Case 7}) in green, and rupture without fracture network (main fault only, \textit{Case 8}, in blue). b) Fourier amplitude spectra (in [m/s Hz$^{-1}$]) of the three different cases. The black line indicates an $\omega^{-1}$ spectral decay. The resolved frequencies reach $\sim 7$~Hz within a refined mesh region and reach up to 13~Hz closest to the fracture network, at station 81.}
    \label{fig:WaveformSpectral}
\end{figure}

For high-resolution wavefield modeling, we reduce the mesh element edge length to at most 250~m within a refinement volume (26 $\times$ 26 $\times$ 10 km$^3$). The resulting 88 million element mesh increases the computational cost of each forward simulation to 18h on 512 nodes (295,000 CPUh). This mesh resolves frequencies up to $\sim 7$~Hz within the refined volume and for constant $c_s = 3464$~m/s. Due to the statically adaptive mesh, resolution reaches up to $\sim 13$~Hz at station 81, closest to the fracture network (Figure \ref{fig:WaveformSpectral}). 
We analyze synthetic seismograms and Fourier amplitude spectra of ground velocity for three different scenarios using the high-resolution computational models: (1) cascading rupture within the fracture network (\textit{Case 6}: initial stress  as \textit{Case 2}, $\Psi=65^\circ$), (2) main fault rupture with off-fault fracture slip (\textit{Case 7}: initial stress as \textit{Case 3}, $\Psi=120^\circ$), and (3) main fault rupture without the fracture network (\textit{Case 8}: initial stress as \textit{Case 3}, $\Psi=120^\circ$). In \textit{Case 8}, we exclude the fracture network in the simulation and consider only the listric main fault. We save synthetic seismograms at 81 near-field stations, evenly spaced across all azimuths and located within a radius of 12.5~km from the center-top of the main fault. 

Figure \ref{fig:WaveformSpectral}a shows synthetic three-component velocity seismograms (in [m/s]) up to $ t = 9$~s simulation time for three high-resolution simulations \textit{Cases 6-8}, ordered from farthest North (station 5) to farthest South (station 45). Station 81 is located above the main fault.
Waveforms for \textit{Case 6} (cascading dynamic rupture within fracture network) have distinct signatures that differ from the other two cases, showing short wavelength amplitude variations (red in Figure \ref{fig:WaveformSpectral}a). The shaking duration at all stations is $\sim t = 4$~s. Qualitatively, waveforms for \textit{Case 6} show larger amplitudes on the east-west (fault-parallel) component than on the north-south (fault-normal) component, indicating predominant seismic radiation on the fault-parallel component, as expected from the strike-slip faulting mechanism. The cascading rupture shows unclear \textit{S-}wave onsets and non-typical coda waves (i.e., station 81 at $t>3$~s, Figure \ref{fig:WaveformSpectral}a) due to the continuous slip activation within the fracture network.

Seismograms for \textit{Cases 7} and \textit{8} are generally similar, on all components, however, \textit{Case 8} (blue in Figure \ref{fig:WaveformSpectral}) has lower higher-frequency components. The absence of higher-frequency components is explained by the rupture occurring only on the listric fault, without dynamic rupture complexities induced by off-fault fractures. In both cases, higher amplitudes are observed on the vertical (UD) components for stations located near the hanging wall of the listric fault (stations 2 and 5 in Figure \ref{fig:WaveformSpectral}a), in agreement with previous studies by \citeA{ofoegbu1998mechanical,passone2017kinematic,rodgers2019effect,moratto2023near}.

Figure \ref{fig:WaveformSpectral}b shows the Fourier velocity spectra (in [m/s Hz$^{-1}$]) for stations 81, 42, and 45. The spectra exhibit an $\omega^{-1}$ high-frequency decay up to the highest resolved frequency. The \textit{Case 6} produces stronger high-frequency radiation than the other two cases (noticeable already in the waveforms), which in contrast have higher energy in the low-frequency band. This suggests that cascading rupture may generate stronger high-frequency content due to the dynamic complexities governing its rupture processes, including abrupt rupture termination, acceleration, and deceleration of rupture fronts when branching and jumping across multiscale fractures.

\section{Discussion}\label{discussion}

Our 3D dynamic rupture simulations in a geometrically complex fault network generate a rich set of results that raise a number of questions and implications. Below, we discuss conditions that lead to cascading rupture, implications of the connectivity and distribution of fractures, and overall source characteristics.

\subsection{Static and dynamic conditions leading to cascading ruptures}

Our study identifies at least three conditions that promote cascading dynamic rupture within the fracture network. First, the state evolution slip distance must scale with fracture and fault size to furnish the minimum fracture energy needed for rupture growth \cite{garagash2022fracture}. Second, at least one fracture family must have a favorable relative pre-stress ratio ($\mathcal{R} \geq 0.6$), while the other family of fractures must have at least a conditionally favorable pre-stress ratio ($0.3 \leq \mathcal{R} < 0.6$). Higher $\mathcal{R}$ implies a more favorable orientation towards the ambient stress and initial shear stresses closer to critical. Third, fractures must be connected or densely packed to allow for sufficient stress transfer. Finally, when considering only one fracture family, the dynamic rupture cascade is suppressed (Figure \ref{fig:S_oneFamily}).

Cascading dynamic rupture can occur independent of the prescribed hypocenter location if all of the above conditions are fulfilled. The hypocenter may be located near the main fault or at the periphery of the fracture network and generate equally sustained rupture cascades. We show two hypocenter locations producing volumetric earthquakes of moment magnitude $\Mw \approx 5.5$.  The cascading earthquake magnitude is restricted by the dimension and distribution of the fracture network. Hence, rupture size may grow for a larger fault and its associated fracture network. 

\subsection{Multiple rupture fronts and re-nucleation}

We observe that during fracture network rupture cascades, fractures may experience repeated nucleation while being ruptured by multiple rupture fronts, particularly if a fracture intersects with more than one other fracture (see Figure \ref{fig:S4}, Movies S3b, S4b, S5b, S6b).
We identify three distinct dynamic mechanisms (1) near-simultaneous nucleation at two or more locations on a single fracture (Figure \ref{fig:S4}a), (2) sequential nucleation at two or more locations on a single fracture (Figure \ref{fig:S4}b), and (3) repeated nucleation at the same point on a fracture (Figure \ref{fig:S4}c). In the first mechanism, rupture fronts from two neighboring fractures simultaneously reach the edge of the same intersecting fracture. In the second mechanism, rupture initiates at different times in response to the timing of neighboring intersecting fractures slipping. The last mechanism occurs when two or more rupture fronts sequentially pass through the same fracture-fracture intersection. Mechanisms 2 and 3 often occur together at the same fracture. In rare cases, for $\sim 1\%$ of the ruptured fractures during a cascade, we observe more than two rupture initiations on a single fracture.

\subsection{Fracture connectivity}

Our simulations demonstrate that connected fractures facilitate cascading rupture. Compared to rupture jumping, rupture branching is significantly more effective: $> 90\%$ of all fractures that slip during a rupture cascade connect by direct branching. As a result, dynamic rupture cascades arrest when encountering unconnected fractures, even though self-sustained rupture can continue toward connected fractures (Figure \ref{fig:S4}d). In our model, dynamic rupture jumping does not occur beyond an inter-fracture spacing of 80~m. This distance is observed as the largest fracture-fracture distance where wave-transmitted dynamic stresses from a neighboring unconnected fracture are sufficient to nucleate self-sustained rupture, i.e., sufficiently overstress an area matching the critical nucleation size of a distant unconnected fracture. While this distance depends on our frictional and geometrical model parameterization, we infer that densely spaced fractures favor dynamic rupture cascades. 

\subsection{Distribution of slip within a fracture network rupture cascade}

The two conjugate fracture families produce left- and right lateral slip, with slip-direction (rake angle) $0^\circ$ and $180^\circ$ for fracture families 1 and 2, respectively. For main fault rupture-induced slip within the fracture network, slip accumulates on the dilatational side of the rupture propagation direction (Figure \ref{fig:S3_3}). This slip distribution resembles deformation patterns occurring in dynamic rupture simulations due to off-fault damage or off-fault plastic yielding (e.g., \citeA{dalguer2003simulation, ando2007effects, okubo2019dynamics, gabriel2021unified, andrews2005rupture, templeton2008off, gabriel2013source}). 
Slip across the fracture network is driven by the intricate interaction of static and dynamic factors, which include dynamic stresses due to seismic waves, static Coulomb stresses due to evolving slip, and clamping and unclamping due to variations in normal stress. Hence, different realizations of fracture network dynamic rupture models regarding distribution, size, spacing, and connectivity of fractures may result in different slip patterns. However, we expect the major slip pattern characteristics of the cascading rupture to remain the same for a different but statistically similar fracture network model, such as in \citeA{garagash2022fracture}, which includes a planar, vertical main fault under strike-slip loading.

\subsection{Source characteristics of cascading ruptures}

The equivalent point-source moment tensor representation of cascading and non-cascading ruptures changes from strike-slip to thrust faulting. Earthquakes across multiple large faults can promote non-DC moment tensor solutions by superposing distinct DC components \cite{julian1998non, palgunadi2020dynamic}. Combining several focal mechanisms may still produce a DC moment tensor solution \cite{julian1998non}. For our set of cascading ruptures within the fracture network, we find that the equivalent moment tensors are characterized by insignificant non-DC components ($\sim -1.8\%$ to $-5\%$). We note that both fracture families are conjugated, and therefore have similar average focal mechanisms. However, if the main fault is activated during cascading rupture a significant non-DC component emerges.   

Complex moment rate functions have been interpreted as signatures of large-scale multi-fault ruptures \cite{vallee2011scardec, holden20172016, ando2018dynamic, wollherr2019landers, ulrich2019dynamic}. Multiple peaks in the MRF have also been attributed to other seismic source complexities either due to heterogeneous pre-stress on a planar fault \cite{ripperger2007earthquake}, fractally rough fault surfaces \cite{ shi2013rupture, zielke2017fault, tal2018effects, danre2019earthquakes}, or non-uniform frictional parameters (e.g., variable characteristic slip distance ($D_c$) on a single fault plane \cite{renou2022deciphering}). Our study reveals that a small-scale fracture network can also generate multi-peak MRFs corresponding to multiple sub-events on fractures.

One of the main features of cascading rupture is the slow cascading speed $v_C$, despite localized occurrences of supershear rupture speed within the fracture network. We hypothesize that the cascading speed $v_C$ may appear as the ``true" single-fault rupture speed $v_{R, tot}$ if observed from a distance, i.e., observations of low rupture speeds may be at least partially explained by cascading rupture on a complex fracture network. The 2019 $\Mw$ 7.1 Ridgecrest earthquake is a good example. Its inferred low rupture speed (1.8 - 2.0~km/s, \cite{chen2020cascading}, which is still faster than our pure cascading speed) could be affected by the fact that rupture propagated through a geometrically complex fault system and activated several off-fault fractures \cite{ross2019hierarchical, xu2020coseismic, taufiqurrahman2023dynamics}.
We expect that $v_C$ may differ for other fracture-network geometries, distributions in fracture size and density. However, the analysis of alternative fracture-network configurations is beyond the scope of this study. Intuitively, we expect cascading rupture involving complicated fault geometries to generate lower $v_C$ than the surface-averaged rupture speed of the slipping fractures and main fault ($v_{R,ave}$), as rupture accelerates and decelerates during branching and jumping across many fractures.

\subsection{Limitations}

Due to the high computational demands of each simulation and the challenges of generating the 3D computational mesh accounting for multiscale intersecting fractures and faults, the results in this study are limited to one specific fracture network configuration, including fixed fracture orientation, distribution and size. Determining an exact fracture network parameterization for a particular rock volume or geo-reservoir is challenging. Often, statistical methods are used to constrain a fracture network because there is no reliable direct method for measuring the field-scale 3D fracture distribution. The geometry and distribution of fractures can vary significantly based on geological, tectonic, and mechanical factors. These factors likely affect the dynamics of a cascading earthquake. For example, removing one fracture family causes a cascading rupture to cease prematurely (Figure \ref{fig:S_oneFamily}).

For simplicity, we consider a homogeneous elastic-isotropic material. We here want to ensure that the subsurface structure does not affect the rupture dynamics and seismic radiation and does not mask the first-order physical signatures we intend to examine. However, natural fault zones comprise not only multi-scale fractures but also a damage zone around the fault core with lower elastic moduli and seismic wave speeds than the surrounding primary host rock, all of which may affect dynamic rupture and seismic-wave radiation \cite{harris1997effects, huang2011pulse, huang2014earthquake}. We do not account for co-seismically induced distributed damage or off-fault plastic deformation that may interact with discrete fractures \cite{andrews2005rupture, gabriel2013source, xu2015dynamic, gabriel2021unified}. 

We also ignore aseismic and poroelastic processes that may additionally affect static and dynamic stress conditions in a fault zone, particularly for overpressurized fluid conditions (e.g., \citeA{segall2015injection, eyre2019role}). Exploring the effects of additional physics, such as viscoelastic attenuation, fault zone anisotropy, fault roughness, or off-fault plasticity (e.g., \citeA{ wollherr2018off, wolf2020optimization, taufiqurrahman2022broadband}),  in our simulation framework will be readily possible in future work. 

\section{Conclusions}

We present eighteen 3D dynamic rupture simulations within a complex fracture network of more than 800 intersected multiscale fracture planes surrounding a listric main fault. We vary prestress conditions, hypocenters, and fluid overpressure and analyze general aspects of seismic wave radiation. 
Our dynamic models reveal three mechanisms that promote cascading rupture: (1) the state evolution slip distance scales with fracture size \cite{garagash2022fracture}, (2) at least one fracture family should have a favorable relative pre-stress ratio ($\mathcal{R} \geq 0.6$), and the other fracture family should have at least a conditionally to favorable pre-stress ($0.3 \leq \mathcal{R} < 0.6$), and (3) fractures within the fracture network are connected or densely packed. 

Our simulations demonstrate the possibility of pure dynamic rupture cascades sustained within the fracture network that cannot trigger self-sustained runaway rupture on an unfavorably oriented main fault. Sustained cascading ruptures pertain under earthquake initiation at a single fracture distant from the main fault. Dynamic rupture on the main fault can promote limited off-fault fracture slip, even on unfavorably oriented fractures. 
Our modeled cascading ruptures within the 3D fracture network can generate moment magnitudes up to $\Mw \approx 5.6$, without activating the main fault. Our study thus has important implications for estimates of seismic hazard of a known fault system and for multiscale fracture networks in actively exploited geo-reservoirs.

We identify potentially observable characteristics of sustained cascading rupture within a fracture network as: (1) multiple peaks in the moment rate function, (2) equivalent moment tensor misaligned with respect to the strike of a known fault, (3) slow cascading speed, (4) high-stress drop, and (5) seismograms enriched in high frequencies.

\acknowledgments
The authors thank the members of the Computational Earthquake Seismology (CES) group at KAUST for many fruitful discussions and suggestions. We thank SeisSol's core developers (see \url{www.seissol.org}). Computing resources were provided by King Abdullah University of Science and Technology, Thuwal, Saudi Arabia (KAUST, Project k1587, k1488 and k1343 on Shaheen II). The work presented in this article was supported by KAUST Competitive Research Grant (FRacture Activation in Geo-reservoir–physics of induced Earthquakes in complex fault Networks [FRAGEN], URF/1/3389-01-01, and BAS/1339-01-01). 
AAG acknowledges support by the European Union’s Horizon 2020 Research and Innovation Programme (TEAR grant number 852992), Horizon Europe (ChEESE-2P grant number 101093038, DT-GEO grant number 101058129, and Geo-INQUIRE grant number 101058518), the National Aeronautics and Space Administration (80NSSC20K0495), the National Science Foundation (grant No. EAR-2121666) and the Southern California Earthquake Center (SCEC awards 22135, 23121). TU and AAG acknowledge support from the Bavarian State Ministry for Science and Art in the framework of the project Geothermal-Alliance Bavaria. DIG acknowledges support by the Natural Sciences and Engineering Research Council (Discovery Grant 05743). 
Part of the analysis was implemented using Obspy \cite{beyreuther2010obspy}. Figures were prepared using Paraview \cite{ahrens2005paraview} and Matplotlib \cite{hunter2007matplotlib}.

\section{Data and Resources}\label{section:data_resources}
The version of SeisSol used in this study is described in \url{https://seissol.readthedocs.io/en/latest/fault-tagging.html#using-more-than-189-dynamic-rupture-tags} with commit version \verb|917250fd|. Another alternative can be retrieved from SeisSol for hundreds of fault tagging in branch SeisSol64FractureNetwork (\url{https://github.com/palgunadi1993/SeisSol/tree/SeisSol64FractureNetwork}). Patched meshing software PUMGen can be cloned from github branch PUMGenFaceIdentification64bit (\url{https://github.com/palgunadi1993/PUMGen/tree/PUMGenFaceIdentification64bit}). Instructions for downloading, installing, and running the code are available in the SeisSol documentation at \url{https://seissol.readthedocs.io/}. Compiling instructions: \url{https://seissol.readthedocs.io/en/latest/compiling-seissol.html}. Instructions for setting up and running simulations: \url{https://seissol.readthedocs.io/en/latest/configuration.html}. All input and mesh files are available in the Zenodo repository at \url{https://doi.org/10.5281/zenodo.8026705}.

\appendix
\counterwithin{figure}{section}

\section{Numerical Method}\label{section:numerical_method}
We solve for spontaneous frictional failure and seismic wave propagation jointly on the listric main fault and all 854 fractures.
We use the open-source software SeisSol (\url{https://github.com/SeisSol/SeisSol}) that couples seismic wave propagation in 3D Earth structure  and frictional fault failure \cite{ pelties2014verification, uphoff2017extreme}. SeisSol uses a flexible nonuniform unstructured tetrahedral mesh with static mesh adaptivity that allows for geometrically complicated fractures and faults embedded in a three-dimensional Earth (\ref{mesh_generation}).
SeisSol employs a Discontinuous Galerkin (DG) method using Arbitrary high-order DERivative (ADER) time stepping \cite{kaser2006arbitrary, dumbser2006arbitrary}. SeisSol is optimized for current multi-petascale supercomputer systems \cite{breuer2014sustained, heinecke2014petascale, rettenberger2016asagi, uphoff2020yet,wolf2020optimization, wolf2022efficient,dorozhinskii2021seissol} utilizing local time stepping \cite{breuer2016petascale} that enables an up to 10-fold simulation speed up for our computational mesh (\ref{mesh_generation}). 
SeisSol has been verified in several community benchmarks, including dynamic rupture simulations with fault branching, dipping faults, and heterogeneous on-fault stresses \cite{harris2011verifying, harris2018suite} and in analytical verification problems for seismic wave propagation \cite{uphoff2016generating,wolf2022efficient}.
Dynamic rupture simulations are sensitive to the geometrical complexity of faults \cite{dunham2011earthquake, kozdon2013simulation, ando2017dynamic, wollherr2018off,  ando2018dynamic, ulrich2019dynamic, kyriakopoulos2019dynamic, zhang2019dynamic, palgunadi2020dynamic, lozos2020dynamic}. We use high-order basis functions of polynomial degree $p=4$ achieving $\mathcal{O}$5-accuracy in wave propagation in space and time for all simulations. Achieving high spatial and temporal resolution is crucial for resolving the detailed spatiotemporal evolution of the rupture processes governed by the variable process zone size in our frictional parameterization and the geometrically complicated fault planes, including the listric main fault geometry and numerous fault-fracture intersections. Since the process zone size varies considerably across our fractures, we measure and ensure to resolve the minimum process zone size following \citeA{wollherr2018off}. Our highest on-fault resolution is 4~m (for the smallest fractures of size 100~m), resolving the minimum cohesive zone of 8~m with on average 2 fifth-order accurate elements, that is 12 Gaussian integration points, ensuring highly accurate results. One high-resolution dynamic rupture simulation with the 45 million cell mesh requires 12.5h on 512 nodes on Shaheen II which is equivalent to 204,800 CPUh.

\section{Equivalent Moment Tensor Calculation}
\label{section:moment_tensor_solution}
We determine an equivalent moment tensor representation for each dynamic rupture simulation. We assume constant rigidity $\mu$ and slip output of dynamic ruptures. On each triangular fault face $i$ which is associated with a dynamic rupture boundary condition within the tetrahedral mesh, we calculate an element-local seismic moment ($M_i$) based on slipped area ($A_i$) and slip ($\delta_i$) as \(M_i = \mu \times A_i \times \delta_i\). The total seismic moment $M_0$ is the summation of all slipped faces $M_i$. For a given triangular fault face $i$ with strike ($\alpha_i$), dip ($\beta_i$), and rake ($\varphi_i$), where $\varphi_i = \text{arctan2}(\text{[slip in dip direction]$_i$}/ \text{[slip in strike direction]$_i$}) \times 180/\pi$, we calculate an element-local moment tensor following \citeA{lay1995modern}. The equivalent moment tensor is then defined as the summation of all fault-local moment tensors of each fault element $i$.

\section{Mesh Generation}\label{mesh_generation}
For this study, we construct the geometry of the fracture network using the third-party commercial software FRACMAN \cite{dershowitz2019fracman}. The boundary of the numerical domain is defined by employing the open-source mesh generator gmsh \cite{geuzaine2009gmsh} in a Cartesian coordinate system. A large numerical domain of 40 $\times$ 40 $\times$ 20 km$^3$ is used to mitigate the effect of expected reflected waves from the absorbing boundary. We select the shortest fracture length to limit small-scale fracture intersections, whose size restricts the smallest time step width. We discretize the unstructured tetrahedral mesh using Simmodeler \cite{simmodeler}. Mesh-element edge lengths vary depending on the size of the process zone to ensure convergence and to numerically resolve the dynamic fault strength drop behind the rupture front. Following \citeA{wollherr2018off}, the on-fault mesh-element edge lengths vary from 4\,m for the smallest ($R$=100\,m) to 45\,m for the largest fracture size ($R$=500\,m). We gradually increase the mesh-element edge size of the tetrahedral mesh by a factor of $6\%$ away from each fracture and fault plane to save computational cost while avoiding reflection from the domain boundary. Consequently, the edge length on the main listric fault reaches a maximum size of 100\,m due to its connected and intersected surface with small fractures. In total, the computational domain is composed of 49 million volume elements for fifteen dynamic rupture simulations (\textit{Cases 1, 1a, 1b, 2, 2a, 3, 4,} and \textit{5}). For \textit{Cases 6, 7,} and \textit{8}, to improve the temporal resolution of seismic wavefield, we reduce the mesh element edge length to a constant value 250~m within refinement volume 26 $\times$ 26 $\times$ 10 km$^3$, resulting in 88 million volume elements.

\section{Rupture Initiation}\label{Appendix: Rupture Initiation}
\begin{figure}
    \centering
    \includegraphics[width=0.95\textwidth]{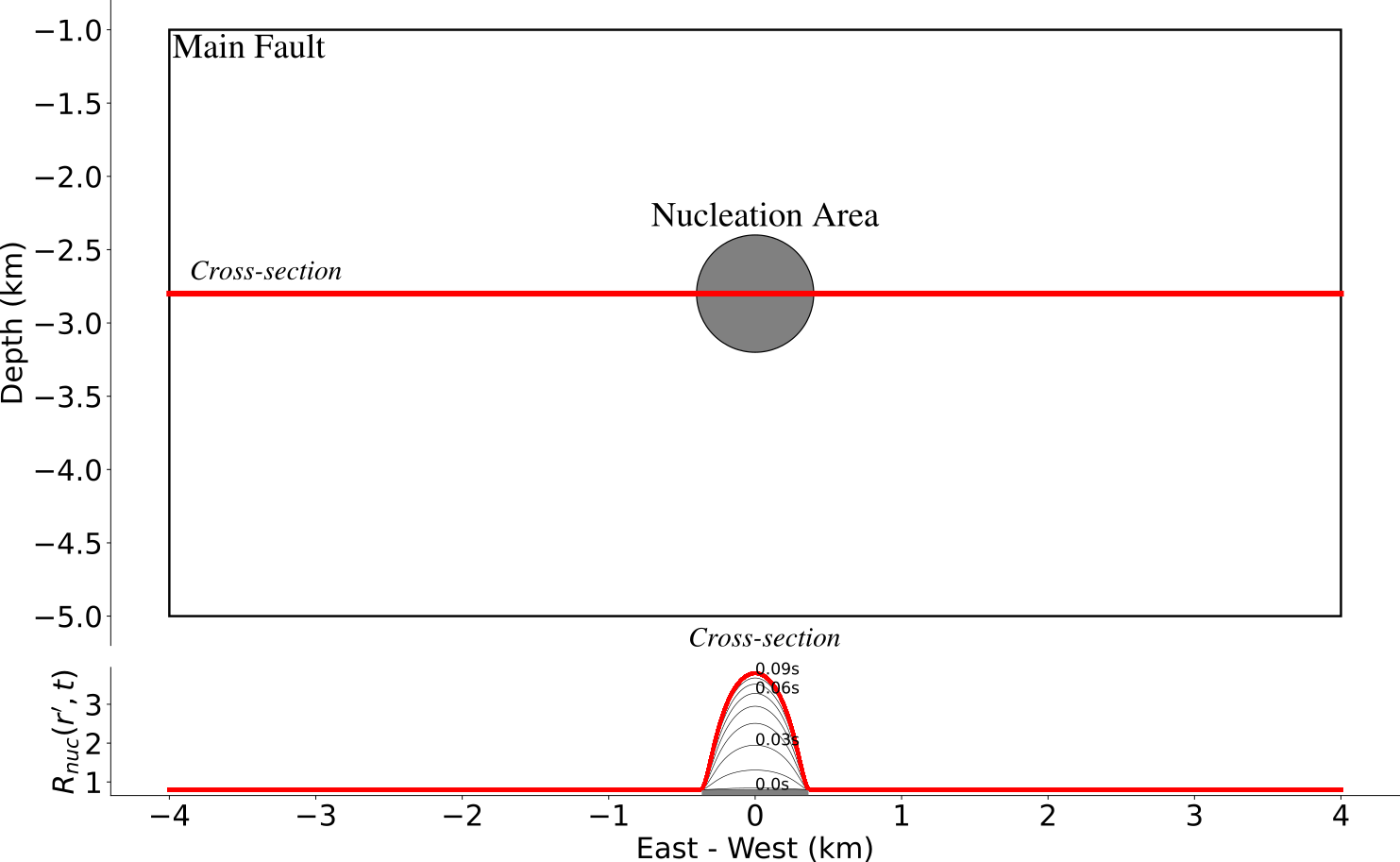}
    \caption{Elevated stress during rupture nucleation. $\mathcal{R}_{nuc}(r',t)$ denotes relative pre-stress ratio with maximum value of 3.5.}
    \label{fig:initial_stress}
\end{figure}

Dynamic earthquake rupture simulations are commonly initiated by assigning a small area on the fault as time-dependent overstressed or reduced in strength; this area is the predefined nucleation zone (or hypocenter). We apply a time-dependent over-stress centered at the hypocenter location selected for each scenario (Figure \ref{fig:initial_stress}). We choose the nucleation radius based on the numerical solution provided by \citeA{galis2015initiation}. Given their mathematical expressions and our initial pre-stress loading conditions, the estimated nucleation radius ($r_{nuc}$) is 400\,m.
The time-dependent stress increase within nucleation area $\mathcal{R}_{nuc}$ is calculated by increasing relative pre-stress ratio $\mathcal{R}_0$ as

\begin{equation}
    \mathcal{R}_{nuc}(t) = \mathcal{R}_0 + \Omega(r') \times S(t)
\end{equation}
where $\Omega(r')$ is a Gaussian step function, $r'$ is the radius from the hypocenter, and $S(t)$ is a smoothed step function. The Gaussian step function is given by

\begin{equation}
\begin{split}
    \Omega(r') &= \xi \exp\left( \frac{r'^2}{r'^2 - r_{nuc}^2} \right), ~ \forall ~ r' < r_{nuc} \\
    \Omega(r') &= 0, ~ \text{otherwise}
\end{split}
\end{equation}
$\xi$ is the initial pre-stress ratio inside the nucleation patch. We set $\xi = 3$. The smoothed step function is formulated as

\begin{equation}
\begin{split}
    S(t) &= \exp \left( \frac{(t-T)^2}{t \times (t - 2\times T)} \right), ~ \text{for}~ 0<t<T \\
    S(t) &= 1, ~ \text{for} ~ t\geq T
\end{split}
\end{equation}
$T$ indicates the nucleation time when the overstress is applied, chosen here as $T=0.1s$. We apply a similar nucleation procedure for \textit{Cases 4} and \textit{5}, but with a smaller nucleation size of $150$~m  (Section \ref{subsec:results:triggering}).

\section{Rupture Kinematics for \textit{Cases 4} and \textit{5}}
\label{section:rupture_kinematics_cases45}

\textit{Case 4} differs from \textit{Case 2} only in the nucleation procedure (on a single fracture vs. volumetric) and nucleation location (distant vs near main fault).
In \textit{Case 4}, the cascading rupture produces kinematic source parameters similar to those of \textit{Case 2}, such as $v_{R,ave}$ and total moment magnitudes. Only slight variations are observed in the average stress drop ($\Delta \tau$) and the equivalent point source moment tensor (Table \ref{table:ruptureKinematicFluidInjection}) with \textit{Case 2}. Thus, even though the cascading rupture occurs at two different locations with identical $\Psi$, the point-source parameters are similar. 

\begin{table}
  \caption{Summary of Rupture Kinematics for \textit{Cases} \textit{4} and \textit{5}. $v_{R,ave}$ represents the average rupture speed. $c_s$ denotes the shear wave speed. $\Delta \tau$ indicates the average stress drop. $M_{w,f}$ is the moment magnitude of the fracture network. $M_{w,F}$ is the moment magnitude of the main fault. $\Mw$ denotes the overall moment magnitude. MTS stands for equivalent moment tensor solution.}\label{table:ruptureKinematicFluidInjection}
  \centering
  \resizebox{\columnwidth}{!}{%
  \begin{tabular}{  c  c  c c c  c  c  c  c }
    \hline
    \multirow{2}{*}{\textbf{Scenario}} & \multirow{2}{*}{\boldmath{$\frac{v_{R,ave}}{c_s}$}} & \multirow{2}{*}{\boldmath{$\Delta\tau$} (MPa)} & \multirow{2}{*}{\boldmath{$M_{w,f}$}} & \multirow{2}{*}{\boldmath{$M_{w,F}$}} & \multirow{2}{*}{\boldmath{$\Mw$}} & \multirow{2}{*}{\textbf{MTS}} & \multicolumn{2}{c}{\textbf{Strike/Dip/Rake (\boldmath{$^\circ$})}}\\ 
    \cline{8-9} 
     & & & & & & & Plane 1 & Plane 2 \\
    \hline
    
    \textit{Case 4}  &  0.90 & 9.1 & 5.51 & 3.84 & 5.52 &
    \begin{minipage}{.1\textwidth}
    \centering
      \includegraphics[width=0.5\linewidth]{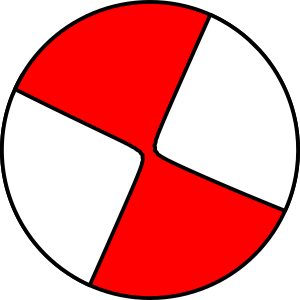}
    \end{minipage}
    & 114/84/2 & 24/88/174
    \\ 
    
    \textit{Case 5}, $\gamma = 0.5$  &  0.90 & 7.3 & 5.43 & 3.66 & 5.44 &
    \begin{minipage}{.1\textwidth}
    \centering
      \includegraphics[width=0.5\linewidth]{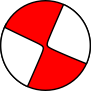}
    \end{minipage}
    & 114/84/3 & 24/97/174
    \\ 
    
    \textit{Case 5}, $\gamma = 0.6$  &  0.90 & 6.3 & 5.13 & 3.16 & 5.14 &
    \begin{minipage}{.1\textwidth}
    \centering
      \includegraphics[width=0.5\linewidth]{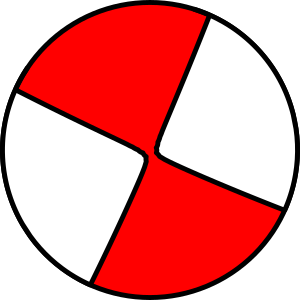}
    \end{minipage}
    & 114/83/2 & 24/88/173
    \\ 
    
    \textit{Case 5}, $\gamma = 0.7$  &  0.89 & 4.4 & 4.87 & 2.96 & 4.87 &
    \begin{minipage}{.1\textwidth}
    \centering
      \includegraphics[width=0.5\linewidth]{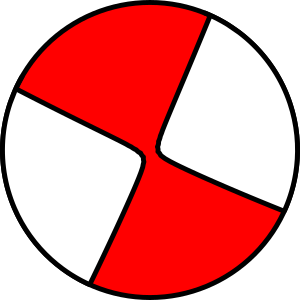}
    \end{minipage}
    & 114/84/1 & 24/89/174
    \\ 
    
    \textit{Case 5}, $\gamma = 0.8$  &  0.88 & 2.6 & 4.32 & 2.5 & 4.32 &
    \begin{minipage}{.1\textwidth}
    \centering
      \includegraphics[width=0.5\linewidth]{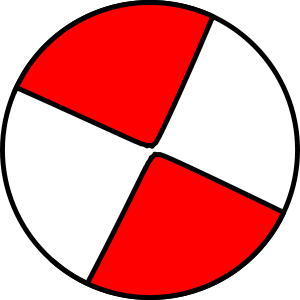}
    \end{minipage}
    & 115/87/3 & 25/87/177
    \\ 
    
    \hline
  \end{tabular}
  }
\end{table}

In the five \textit{Case 5} simulations, we increase the fluid pressure ratio $\gamma$, leading to decreasing $v_{R,ave}$ and $\Delta \tau$ (Table \ref{table:ruptureKinematicFluidInjection}). All \textit{Case 5} simulations yield similar focal mechanisms, despite different numbers and locations of slipped fractures. All moment tensors are essentially double couple solutions, with $M_{DC} \geq 97\%$, $M_{CLVD}<3\%$ and $M_{ISO} \approx 0\%$.

\begin{figure}
    \centering
    \includegraphics[width=0.95\textwidth]{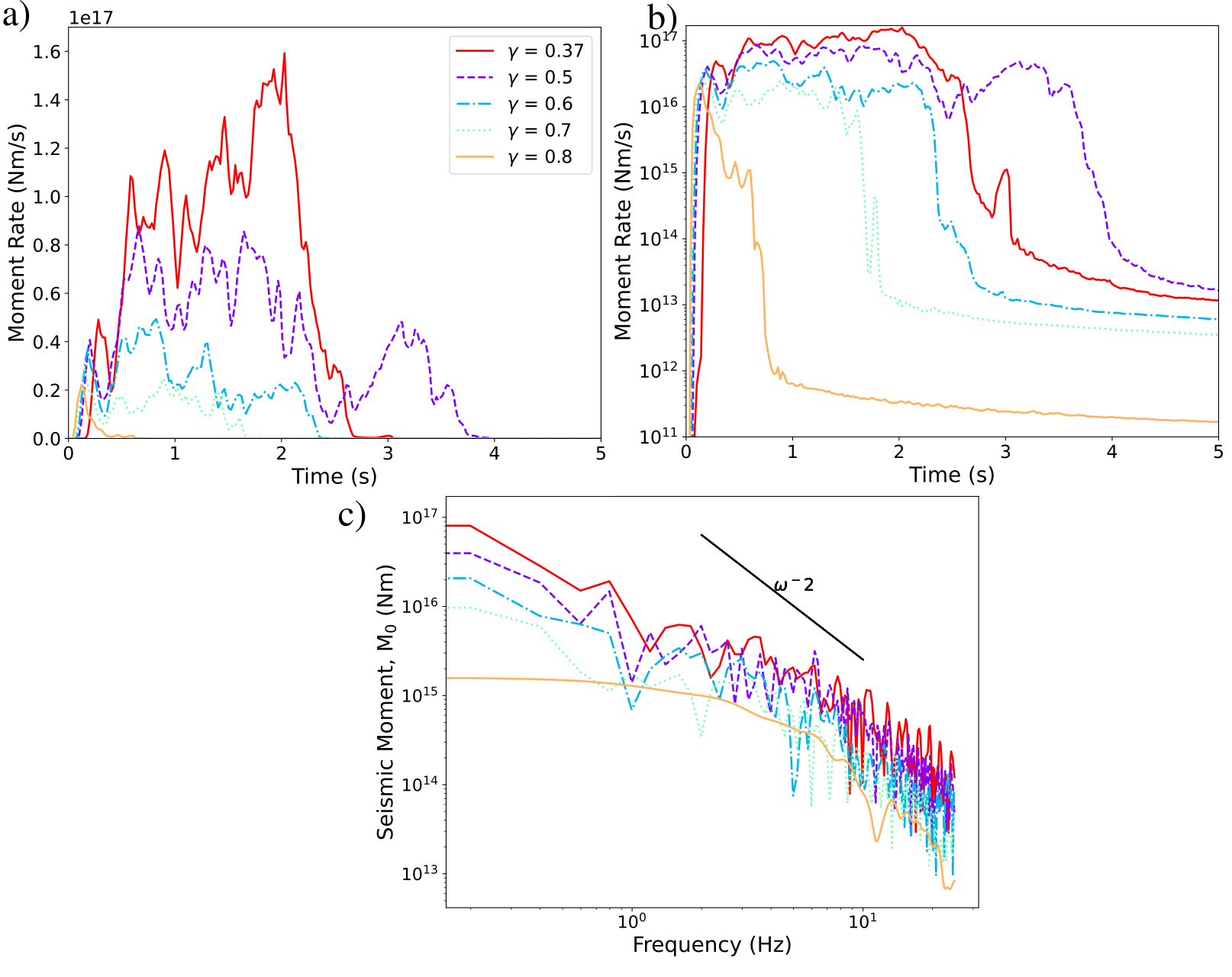}
    \caption{Moment rate function (a), moment rate function in logarithmic scale (b), and seismic moment spectra (c) of scenarios with remote nucleation, which may resemble induced events, nucleated by fluid injection and migrating towards a main fault. The red line indicates \textit{Case 4}. The dashed violet line demonstrates \textit{Case 5} for $\gamma=0.5$. The dash-dotted blue line denotes \textit{Case 5} for $\gamma=0.6$. The dotted green line illustrates \textit{Case 5} for $\gamma=0.7$. The orange line shows \textit{Case 5} for $\gamma=0.8$. The black line shows $\omega^{-2}$ spectral decay.}
    \label{fig:MomentRateFluidInjection}
\end{figure}

The analysis of both scenarios reveals multiple peaks in the moment rate function, as shown in Figures \ref{fig:MomentRateFluidInjection}a and \ref{fig:MomentRateFluidInjection}b. \textit{Case 4} generates higher seismic moment spectra than the scenarios of \textit{Case 5}, and the seismic moment decreases as $\gamma$ increases (see Figure \ref{fig:MomentRateFluidInjection}c). \textit{Case 4} also releases a higher seismic moment in a shorter time than \textit{Case 5} with $\gamma=0.5$. As the value of $\gamma$ increases, the moment rate function becomes lower in amplitude and shorter in time, eventually yielding a simple triangular-shaped function for $\gamma = 0.8$ (orange line in Figure \ref{fig:MomentRateFluidInjection}). The seismic moment spectrals also show $\omega^{-2}$ decays for their high-frequency part.

\appendix

\clearpage
\begin{glossary}
    \term{$\alpha$}
    Strike angle
    \term{$A$}
    Slipped area
    \term{$\Sigma$}
    Domain area of slipped fracture or fault
    \term{$A_\phi$}
    Relative stress magnitude
    \term{$a$}
    Direct effect for rate and state friction law
    \term{$\beta$}
    Dip angle
    \term{$b$}
    Evolutional effect for rate and state friction law
    \term{$c$}
    Frictional cohesion
    \term{$C$}
    Fault-specific constant to define fracture density
    \term{$c_s$}
    $S$-wave velocity
    \term{$d$}
    Distance from the main fault
    \term{$\delta$}
    Absolute slip on fault/fracture
    \term{$D_c$}
    Critical slip distance
    \term{$f_0$}
    Steady-state friction coefficient at $V_0$
    \term{$f_w$}
    Fully weakened friction coefficient
    \term{$f_p$}
    Peak friction coefficient
    \term{$f$}
    Frequency
    \term{$f_{SS}$}
    Steady-state friction coefficient
    \term{$f_{LV}$}
    Low-velocity steady-state friction coefficient
    \term{$\gamma$}
    Fluid pressure ratio
    \term{$g$}
    Gravitational force
    \term{$G$}
    Fracture energy
    \term{$G_{c}(R)$}
    Minimum fracture energy linked to fault size
    \term{$L$}
    Fracture-size-dependent evolution slip distance
    \term{$\lambda$}
    Lam\'e constant
    \term{$m$}
    Power-law constant
    \term{$M_{w,f}$} 
    Moment magnitude of slip in the fracture network
    \term{$M_{w,F}$}
    Moment magnitude of slip on the main fault
    \term{$n$}
    Faulting style
    \term{$\phi$}
    Stress shape ratio
    \term{$\Psi$}
    $\Shmax$ orientation
    \term{$\Omega(r')$}
    Radius-dependent Gaussian step function
    \term{$\rho$}
    Bulk density
    \term{$\mathcal{R}_0$}
    Maximum pre-stress ratio
    \term{$\mathcal{R}$}
    Relative pre-stress ratio
    \term{$S(t)$}
    Time-dependent smoothed step function
    \term{$\sigma_{1,2,3}$}
    Principal stress for: 1. Maximum, 2. Intermediate, 3. Minimum
    \term{$\sigma_n$}
    Normal stress
    \term{$\sigma_n'$}
    Effective normal stress
    \term{$t$}
    time
    \term{$T$}
    Nucleation time
    \term{$\theta$}
    State variable
    \term{$\theta_{SS}$}
    Steady-state state variable
    \term{$P_f$}
    Pore fluid pressure
    \term{$R$}
    Fracture size
    \term{$r'$}
    Radius from the hypocenter
    \term{$r_{nuc}$}
    Radius of initial nucleation
    \term{$S_v$}
    Overburden stress
    \term{$\Shmax$}
    Maximum horizontal stress
    \term{$\Shmin$}
    Minimum horizontal stress
    \term{$\Delta \tau$}
    Stress drop
    \term{$\Delta \tau_d$}
    Dynamic stress drop
    \term{$\tau$}
    Fault/fracture shear stress
    \term{$\tau_0$}
    Initial shear stress
    \term{$\tau_p$}
    Peak stress
    \term{$\tau_d$}
    Dynamic stress
    \term{$\mu$}
    Material rigidity
    \term{$\varphi$}
    Rake angle
    \term{$v_C$}
    Cascading speed
    \term{$v_{R,tot}$}
    Global rupture speed
    \term{$v_{R,ave}$}
    Average of the average rupture speed over all planes
    \term{$v_p$}
    $P$-wave velocity
    \term{$V_w$}
    Weakening slip velocity
    \term{$V_{ini}$}
    Initial slip rate
    \term{$V_0$}
    Reference slip rate
    \term{$V_{S30}$}
    $S$-wave speed for the first $30m$ depth
    \term{$\xi$}
    Initial pre-stress ratio inside the nucleation patch
    \term{$z$}
    Depth
\end{glossary}

\bibliography{agusample.bib}

%
%
%
%

 \setcounter{figure}{0}
 \renewcommand{\thefigure}{S\arabic{figure}}
%
%


%
%


\title{Supporting Information for "Rupture Dynamics of Cascading Earthquakes in a Multiscale Fracture Network"}

\authors{Kadek Hendrawan Palgunadi\affil{1}, Alice-Agnes Gabriel\affil{2,4}, Dmitry Garagash\affil{3}, Thomas Ulrich\affil{4}, Paul Martin Mai\affil{1}}

\affiliation{1}{Physical Science and Engineering, King Abdullah University of Science and Technology, Thuwal, Saudi Arabia}
\affiliation{2}{Institute of Geophysics and Planetary Physics, Scripps Institution of Oceanography, University of California, San Diego, CA, USA}
\affiliation{3}{Dalhousie University, Department Civil Resource Engineering, Halifax, Canada}
\affiliation{4}{Department of Earth and Environmental Sciences, Geophysics, Ludwig-Maximilians-Universit\"{a}t M\"{u}nchen, Munich, Germany}


%
%

\noindent\rule{\textwidth}{1pt}

\section*{Introduction}

The supplementary material includes figures and videos that provide detailed representations of rupture processes in various scenarios described in the main paper, including variations in the orientation of $S{\!}H_\mathrm{max}$ ($\Psi$) and fluid injection scenarios. Figures present snapshots focusing on the physical processes involved in cascading rupture. Videos illustrate the space-time evolution of the rupture process from two different perspectives, in an ``exploded" view and in the original constellation of the fault network . Supplementary videos can be accessed at the following link: \url{https://bit.ly/FractureNetworkVideoSupps}.

\section*{Contents of this file}
\begin{figure}[htp!]
    \centering
    \includegraphics[width=0.7\textwidth]{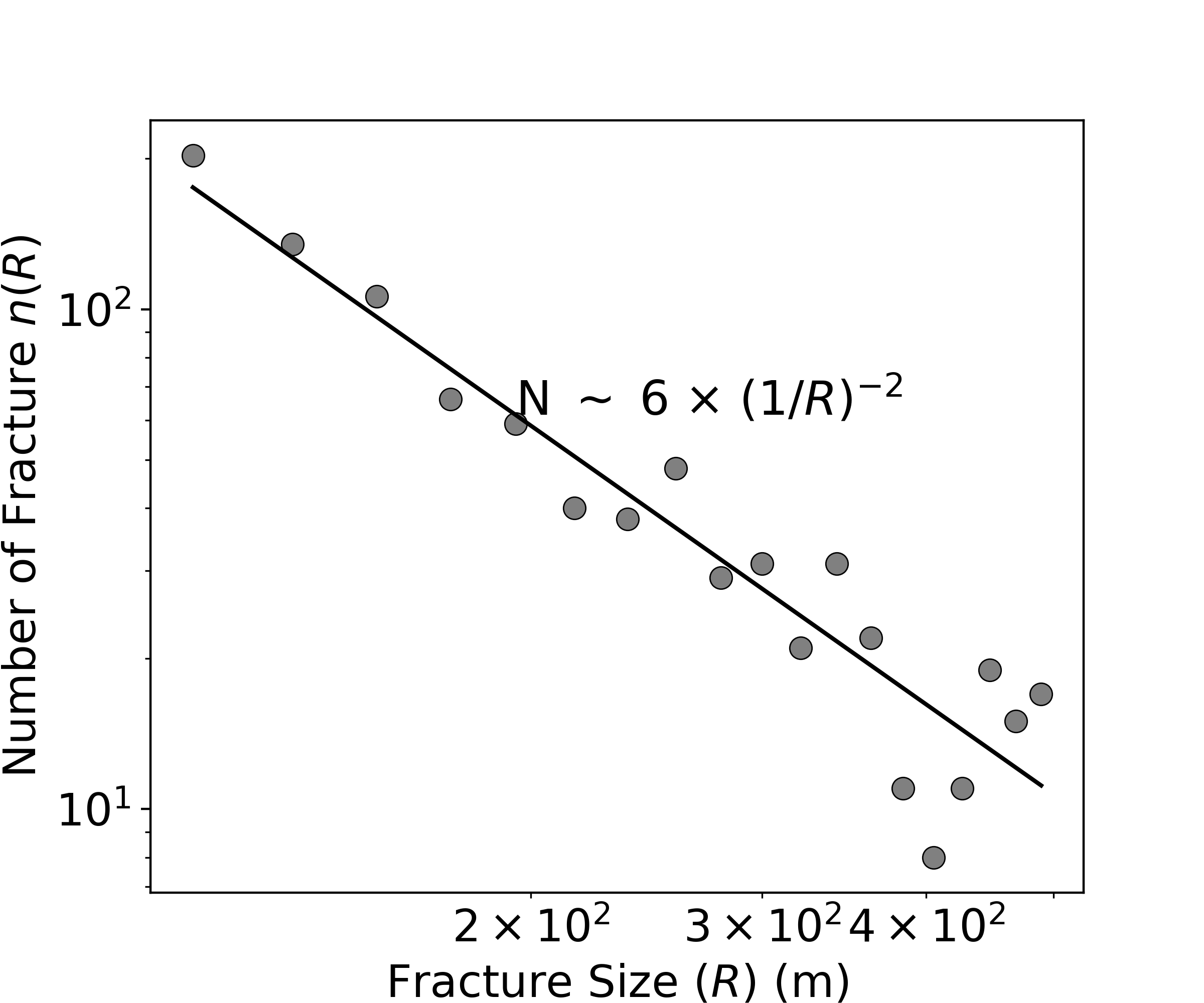}
    \caption{Fracture size distribution considered in this study, following a power-law decay with exponential decay 2 as observed in \cite{lavoine2019density}. The black line depicts the regression line to an exponential fit, $N$ denotes the number of fractures and $R$ is the fracture size. The grey dots mark histogram values at 22~m intervals.} 
    \label{fig:fracture_size}
\end{figure}

\begin{sidewaysfigure}
    \centering
    \includegraphics[width=0.99\textwidth]{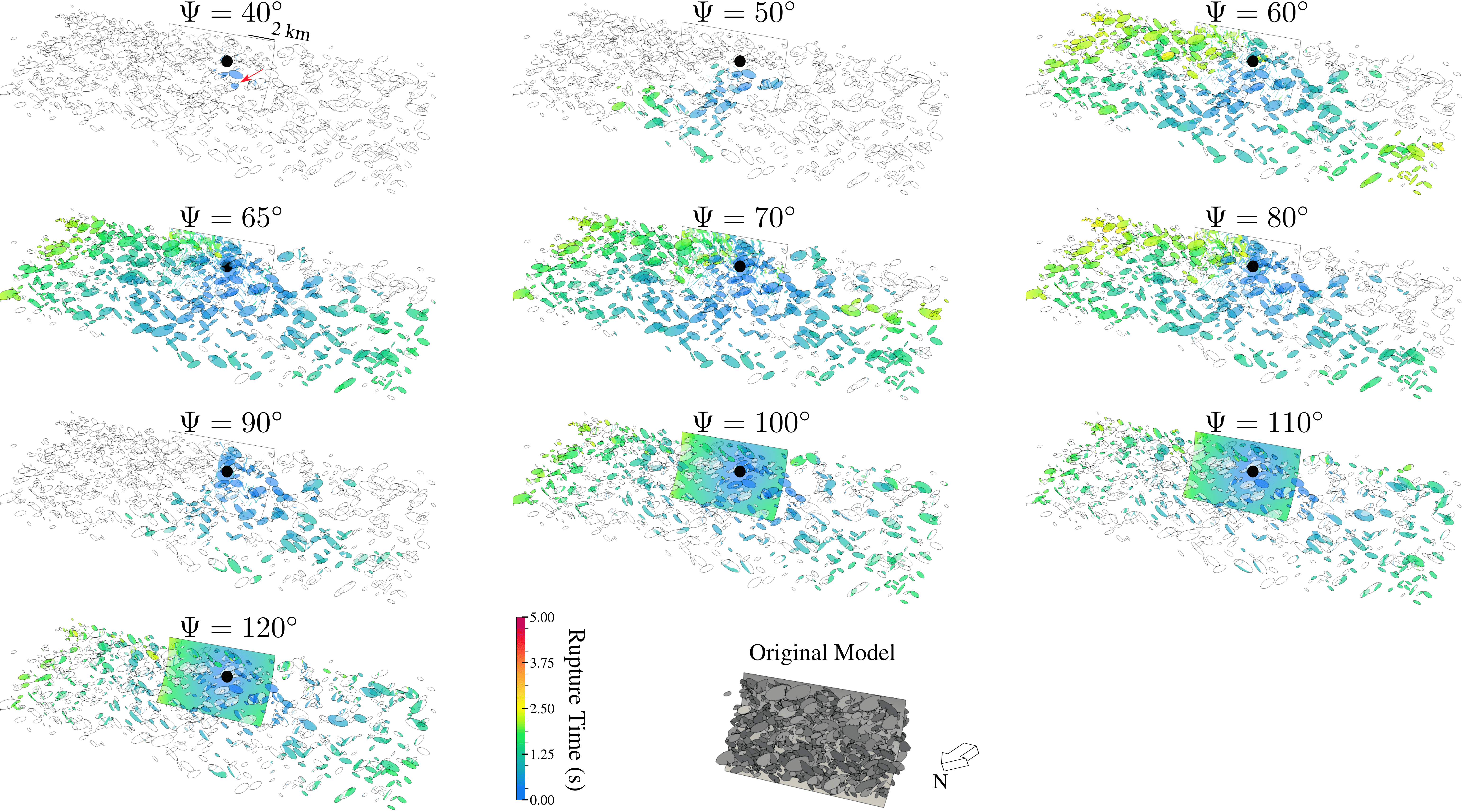}
    \caption{Rupture times for different orientations of $S{\!}H_\mathrm{max}$ ($\Psi$).}
    \label{fig:S1}
\end{sidewaysfigure}

\begin{sidewaysfigure}
    \centering
    \includegraphics[width=0.99\textwidth]{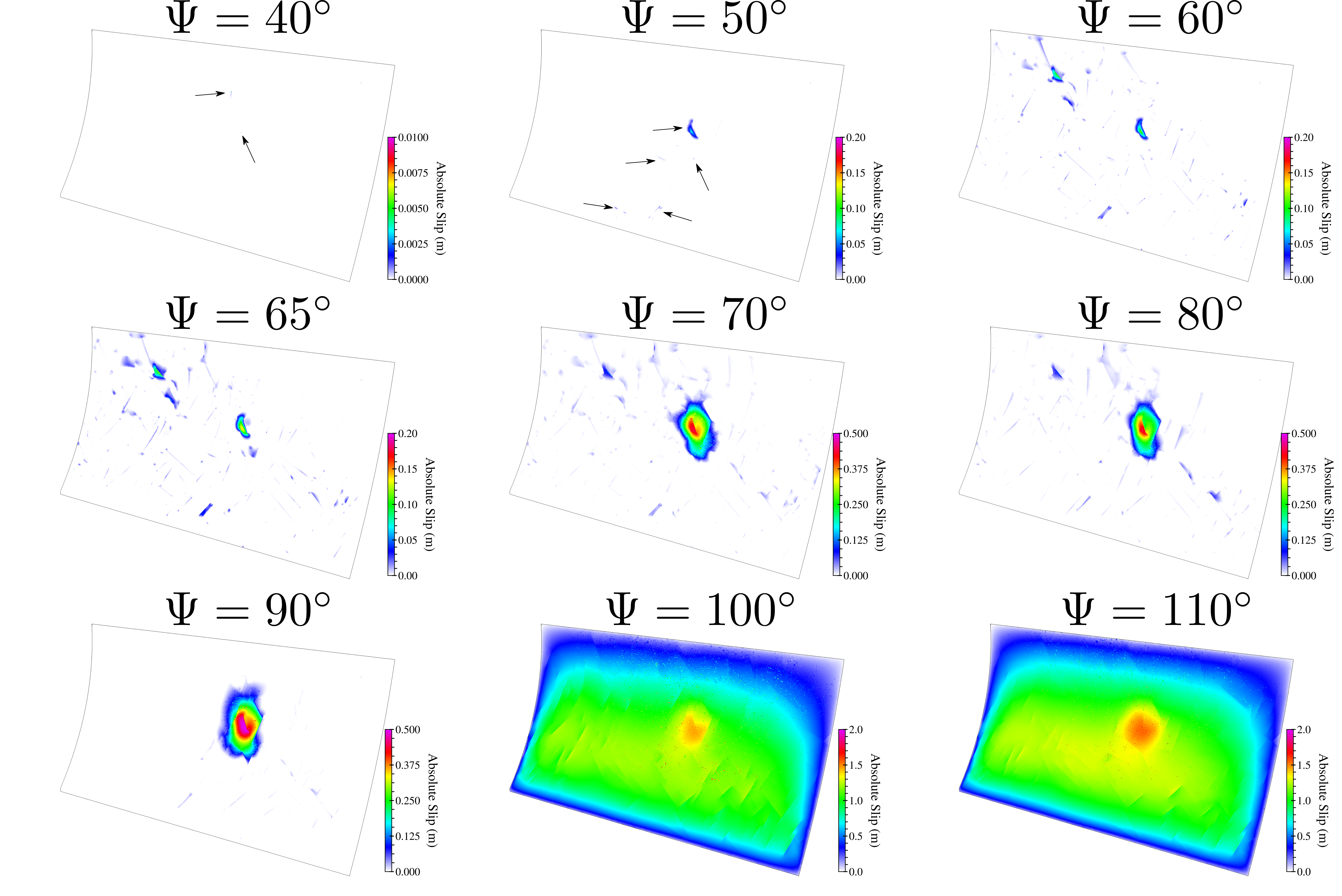}
    \caption{Final slip on the main fault for different orientations of $S{\!}H_\mathrm{max}$ ($\Psi$). The arrows in $\Psi = 50^\circ$ indicate a small amount of slip on the main fault. Please note the colorboar changes across all panels, reflecting smaller values of maximum slip, particularly for $\Psi < 70^\circ$, indicating less slip on the main fault previously not visible in the colorbar of the main manuscript.}
    \label{fig:S_main_fault_slip}
\end{sidewaysfigure}

\begin{sidewaysfigure}
    \centering
    \includegraphics[width=0.99\textwidth]{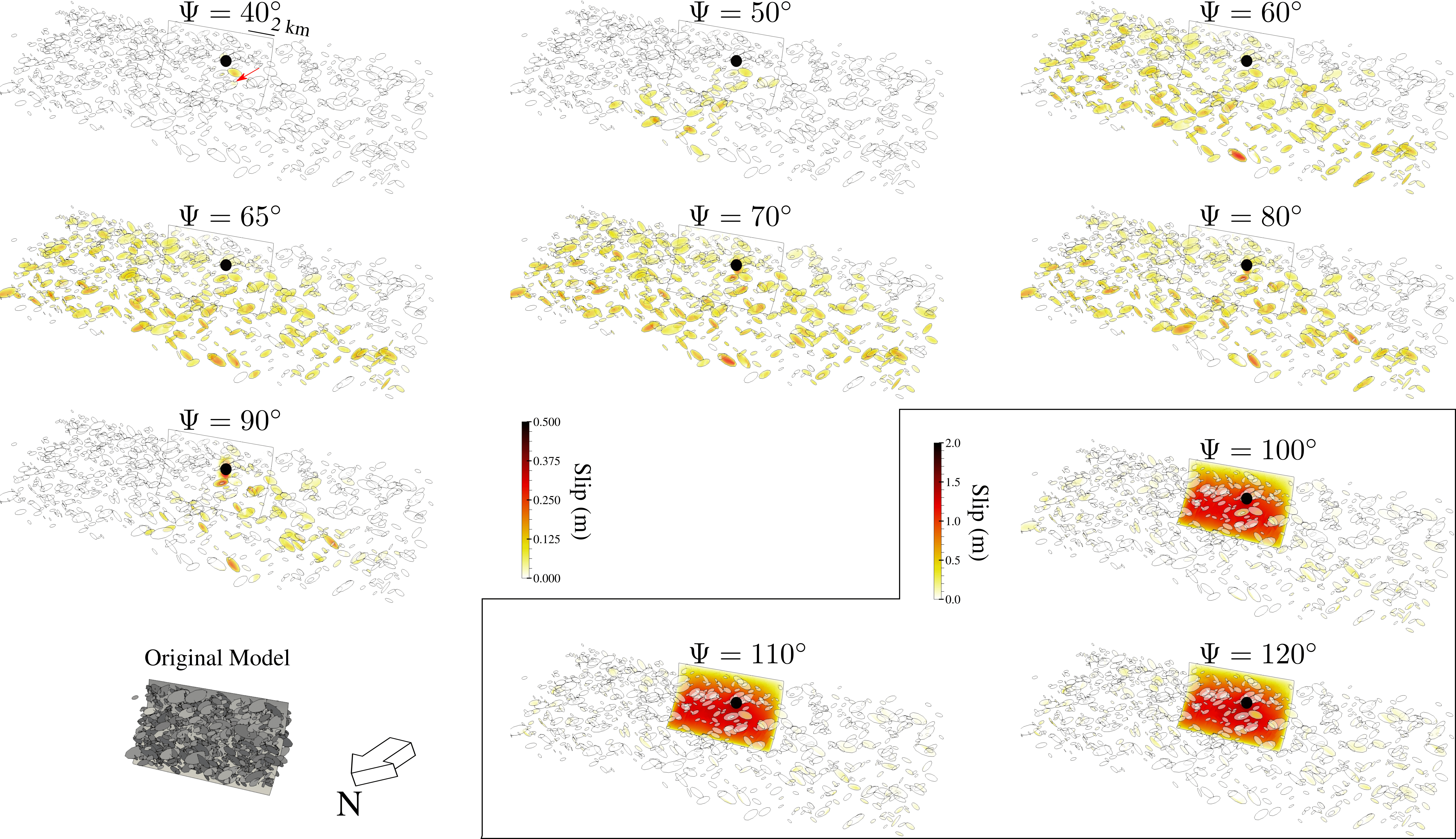}
    \caption{Final fault slip for different orientations of $S{\!}H_\mathrm{max}$ ($\Psi$).}
    \label{fig:S2}
\end{sidewaysfigure}

\begin{figure}
    \centering
    \includegraphics[width=0.95\textwidth]{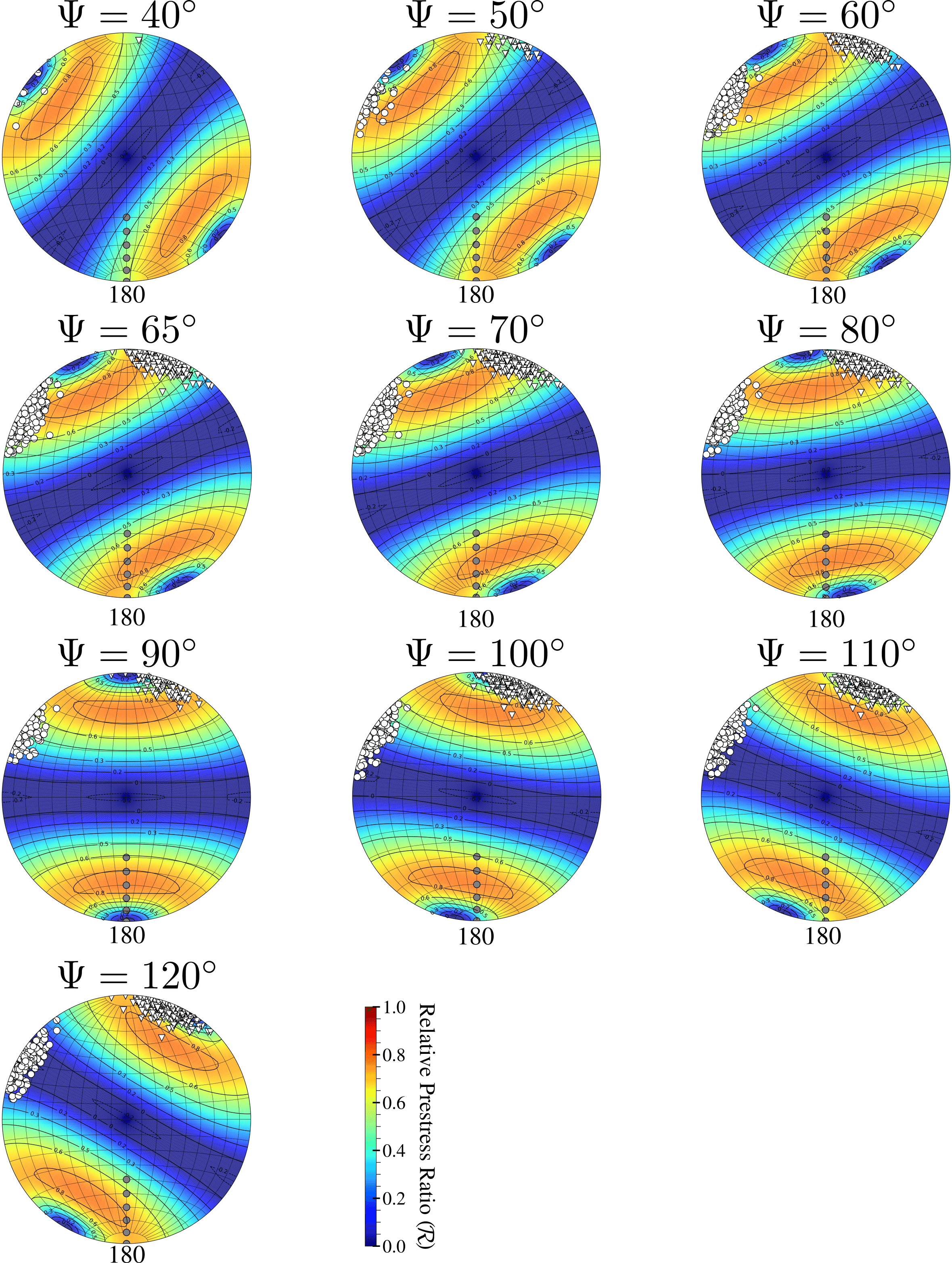}
    \caption{Stereonet plot displaying slipped fractures superimposed on the relative prestress ratio ($\mathcal{R}$) for different $\Psi$ values, shown in a lower hemisphere projection. White triangles represent individual slipped fractures from fracture family 1 (strike $120^\circ$), while white dots indicate fractures from family 2 (strike $20^\circ$). Grey dots represent the listric fault.}
    \label{fig:S3}
\end{figure}

\begin{figure}
    \centering
    \includegraphics[width=0.95\textwidth]{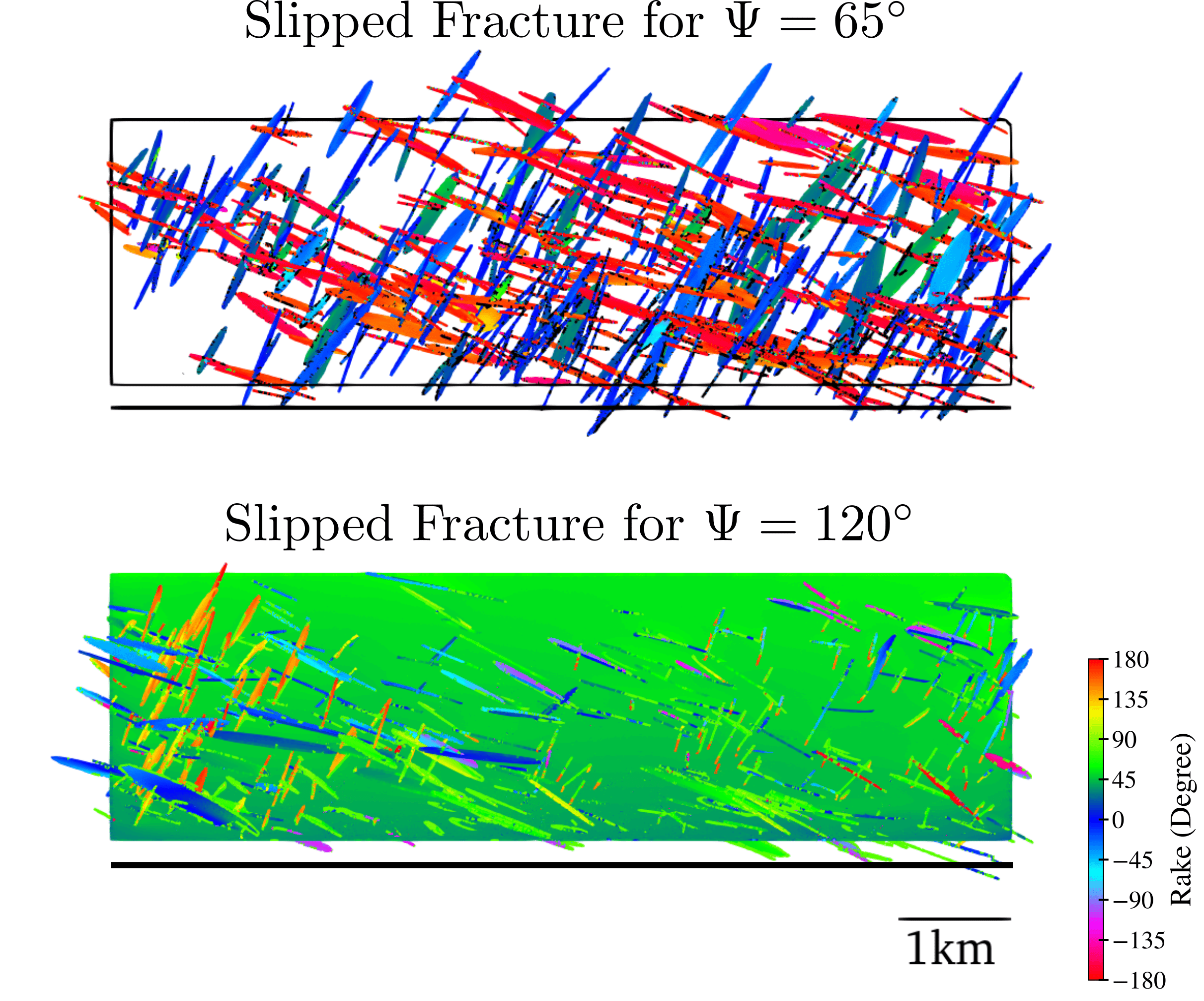}
    \caption{Map view illustrating the slipped fractures and the main fault for two different orientations: $\Psi = 65^\circ$ (top panel) and $\Psi= 120^\circ$ (lower panel). The solid black line represents the top of the main fault. Red colors indicate left-lateral, while blue color denotes right-lateral strike-slip faulting.}
    \label{fig:S3_1}
\end{figure}

\begin{figure}
    \centering
    \includegraphics[width=0.95\textwidth]{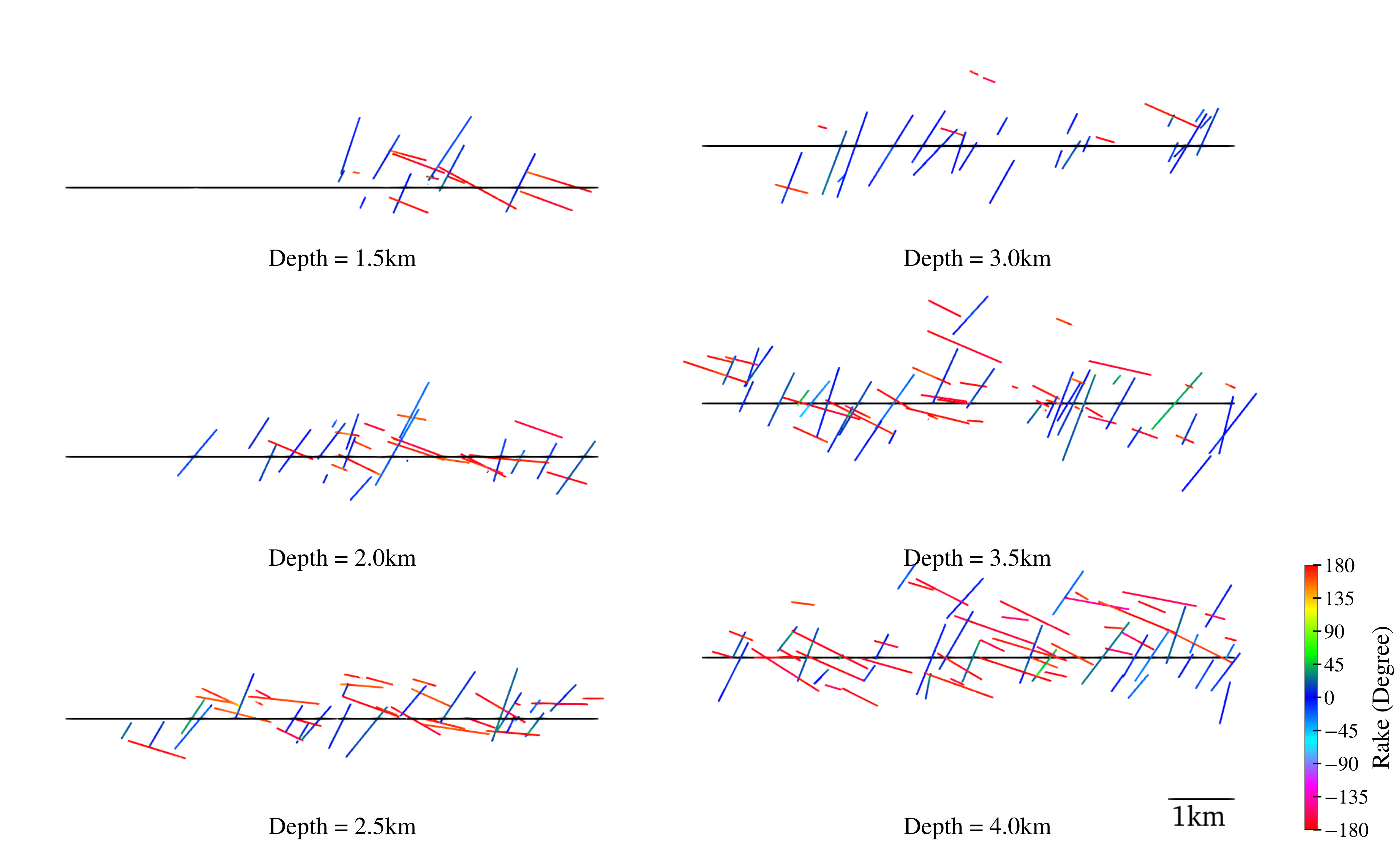}
    \caption{Depth slices (at 0.5 km intervals) showing the distribution of slipped fractures  for $\Psi = 65^\circ$. Red colors indicate left-lateral, while blue color denotes right-lateral strike-slip faulting. The solid black line corresponds to the unslipped main fault.}
    \label{fig:S3_2}
\end{figure}

\begin{figure}
    \centering
    \includegraphics[width=0.95\textwidth]{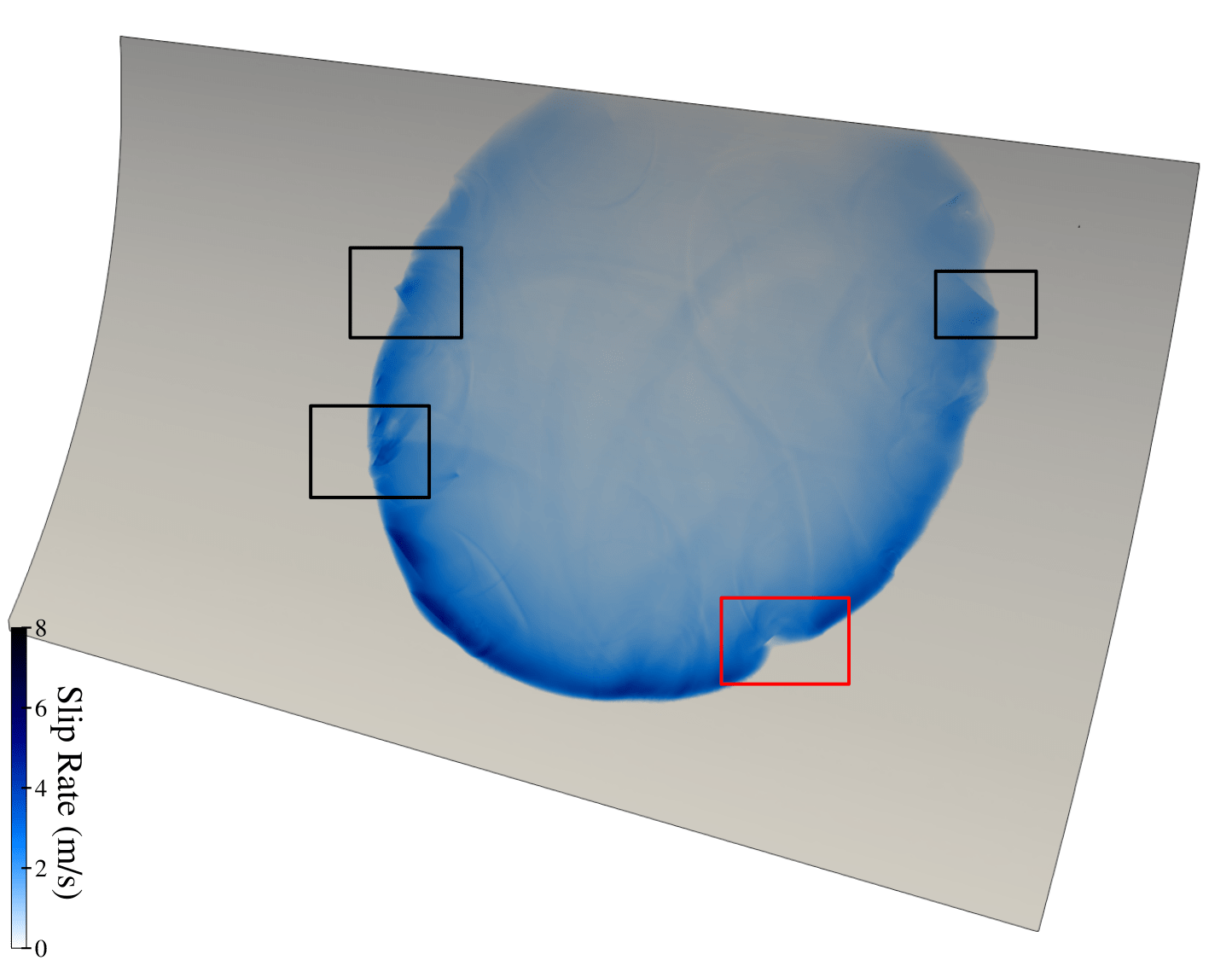}
    \caption{This snapshot focuses on the rupture front of the main fault in scenario $\Psi=120^\circ$, without displaying the small-scale fracture network. Three black rectangles highlight areas with faster rupture speeds. The red rectangle indicates an example of unfavorably oriented fractures that act as barriers to the propagation of the rupture.}
    \label{fig:S5}
\end{figure}

\begin{figure}
    \centering
    \includegraphics[width=0.95\textwidth]{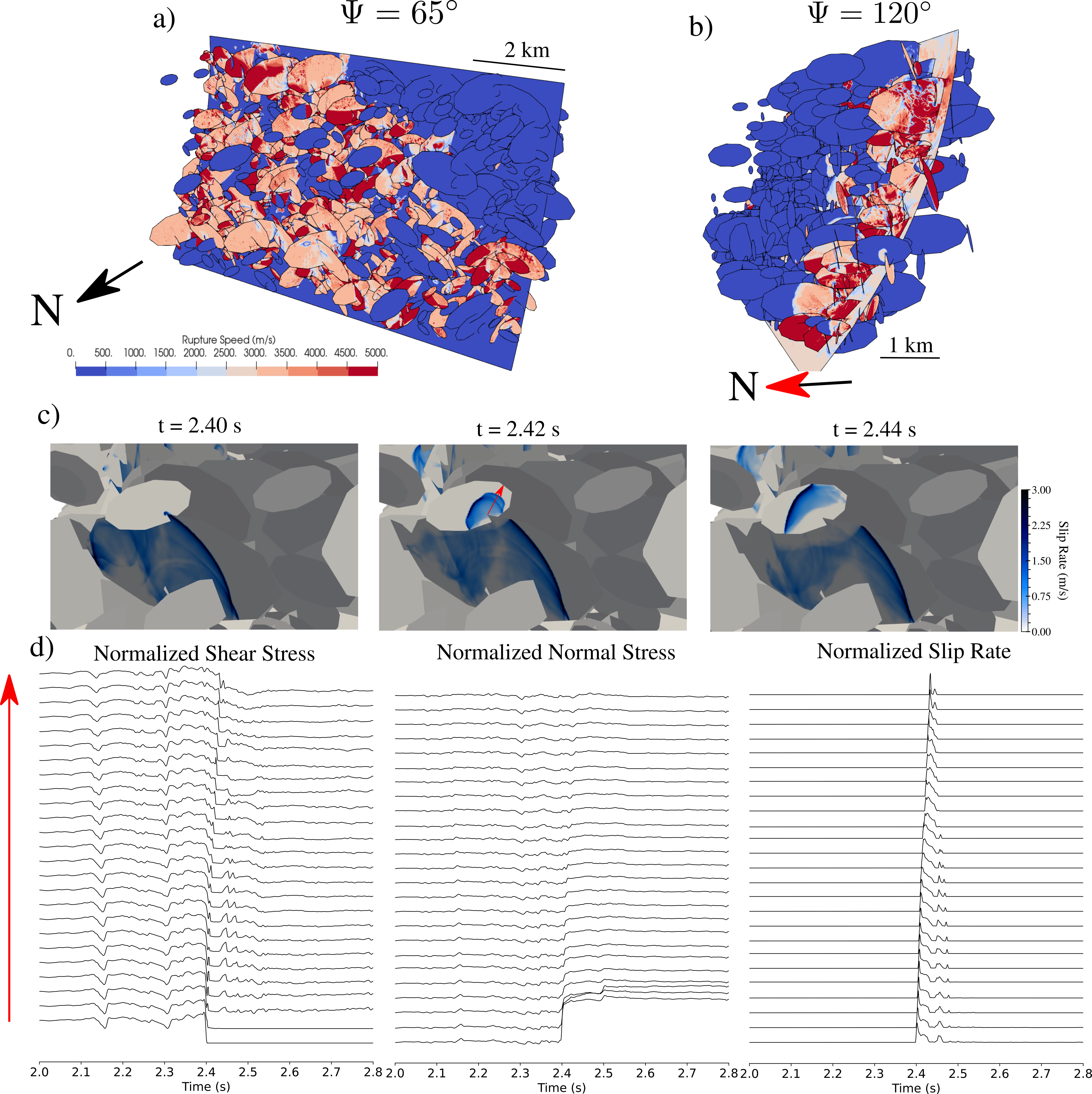}
    \caption{Supershear rupture speed for (a) pure rupture cascade ($\Psi = 65^\circ$) and (b) rupture with off-fault fracture slip ($\Psi = 120^\circ$). The color range for rupture speed values is saturated at 5000~m/s. The arrow is pointing North. c) Slip velocity evolution of one of the supershear rupture speed transitions on a small fracture in (a). d) Normalized shear stress, normal stress, and slip rate on a small fracture in (c).  Shear stress, normal stress, and  slip rate are normalized at 10~MPa, 20~MPa, and 3~m/s, respectively. The red arrow indicates an increasing distance from the nucleation point as shown in (c) at $t=2.42$~s.}
    \label{fig:S_ruptureSpeed}
\end{figure}

\begin{sidewaysfigure}
    \centering
    \includegraphics[width=0.99\textwidth]{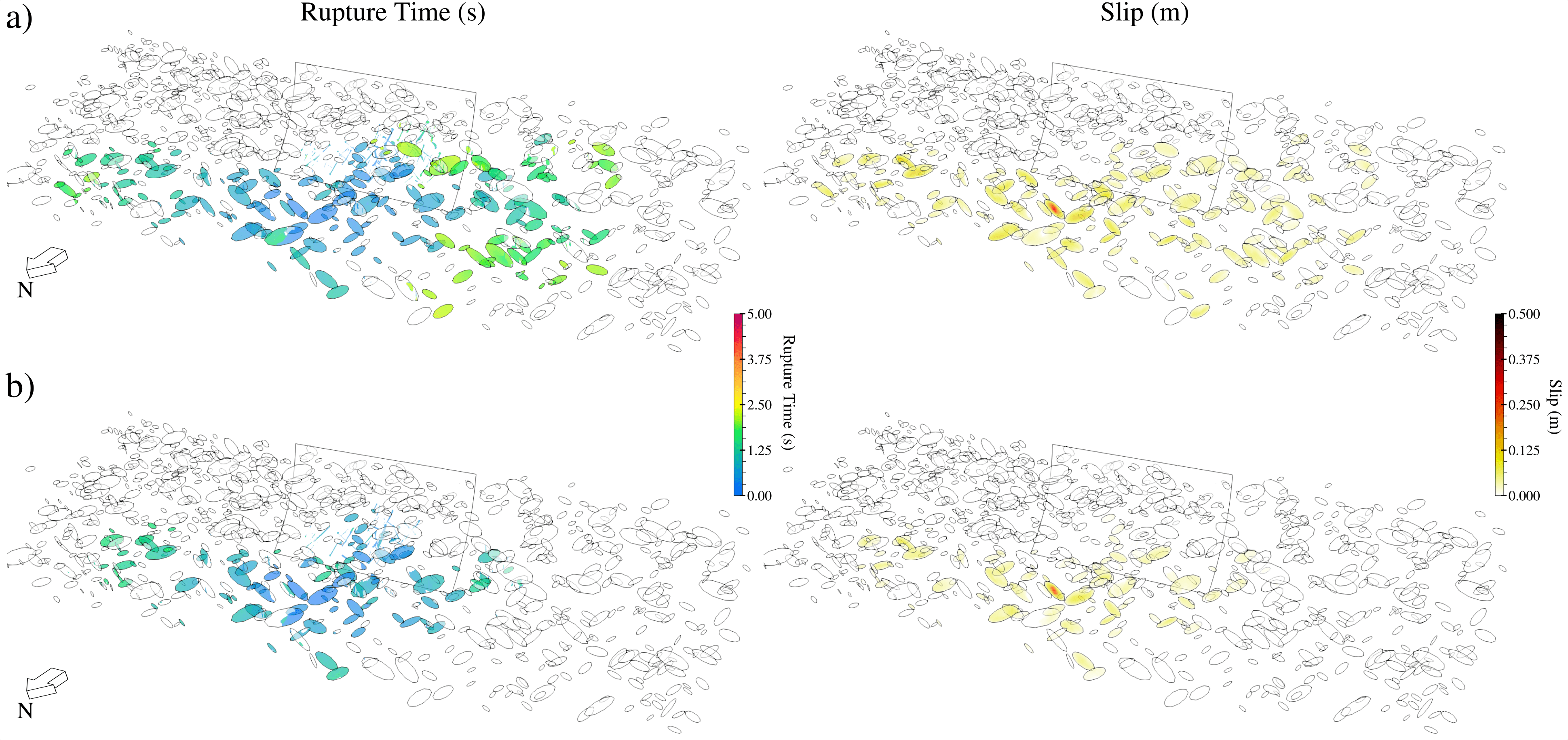}
    \caption{Exploded view of rupture time evolution and final slip of Scenario 2 for two examples: a) $\gamma = 0.6$ and b) $\gamma=0.7$.}
    \label{fig:S6}
\end{sidewaysfigure}

\begin{figure}
    \centering
    \includegraphics[width=0.99\textwidth]{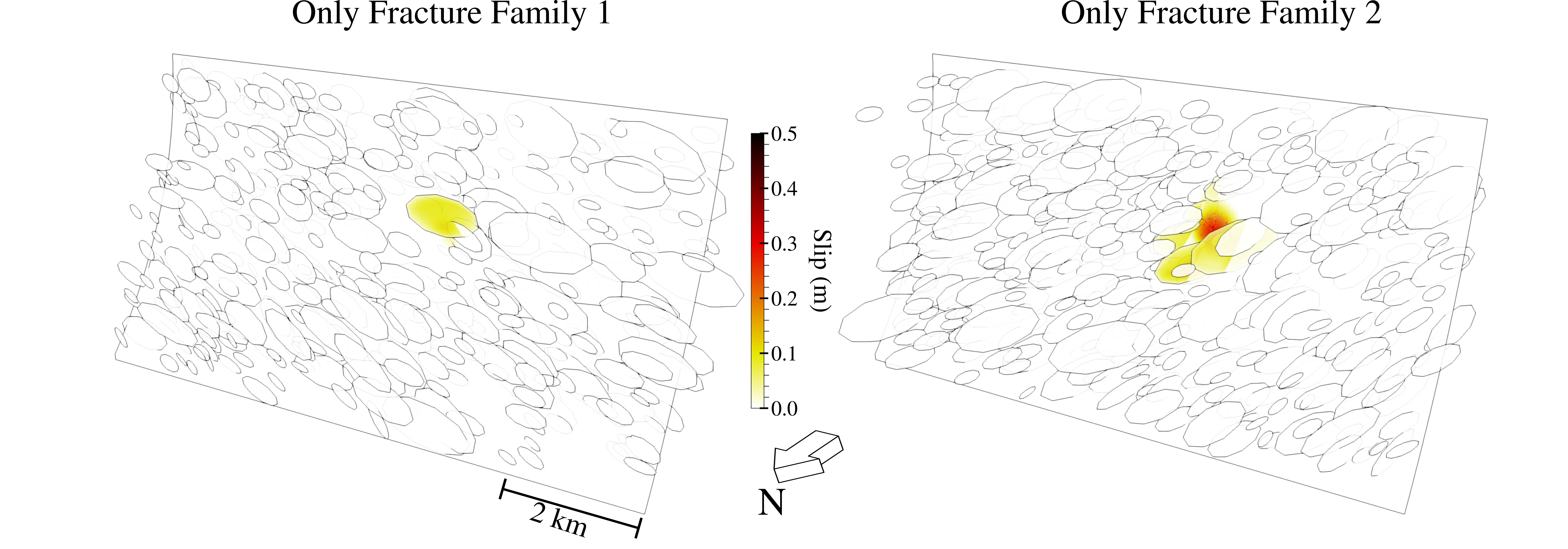}
    \caption{Final slip distribution considering only one fracture family. All friction parameters, initial stress (\textit{Case 2}, $\Psi=65^\circ$), and nucleation procedures remain the same. One fracture family is removed in each simulation. The left panel displays the slip distribution on fracture family 1 (N120E$\pm10^\circ$), indicating the absence of a cascade. The right panel illustrates the absence of a cascade when only considering fracture family 2 (N20E$\pm10^\circ$).}
    \label{fig:S_oneFamily}
\end{figure}

\begin{figure}
    \centering
    \includegraphics[width=1\textwidth]{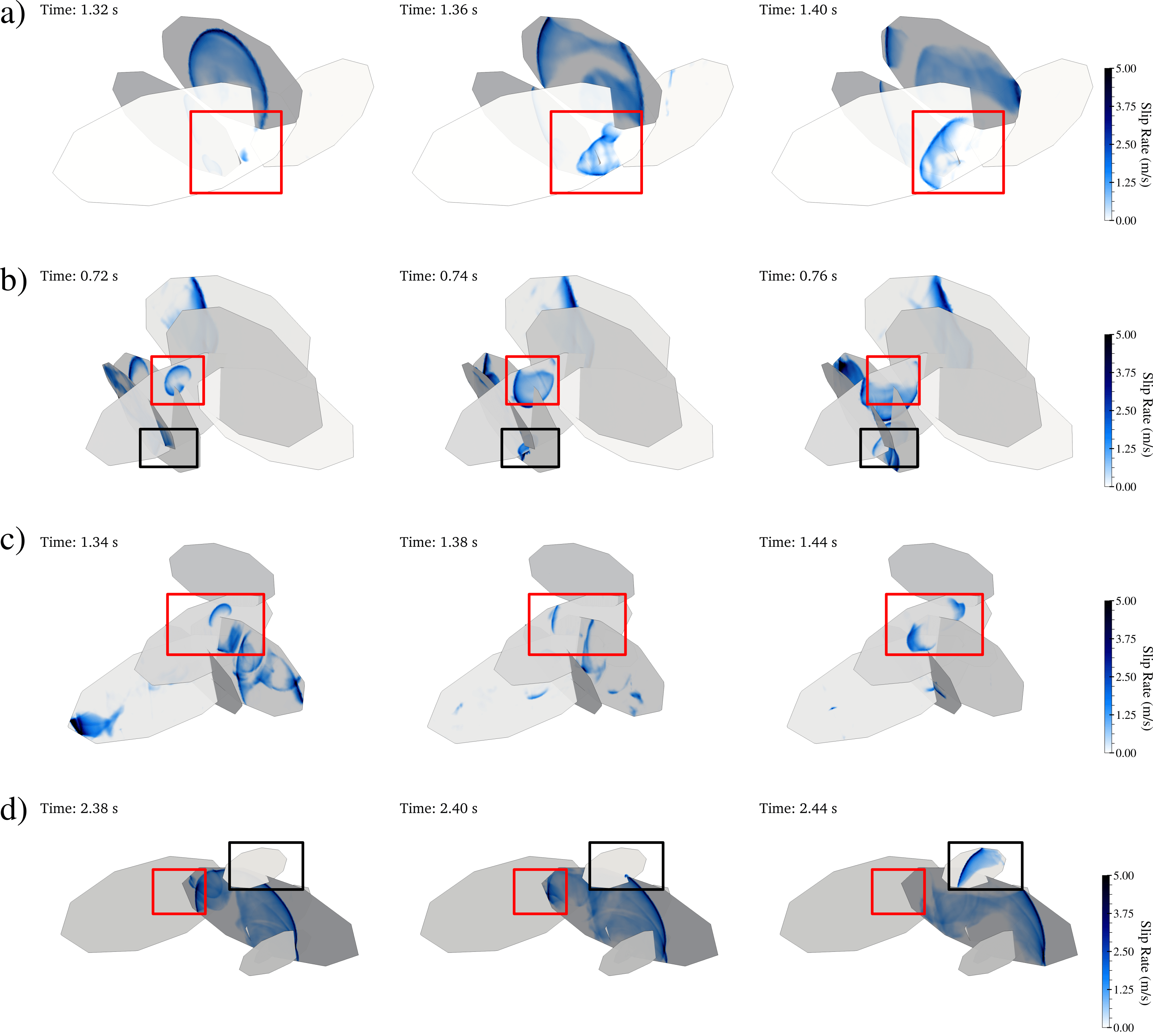}
    \caption{Examples of four subsets of fracture-fracture interaction scenarios. Slip rate is represented by the blue color, ranging from 0 -- 5 m/s (maximum value reaches is 6 m/s). a) Illustration of rupture nucleation in two locations on a fracture. The red rectangle highlights the two nucleation points. b) Consecutive rupture nucleation in two or more positions on a fracture. The first nucleation is emphasized within the red rectangle, while the second nucleation is highlighted by the black rectangle. c) Repeated nucleation at a single point on a fracture. The red rectangle indicates the nucleation area at t = 1.34~s and t = 1.42~s (the figure shows rupture at t = 1.44~s after the second nucleation occurs). d) Rupture on a fracture fails to transfer sufficient dynamic stress to a nearby fracture ($< 80$~m in the distance, red rectangle). However, it still manages to propagate dynamically through rupture branching to an interconnected fracture (black rectangle).}
    \label{fig:S4}
\end{figure}

\begin{figure}
    \centering
    \includegraphics[width=0.95\textwidth]{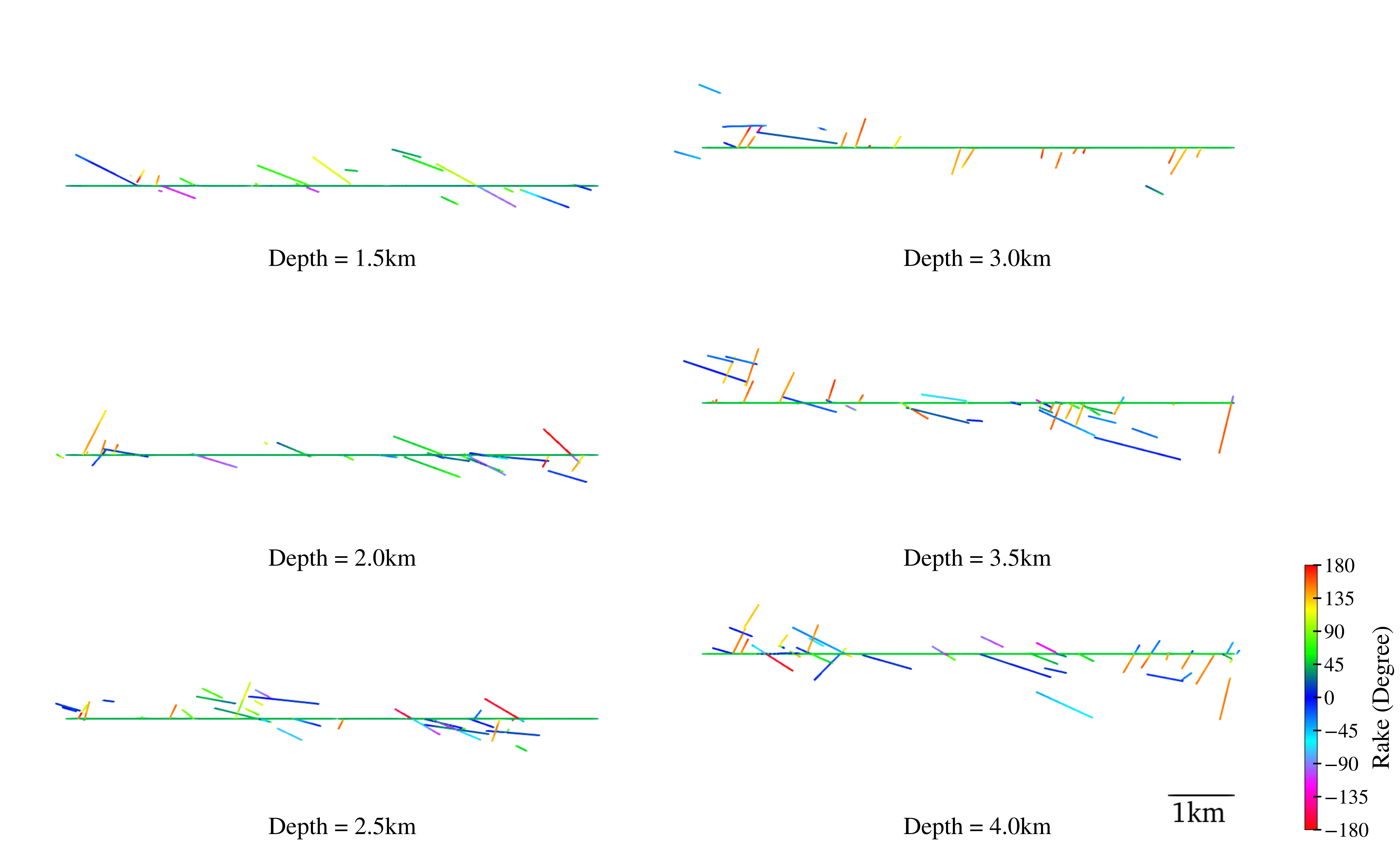}
    \caption{Depth slices (at 0.5 km intervals) showing the distribution of slipped fractures for $\Psi = 120^\circ$. The red color indicates strike-slip left-lateral faulting, while the blue color represents strike-slip right-lateral faulting.}
    \label{fig:S3_3}
\end{figure}

\begin{enumerate}
    \item Figure \ref{fig:fracture_size}: Fracture size distribution.
    \item Figure \ref{fig:S1}: Rupture time of different orientation of $S{\!}H_\mathrm{max}$ ($\Psi$).
    \item Figure \ref{fig:S_main_fault_slip}: Slip only on the main fault.
    \item Figure \ref{fig:S2}: Slip of different orientation of $S{\!}H_\mathrm{max}$ ($\Psi$).
    \item Figure \ref{fig:S3}: Stereonet plot of slipped fractures overlain by relative prestress ratio ($\mathcal{R}$) for different $\Psi$ shown in lower hemisphere projection. 
    \item Figure \ref{fig:S3_1}: Map view of the slipped fractures and the main fault for $\Psi = 65^\circ$ (top panel) and $\Psi= 120^\circ$ (lower panel). 
    \item Figure \ref{fig:S3_2}: Depth slice of the slipped fractures every 0.5~km for $\Psi = 65^\circ$. 
    \item Figure \ref{fig:S5}: Snapshot focuses on the rupture front of the main fault without showing fractures for scenario $\Psi=120^\circ$.
    \item Figure \ref{fig:S_ruptureSpeed}: Supershear rupture speed.
    \item Figure \ref{fig:S6}: Exploded view of rupture time evolution and final slip of Scenario 2 for two examples.
    \item Figure \ref{fig:S_oneFamily}: Slip distribution if only considering one fracture family.
    \item Figure \ref{fig:S4}: Overview examples from 4 subsets of fracture-fracture interaction scenarios.
    \item Figure \ref{fig:S3_3}: Depth slice of the slipped fractures every 0.5~m for $\Psi = 120^\circ$. 
\end{enumerate}


\section*{Additional Supporting Information (Files uploaded separately)}
The separately uploaded files contain movies of different cases and scenarios explained in the main paper. The files comprise 26 movies showing the spatiotemporal evolution of slip rate.

\begin{enumerate}
    \item Case $\Psi = 40^\circ$:
    \begin{itemize}
        \item Movie S1a (\texttt{SR\_E40}): Slip rate (in [m/s]) presented in exploded view.
        \item Movie S1b (\texttt{SR\_N40}): Slip rate (in [m/s]) presented in original view.
    \end{itemize}
    
    \item Case $\Psi = 50^\circ$:
    \begin{itemize}
        \item Movie S2a (\texttt{SR\_E50}): Slip rate (in [m/s]) presented in exploded view.
        \item Movie S2b (\texttt{SR\_N50}): Slip rate (in [m/s]) presented in original view.
    \end{itemize}
    
    \item Case $\Psi = 60^\circ$:
    \begin{itemize}
        \item Movie S3a (\texttt{SR\_E60}): Slip rate (in [m/s]) presented in exploded view.
        \item Movie S3b (\texttt{SR\_N60}): Slip rate (in [m/s]) presented in original view.
    \end{itemize}
    
    \item Case $\Psi = 65^\circ$:
    \begin{itemize}
        \item Movie S4a (\texttt{SR\_E65}): Slip rate (in [m/s]) presented in exploded view.
        \item Movie S4b (\texttt{SR\_N65}): Slip rate (in [m/s]) presented in original view.
    \end{itemize}
    
    \item Case $\Psi = 70^\circ$:
    \begin{itemize}
        \item Movie S5a (\texttt{SR\_E70}): Slip rate (in [m/s]) presented in exploded view.
        \item Movie S5b (\texttt{SR\_N70}): Slip rate (in [m/s]) presented in original view.
    \end{itemize}
    
    \item Case $\Psi = 80^\circ$:
    \begin{itemize}
        \item Movie S6a (\texttt{SR\_E80}): Slip rate (in [m/s]) presented in exploded view.
        \item Movie S6b (\texttt{SR\_N80}): Slip rate (in [m/s]) presented in original view.
    \end{itemize}
    
    \item Case $\Psi = 90^\circ$:
    \begin{itemize}
        \item Movie S7a (\texttt{SR\_E90}): Slip rate (in [m/s]) presented in exploded view.
        \item Movie S7b (\texttt{SR\_N90}): Slip rate (in [m/s]) presented in original view.
    \end{itemize}
    
    \item Case $\Psi = 100^\circ$:
    \begin{itemize}
        \item Movie S8a (\texttt{SR\_E100}): Slip rate (in [m/s]) presented in exploded view.
        \item Movie S8b (\texttt{SR\_N100}): Slip rate (in [m/s]) presented in original view.
    \end{itemize}
    
    \item Case $\Psi = 110^\circ$:
    \begin{itemize}
        \item Movie S9a (\texttt{SR\_E110}): Slip rate (in [m/s]) presented in exploded view.
        \item Movie S9b (\texttt{SR\_N110}): Slip rate (in [m/s]) presented in original view.
    \end{itemize}
    
    \item Case $\Psi = 120^\circ$:
    \begin{itemize}
        \item Movie S10a (\texttt{SR\_E120}): Slip rate (in [m/s]) presented in exploded view.
        \item Movie S10b (\texttt{SR\_N120}): Slip rate (in [m/s]) presented in original view.
    \end{itemize}
    
    \item \textit{Case 4}, rupture nucleation on a fracture in damage zone at distance $1km$ from the main fault:
    \begin{itemize}
        \item Movie S11a (\texttt{SR\_FarSourceE65}): Slip rate (in [m/s]) presented in exploded view.
        \item Movie S11b (\texttt{SR\_FarSourceN65}): Slip rate (in [m/s]) presented in original view.
    \end{itemize} 
    
    \item \textit{Case 5}, similar to \textit{Case 4} with varying fluid pressure ratio ($\gamma$):
    \begin{itemize}
        \item Movie S12 (\texttt{SR\_Gamma05E65}): Slip rate (in [m/s]) presented in exploded view for $\gamma=0.5$.
        \item Movie S13 (\texttt{SR\_Gamma06E65}): Slip rate (in [m/s]) presented in exploded view for $\gamma=0.6$.
        \item Movie S14 (\texttt{SR\_Gamma07E65}): Slip rate (in [m/s]) presented in exploded view for $\gamma=0.7$.
        \item Movie S15 (\texttt{SR\_Gamma08E65}): Slip rate (in [m/s]) presented in exploded view for $\gamma=0.8$.
    \end{itemize}
\end{enumerate}

\clearpage

\section*{Figures:} 

\end{document}